\documentclass[a4paper,11pt]{article}
\usepackage{jheppub} 
\usepackage{lineno}
\usepackage{float}
\usepackage{slashed}
\usepackage{xfrac,bigints}
\usepackage{physics}
\usepackage{subfigure}
\usepackage{booktabs} 
\usepackage{mathtools}
\newcommand{\ki}[1]{{\color[rgb]{0,0,1}{[K.I.: #1]}}}

\title{Quantum Simulation of Fermions in $AdS_2$ Black Hole: Chirality, Entanglement, and Spectral Crossovers}

\author[a,b]{Kazuki Ikeda}
\emailAdd{kazuki.ikeda@umb.edu}
\affiliation[a]{Department of Physics, University of Massachusetts Boston, Boston, MA 02125, USA}
\affiliation[b]{Center for Nuclear Theory, Department of Physics and Astronomy, Stony Brook University, Stony Brook, NY 11794-3800, USA}
\author[c]{Yaron Oz}
\affiliation[c]{School of Physics and Astronomy, Tel-Aviv University,
Tel-Aviv 69978, Israel}
\emailAdd{yaronoz@tauex.tau.ac.il}

\abstract{
We consider free Dirac fermions on a discretized $AdS_2$ black hole background, and analyze how curved space redshift, horizons, and the spin connection induced chiral gravitational effect shape spectral, transport, and scrambling phenomena. The system is discretized via staggered fermions followed by the Jordan–Wigner transform to encode the model in qubit degrees of freedom, whose Hamiltonian carries site dependent warp factors and bond chirality terms encoding the redshift and spin connection effects.
We calculate the ground state and first excited states energies, their local charge profiles, and their half‐chain entanglement entropies, showing how redshift and chirality affect the transition from criticality to a gapped regime. Probing operator growth via out‐of‐time‐order correlators, we find that horizons and the chiral coupling accelerate scrambling, yet remain within a non‐chaotic regime. Finally, we map out an integrable to ergodic crossover via level‐spacing statistics and Brody fits, and introduce on‐site disorder to drive a many body localization transition. 
}

\begin{document}
\maketitle
\flushbottom

\section{Introduction}
Understanding how quantum matter behaves in curved spacetime lies at the heart of many frontier problems in theoretical physics, from holographic dualities in quantum gravity \cite{Maldacena1998,GubserKlebanovPolyakov1998,Witten1998, AharonyGubserMaldacenaOoguriOz2000} to quantum simulation of gravitational phenomena in tabletop platforms \cite{Steinhauer2016}.  In particular, the dynamics of fermionic degrees of freedom in an $AdS$ black hole background encapsulate essential features of near-horizon physics, including gravitational redshift, spin connection effects, and quantum anomalies. Fermions in $AdS$ black hole backgrounds, particularly in the context of holographic non‐Fermi liquids, have been studied in \cite{FaulknerLiuMcGreevyVegh2011,IqbalLiu2009,CubrovicZaanenSchalm2009}.
While continuum analyses, such as Jackiw–Teitelboim gravity \cite{MertensTuriaci2022} and SYK duality \cite{Kitaev2015}, reveal deep insights into maximal chaos and boundary reparameterization modes, a complementary, fully controllable lattice model can act as a qubit testbed for digital or analog quantum simulation.
Such a microscopic model allows to dissect the interplay of redshift, chirality, and disorder,
as well as a bridge between free fermion integrability and emergent quantum chaotic signatures.

In this work, we construct a staggered fermion discretization of the two-dimensional Dirac theory \cite{Kogut:1974ag,Susskind:1976jm}
on an $AdS_2$ black hole geometry \cite{Teitelboim1983,Jackiw1984}, and perform a Jordan–Wigner (JW) transformation \cite{Jordan:1928wi} to map the model onto qubit degrees of freedom with site‐dependent hopping amplitudes and bond‐chirality operators.  The resulting qubit Hamiltonian encodes
the warp factor weights from the redshift function, 
a spin connection induced chirality via a two‐site antisymmetric hopping (gravitational Chern–Simons \cite{DeserJackiwTempleton1982,DeserJackiwTempleton1982Annals} analogue), and a chemical potential filling that populates redshifted energy levels.

We derive closed‐form single‐particle dispersion relations and compute the ground state 
and first excited state energies, their local charge profiles, and their
half‐chain entanglement entropies, showing the effects of warp factor and chirality on the critical to gapped transition \cite{Kitaev_2001}.
We reveal the existence of a current induced by the spatial curvature, reminiscent of the chiral vortical effect \cite{PhysRevLett.106.062301,PhysRevD.98.096011} and the gravitational spin Hall effect \cite{GosselinBérardMohrbach2007}, where the curved geometry sources an equilibrium spin current.
We analyze the operator dynamics, compute the out‐of‐time‐order correlators \cite{LarkinOvchinnikov1969,MaldacenaShenkerStanford2016}, and demonstrate that black hole horizons and spin connection couplings accelerate scrambling, although within a quadratic (integrable) framework without exponential Lyapunov growth.
We analyze level‐spacing ratios \cite{OganesyanHuse2007} and fit Brody distributions \cite{Brody1973} to reveal a crossover from Poisson‐like integrability toward Wigner–Dyson–style level repulsion as the horizon grows, yet never reaching full chaos in the free model.
Finally, by introducing on‐site random fields, we track the decay of the Néel‐state imbalance to identify an ergodic to many body localized crossover
\cite{AbaninAltmanBlochSerbyn2019,NandkishoreHuse2015}.  Intrinsic warp‐factor inhomogeneity from a large  supplements this localization, effectively lowering the disorder threshold.

The paper is organized as follows.
In Section \ref{sec:Curved-Spaces} we briefly review the continuum framework of fermions in curved space-time, staggered fermions and their qubit representation.
In Section \ref{sec:BH} we consider fermions in $AdS_2$ black hole background,
the staggered fermion discretization, JW mapping to qubits, and provide derivations of dispersion, energy, charge, and gap formulas.
In Section \ref{sec:simulation} we analyze in detail the energy, local charge distribution and entanglement entropy of the system.
In Section \ref{sec:CGEOTOC} we investigate operator scrambling, spectral statistics ($r$‑statistic and Brody fits), and disorder‐induced localization.
Section \ref{sec:discussion} concludes with a discussion of quantum simulation prospects and extensions to interacting theories.

\section{\label{sec:Curved-Spaces}Fermions in Curved Spaces and their Qubit Representation}

\subsection{Fermions in Two-dimensional Curved Space}

We will work in two-dimensional curved space-time with metric $g_{\mu\nu}$, related to 
the flat Minkowski metric 
$\eta_{ab}$ by the vielbein $e_\mu^a$ by:
\begin{equation}
g_{\mu\nu} = e_\mu^a e_\nu^b \eta_{ab} \ .
\label{VB}
\end{equation}
The gamma matrices in curved space $\gamma^{\mu}$ are  
related to the flat space gamma matrices 
$\gamma^a$ by:
\begin{equation}
\gamma^\mu = e_a^\mu \gamma^a \ ,
\end{equation}
where $e_\mu^a e_b^\mu = \delta^a_b$.
We consider a massive Dirac fermion in the presence of a chemical potential $\mu$, whose action reads:
\begin{equation}
S = \int d^2x \, \sqrt{-g} \, \bar{\psi} \left( i \gamma^\mu D_\mu - m + \mu A_\mu \gamma^\mu \right) \psi \ ,
\label{S}
\end{equation}
where the adjoint spinor is defined as $\bar{\psi}= \psi^{\dagger}\gamma^0$, where $\gamma^0$ is
the flat space gamma matrix, and 
 $A = (A_t,0)$ is a time-like vector field.
$D_{\mu}$ is the fermionic covariant derivative:
\begin{equation}
D_\mu \psi = \partial_\mu \psi - \frac{1}{4} \omega_\mu^{ab} \gamma_a \gamma_b \psi \ ,    \end{equation}
where the torsionless spin connection $\omega^a{}_b = \omega^a{}_{b\mu}dx^{\mu}$ satisfies the 
Cartan equation:
\begin{equation}
de^a + \omega^a{}_b\wedge e^b=0 \ ,  
\end{equation}
where $e^a = e^{a}_{\mu}dx^{\mu}$.

The chemical potential term can be written as:
\begin{equation}
S_{chemical} = \mu \int d^2x \, \sqrt{-g} \, \psi^{\dagger}  \psi \ .
\label{smu}
\end{equation}
We will use the two-dimensional representation of the Clifford algebra:
\begin{equation}
\gamma^0 = \sigma_z, \gamma^1 = i \sigma_y, \gamma^5=\gamma^0\gamma^1 = \sigma_x \ ,
\label{gnotation}
\end{equation}
and $(\gamma^0)^{\dagger} = \gamma^0,  (\gamma^1)^{\dagger} = -\gamma^1, 
(\gamma^5)^{\dagger} = \gamma^5$.

The field equations read:
\begin{equation}
\left(i\gamma^{\mu}D_{\mu} - m + \mu\gamma^0\right)\psi = 0,~~~ \bar{\psi}\left(i\gamma^{\mu}\overset{\leftarrow}{D}_{\mu} + m - \mu\gamma^0\right) = 0  \ .
\end{equation}
The action is invariant under the constant phase rotation:
\begin{equation}
\psi\;\mapsto\;e^{i\alpha}\psi,
\qquad
\bar\psi\;\mapsto\;\bar\psi\,e^{-i\alpha} \ ,
\end{equation}
and the conservation law reads:
\begin{equation}
D_\mu J^\mu = 0,~~~~
J^{\mu} = \bar\psi\,\gamma^\mu\psi \ .
\label{Con}
\end{equation}
The conserved charge $Q$ associated with the conserved current (\ref{Con}) is
\begin{equation}
Q = \int_\Sigma \sqrt{h}\, n_\mu\, J^\mu\, dr \ ,    
\end{equation}
where  $\sqrt{h}d r$ is the induced measure on the spatial slice $\Sigma$,
and $n_\mu$ is its normal. This gives:
\begin{equation}
Q = \int dr\, J^t(r)  =  \int d r~
e^t_a \bar\psi \gamma^a \psi \ .
\label{QC_conti}
\end{equation}

\if{
Under axial rotation:
\begin{equation}
\psi \;\mapsto\;  e^{\,i\alpha\gamma^5}\,\psi\qquad
 \bar\psi \;\mapsto\; \bar\psi \,e^{\,i\alpha\gamma^5} \ ,
\end{equation}
the classical non-conservation law with mass and chemical potential reads:
\begin{equation}
 D_\mu J_5^\mu
=2i\left(\,m\,\bar\psi\,\gamma^5\psi
- \mu \bar\psi\,\gamma^0\gamma^5\psi\right),~~~~
J^{\mu}_5 = \bar\psi\,\gamma^\mu\gamma^{5}\psi \  ,
\label{Axial}
\end{equation}
and the axial charge is:
\begin{equation}
Q_5 = \int dr\, J^t(r)  =  \int d r~
e^t_a \bar\psi \gamma^a \gamma^5\psi \ .
\label{QCA}
\end{equation}

}\fi

The Hamiltonian density is given by the Legendre transform:
\begin{equation}
{\cal H} =   \Pi_{\psi}\partial_t \psi - {\cal L}  \ ,
\end{equation}
where 
$\Pi_{\psi} \equiv \frac{\partial{\cal L}}{\partial (\partial_t \psi)}$ is the conjugate momentum to $\psi$.
The Hamiltonian $H$ commutes with the charge $Q$ associated with the conserved current (\ref{Con}),
and it can be decomposed as a sum of the different charge sectors $H=\sum_{Q} H_{Q}$.

Define the charge‐conjugate spinor by:
\begin{equation}
\psi^c = C\,\bar\psi^{\,T} \ ,
\end{equation}
where the charge‐conjugation matrix \(C\) satisfies
\begin{equation}
C\,\gamma^a\,C^{-1} = +(\gamma^a)^T,
\quad
C^T = -\,C \ .
\end{equation}
Under \(\psi\to\psi^c\), one finds that the Dirac operator in the action (\ref{S}) flips the signs 
of the mass $m$ and the chemical potential $\mu$.
The Hamiltonian \(H(m,\mu)\) satisfies:
\begin{equation}
H(m,\mu)\,\psi = E\,\psi
\quad\Longleftrightarrow\quad
H(-m,-\mu)\,\psi^c = E\,\psi^c \ .
\label{HC}
\end{equation}
Hence each eigenvalue \(E\) of \(H(m,\mu)\) is also an eigenvalue of \(H(-m,-\mu)\), showing that the spectrum is identical:
\begin{equation}
\mathrm{Spec}\bigl(H(m,\mu)\bigr)
\;=\;
\mathrm{Spec}\bigl(H(-m,-\mu)\bigr) \ .
\label{spec}
\end{equation}
Filling all negative-energy modes yields the ground‐state energy \(E_0(m,\mu)\).  Up to an overall constant shift (choice of zero of energy), the symmetry of the spectrum implies:
\begin{equation}
E_0(m,\mu) \;=\; E_0(-m,-\mu) \ .
\label{gse}
\end{equation}
The charge $Q$ (\ref{QC_conti}) flips sign under charge conjugation, hence ${\cal O} = \mu Q$ exhibits the symmetry
\begin{equation}
{\cal O} (m,\mu) \;=\; {\cal O}(-m,-\mu) \ .    
\label{chargesymmetry}
\end{equation}

\subsection{The Qubit Representation}

To convert the Hamiltonian into the lattice Hamiltonian, we use the staggered fermion $\chi_n$, a single component Grassmann field, at each lattice site $n$~\cite{Kogut:1974ag,Susskind:1976jm}:
\begin{equation}
    \psi(t=0,x=na)=\frac{1}{\sqrt{a}}\begin{pmatrix}
        \chi_{2n}\\
        \chi_{2n+1}
    \end{pmatrix} \ ,
\end{equation}
where $a$ is the lattice size. They satisfy:
\begin{equation}
\{\chi_n,\chi_m\} = 0,~~~~\{\chi_n^{\dagger},\chi_m\} = \delta_{nm} \ .    
\end{equation}

The qubit-representation of the lattice Hamiltonian is obtained by Jordan-Wigner transformation~\cite{Jordan:1928wi}:
\begin{align}
 \chi_n \;=\; \frac{X_n-i Y_n}{2}\prod_{i=1}^{n-1}(-i Z_i) \ ,\quad
 \chi^\dag_n \;=\; \frac{X_n+i Y_n}{2}\prod_{i=1}^{n-1}(i Z_i) \ ,
\label{JW}
\end{align}
where $X_n,Y_n,Z_n$ are the Pauli matrices at the $n$-th site. For $n=1$, it is defined as $\chi_1=\frac{X_1-iY_1}{2}$. 

For instance, 
\begin{equation}
    \int \overline{\psi}\psi(x)dx=\sum_n\frac{1}{a}(\chi^\dagger_{2n}\chi_{2n}-\chi^\dagger_{2n+1}\chi_{2n+1})=\sum_n\frac{(-1)^n(Z_{n}+1)}{2a} \ .
\end{equation}
A straightforward computation yields:
\begin{eqnarray}
    \int \overline{\psi}\gamma_0\psi(x)dx=\sum_n\frac{1}{a}(\chi^\dagger_{2n}\chi_{2n}+\chi^\dagger_{2n+1}\chi_{2n+1})=\sum_n\frac{Z_{n}+1}{2a} \ .
\end{eqnarray}
However, the constant term gives the volume term $N/2a$, which diverges as $N\to \infty$. To remove the divergence 
we regularize the fermions bilinear:
\begin{equation}
    :\overline{\psi}\gamma_0\psi(x):=\overline{\psi}\gamma_0\psi(x)-\langle\overline{\psi}\gamma_0\psi(x)\rangle \ ,
\end{equation}
which gives $\chi^\dagger_{n}\chi_n=\frac{Z_n+(-1)^n}{2a}$.

While a naive computation yields $\overline{\psi}\gamma_1\psi_n=\frac{\chi^\dagger_{2n}\chi_{2n+1}+\chi_{2n}\chi^\dagger_{2n+1}}{a}$ for $n=1,2,\cdots$, from a continuous perspective, a vector current $\overline{\psi}\gamma_1\psi(x)$ measures the flow across a tiny interval $dx$, so its natural lattice home is the mid-point of that interval. The operator therefore sits on the bond between $n,n+1$, i.e. at the half-integer position $x=(n+1/2)a$. Hence, we define the current density as:
\begin{equation}
    J_{n}=\frac{\chi^\dagger_{n}\chi_{n+1}+\chi_{n}\chi^\dagger_{n+1}}{2a}=\frac{X_{n}Y_{n+1}-Y_{n}X_{n+1}}{4a} \ .
\end{equation}
The methodology is applied to define terms that involve mixing the left and right components $\psi_L$ and $\psi_R$ of the Dirac spinor $\psi=\binom{\psi_L}{\psi_R}$, such as $\overline{\psi}\gamma_5\psi(x)$ and $\overline{\psi}\gamma_1\partial_1\psi(x)$. 
In summary, a dictionary to translate the fields and staggered fermions into Pauli operators is provided in Table \ref{tab:dic}.
\begin{table}[H]
\begin{center}
\begin{tabular}{c|c|c}
Dirac Fermion Bilinears& Staggered  & Pauli \\\hline
     $\overline{\psi}\psi$ & $\frac{(-1)^n}{a}\chi^\dagger_n\chi_n$ &  $\frac{(-1)^n}{2a}(Z_n+1)$ \\
     $\overline{\psi}\gamma_0\psi$ & $\frac{1}{a}\chi^\dagger_n\chi_n$ &  $\frac{1}{2a}(Z_n+(-1)^n)$ \\
     $\overline{\psi}\gamma_1\psi$ & $\frac{1}{2a}(\chi^\dagger_n\chi_{n+1}+\chi^\dagger_{n+1}\chi_{n})$ &  $\frac{1}{4a}(X_nY_{n+1}-Y_nX_{n+1})$ \\
    $\overline{\psi}\gamma_5\psi$ & $\frac{(-1)^n}{2a}(\chi^\dagger_n\chi_{n+1}-\chi^\dagger_{n+1}\chi_{n})$ &  $-\frac{i(-1)^n}{4a}(X_nX_{n+1}+Y_nY_{n+1})$ \\
    $\overline{\psi}\gamma_1\partial_1\psi$ & $-\frac{1}{2a^2}(\chi^\dagger_n\chi_{n+1}-\chi^\dagger_{n+1}\chi_{n})$ &  $-\frac{i}{4a^2}(X_nX_{n+1}+Y_nY_{n+1})$ \\
\end{tabular}
\end{center}
    \caption{The three different representations of the Dirac fermion field in the flat background. To reflect the AdS black hole background, the redshift factor should be multiplied to the operators accordingly. For the details, see the following sections.}
    \label{tab:dic}
\end{table}

\section{\label{sec:BH}Fermions in Schwarzschild-Like Black Hole}

\subsection{Lagrangian and Hamiltonian}

\subsubsection{Continuum}
Consider Schwarzschild black hole solution in AdS$_2$ with radius $L$:
\begin{equation}
\mathrm{d}s^2 
\;=\;
-\,f(r)\,\mathrm{d}t^2 
\;+\;
\frac{1}{f(r)}\,\mathrm{d}r^2,
\quad
\text{where}
\quad
f(r) 
\;=\;
\frac{\,r^2 - r_h^2\,}{\,L^2\,} \ .
\label{AdSBH}
\end{equation}
Here $r_h$ is the horizon radius. 
In the units $16\pi G_{2}=1$ the mass of the black hole is $M=r_h^2$  and its temperature $T= \frac{r_h}{2\pi}$.
The zweibein read:
\begin{equation}
e^a_{\;\mu} \;=\;
\begin{pmatrix}
\sqrt{f(r)} & 0\\[4pt]
0 & \frac{1}{\sqrt{f(r)}}
\end{pmatrix},~~~
e_a^{\;\mu} \;=\;
\begin{pmatrix}
\frac{1}{\sqrt{f(r)}} & 0\\[4pt]
0 & \sqrt{f(r)} 
\end{pmatrix}.
\end{equation}
The nonzero spin connection is:
\begin{equation}
\omega_{t}^{01}
\;=\; -\frac{r}{L^2} \ .
\end{equation}

When $r_h \to 0$, we have $f(r)
\;\longrightarrow\;
\frac{r^2}{\,L^2\,}$, and the metric becomes:
\begin{equation}
\mathrm{d}s^2 
\;\to\;
-\,
\frac{r^2}{\,L^2\,}\,\mathrm{d}t^2 
\;+\;
\frac{L^2}{\,r^2\,}\,\mathrm{d}r^2.
\end{equation}
Introducing a coordinate $z = -\tfrac{L^2}{r}$, we get the metric:
\begin{equation}
\mathrm{d}s^2
\;=\;
\frac{L^2}{\,z^2\,}
\Bigl(
-\,\mathrm{d}t^2 
\;+\;
\mathrm{d}z^2
\Bigr),
\end{equation}
which is the Poincar\'e AdS$_2$ form.

The Lagrangian density takes the form:\footnote{The operator
$\frac{i}{2}\{A(r),\partial_r\}\;=\;i\,A(r)\,\partial_r\;+\;\frac{i}{2}\,A'(r)$ is hermitian. Here   
$A(r) = \sqrt{f(r)}\,\sigma_x$.}
\begin{equation}
{\cal L} = \psi^{\dagger} \left[ i \frac{1}{\sqrt{f(r})} \partial_t + i \sqrt{f(r)} \sigma_x \partial_r +i \frac{r}{2L^2\sqrt{f(r)}} \sigma_x  - m \sigma_z + \mu \right] \psi \ .
\end{equation}
The conjugate momentum to $\psi$ is:
\begin{equation}
\Pi_{\psi} \equiv \frac{\partial{{\cal{L}}}}{\partial (\partial_t \psi)} =  \frac{i}{\sqrt{f(r)}} \psi^{\dagger} \ ,
\end{equation}
and the canonical anticommutation relations read:
\begin{equation}
\{\psi_\alpha(r,t),\psi_\beta^\dagger(r’,t)\}
=\delta_{\alpha\beta}\,\sqrt{f(r)}\,\delta(r-r’) \ ,
\label{ACR}
\end{equation}
where $\alpha$ and $\beta$ are the spinor indices.

We can flatten the inner product by a local rescaling: $\chi(r)=f(r)^{-1/4}\psi(r)$, hence:
\begin{equation}
\{\chi(r),\chi^\dagger(r’)\}=\delta(r-r’) \ . 
\label{ACRF}
\end{equation}
The Lagrangian density takes the form:
\begin{equation}
{\cal L} = \chi^{\dagger} \left[ i \partial_t + i f(r) \sigma_x \partial_r +i \frac{r}{L^2} \sigma_x  - \sqrt{f(r)}m \sigma_z + \sqrt{f(r)}\mu \right] \chi \ ,
\end{equation}
and the Hamiltonian reads:
\begin{equation}
    H = \int_{r_h}^{\infty} d r  \mathcal{H} = \int d r \chi^{\dagger} \left[
    - i f(r) \sigma_x \partial_r -i \frac{r}{L^2} \sigma_x  + \sqrt{f(r)}m \sigma_z - \sqrt{f(r)}\mu
    \right] \chi \ .
    \label{BHHF}
\end{equation}
In the limit $r_h \to 0$, the Hamiltonian reduces to the 
AdS$_2$ Hamiltonian with the Poincaré coordinates.
Outside the horizon $r>r_h$, $\xi=\partial_t$ is a timelike Killing vector and the Hamiltonian is conserved and corresponds to
the symmetry $\mathcal{L}_\xi$. The vector $\xi$ becomes null at the horizon and space-like inside the horizon, and while it remains a killing vector, we cannot use it to define a Hamiltonian flow. Inside the horizon $\partial_r$ is timelike and generates  evolution along infalling time-like geodesics. This, however, is not associated with a conserved energy measured at infinity, because the usual notion of energy is tied to asymptotic symmetries at the boundary.
Hence, we are restricted to study the system's properties outside the horizon.

\if{
The Hamiltonian density is given by the Legendre transform:
\begin{equation}
{\cal H} =   \Pi_{\psi}\partial_t \psi - {\cal L} =   
\psi^{\dagger} \left[ -i \sqrt{f(r)} \sigma_x \partial_r 
+ \frac{r}{2L^2\sqrt{f(r)}} \sigma_x + m \sigma_z - \mu \right] \psi \ .
\end{equation}
The Hamiltonian is obtained by integrating over a spatial slice:
\begin{equation}
    H = \int_{r_h}^{\infty} \frac{1}{\sqrt{f(r)}} d r  \mathcal{H} = \int d r \psi^{\dagger} \left[ - i \sigma_x \partial_r + \frac{r}{2L^2 f(r)} \sigma_x + \frac{m}{\sqrt{f(r)}}  \sigma_z - \frac{\mu}{\sqrt{f(r)}} \right] \psi \ .
    \label{BHH}
\end{equation}

}\fi


We will consider two conserved charges, the flat charge:
\begin{equation}
Q_{flat}=\int_{r_h}^\infty dr\,\chi^\dagger\chi \ ,     
\label{qflat}
\end{equation}
and the weighted charge that multiplies the chemical potential $\mu$ in the Hamiltonian (\ref{BHHF}):
\begin{equation}
Q_{weighted}=\int_{r_h}^\infty dr\,\sqrt{f(r)}\chi^\dagger\chi \ .  
\label{qweighted}
\end{equation}
Note that the weighted charge arises since 
the chemical potential action (\ref{smu}) takes in the $\chi$ variables the form:
\begin{equation}
S_{\rm chem}=\mu \int dt\,dr\,\sqrt{f(r)}\,\chi^\dagger\chi \ ,
\end{equation}
and  $\sqrt{f(r)}$ is the local redshift converting the flat number density into a proper-energy density which couples to $\mu$. Thus, the weighted charge $-\mu Q_{weighted}$ appears in the Hamiltonian (\ref{BHHF}).

\if{
The conserved charge $Q$ (\ref{QC}) reads:
\begin{equation}
Q =  \int_{r_h}^\infty \frac{d r}{\sqrt{f(r)}} 
\bar\psi \gamma^0 \psi \ ,
\label{QC1}
\end{equation}
and using the fermionic anticommutation relations one has $[H,Q]=0$.

Similarly the axial charge associated with the axial current (\ref{Axial}) is:
\begin{equation}
Q_5 = \int_{r_h}^\infty \frac{d r}{\sqrt{f(r)}} 
\bar\psi \gamma^0 \gamma^5\psi =  \int_{r_h}^\infty \frac{d r}{\sqrt{f(r)}} 
\bar\psi \gamma^1 \psi \ ,
\label{QC_axial}
\end{equation}
and
\begin{equation}
[H, Q_5] = -2m \int dr\, \bar\psi \gamma^5 \psi \;+\; 2\mu \int dr\, \bar\psi \gamma^1 \psi \ .
\label{NC}
\end{equation}

}\fi

The redshift factor is $\alpha(r)=\sqrt{-g_{tt}(r)} = \sqrt{f(r)}$, where $\alpha(r)$ is defined as:
\begin{equation}
\alpha(r)=\sqrt{\frac{r^2 - r_h^2}{L^2}} \ ,    
\label{redshift}
\end{equation}
which is the ratio of boundary time to near-horizon proper time $d\tau = \alpha dt$, and thus measures the gravitational redshift between the $AdS_2$ boundary where we define our field theory Hamiltonian, and the black hole throat where modes live.
This is the gravitational redshift (blueshift), that rescales all near-horizon energies and momenta by $\alpha$.
Thus, a mode of frequency $\omega_{\rm throat}$ and momentum $k_{\rm throat}$ near the horizon is seen at the boundary with frequency
\begin{equation}
\omega_{\rm bdry} \;=\;\alpha\;\omega_{\rm throat},~~~k_{\rm bdry} \;=\;\alpha\;k_{\rm throat} \ .    
\end{equation}
At the horizon $\alpha=0$, hence a finite frequency at the horizon appears infinitely redshifted to the boundary — i.e., it has zero frequency from the boundary perspective.
When going to the boundary we need to normalize the boundary clock so that the physical time is 
$t_{\rm bdy}$, hence at the boundary $\alpha \to 1$.
The energy-momentum dispersion relation reads:
\begin{equation}
\varepsilon(k) = \alpha\sqrt{m^2 + (\alpha k)^2} \  . 
\label{dr}
\end{equation}
Note that the reason for the additional $\alpha$ factor in front of the wave number $k$ in (\ref{dr}) is that it is
the momentum conjugate to $r$, and not to the
proper spatial coordinate $\rho$, $d\rho = \frac{dr}{\alpha(r)}$.

There are two equivalent ways to introduce the chemical potential. In the first, we hold fixed 
a single number $\mu$ measured by the boundary clock $t_{\rm bdry}$.
In this scheme that spatial inhomogeneity due to the redshift factor is included in the charge operator $Q$.
In the second approach, we use a position independent $\mu_{\rm loc}$, measured by the proper time $\tau$. The two quantities are related
by:
\begin{equation}
\mu \;=\; \alpha(r)\,\mu_{\rm loc} \ .    
\end{equation}
Our discussions will be in the first convention, hence in the presence of the chemical potential (\ref{dr}) is modified to
\begin{equation}
E(k) = \alpha\sqrt{m^2 + (\alpha k)^2} - \mu \  . 
\label{drmu}
\end{equation}

\subsubsection{Lattice}
We consider a lattice uniform in the coordinate $r$, with sites
$r_n = r_h + na$, with $n=1,\cdots,N$,
and work in the region outside the horizon $r>r_h$. 
The anticommutation relations (\ref{ACR}) yields:
\begin{equation}
\{\psi_n,\psi_m^\dagger\}=\alpha_n\,\delta_{nm} \ ,    
\end{equation}
where $\alpha_n = \sqrt{f(r_n)}$ is the redshift factor at site $n$:
\begin{equation}
\alpha_n = \sqrt{\frac{r_n^2 - r_h^2}{L^2}} \ .    
\label{redshiftL}
\end{equation}

We can recover the flat anti-commutator by rescaling $\chi_n=\psi_n/\sqrt{\alpha_n}$:
\begin{equation}
\{\chi_n,\chi_m^\dagger\}=\delta_{nm} \ .   
\label{eq:scaled_fields}
\end{equation}
Using the Jordan-Wigner transformation (\ref{JW}), which respects this (flat) anti-commutation relation, the qubit Hamiltonian corresponding to (\ref{BHHF}) reads:
\begin{align}
    \begin{aligned}
        H=&\frac{1}{4a}\sum_{n=1}^{N-1}\alpha_n^2(X_nX_{n+1}+Y_nY_{n+1})+\frac{a}{8L^2}\sum_{n=1}^{N-1}n(X_nY_{n+1}-Y_nX_{n+1})\\
        &+\frac{m}{2}\sum_{n=1}^N(-1)^n\alpha_n(Z_n+1)-\frac{\mu}{2}\sum_{n=1}^N\alpha_n(Z_n+(-1)^n) \ , 
    \end{aligned}
    \label{HBHF}
\end{align}
and the constant terms have been neglected. In the qubit Hamiltonian (\ref{HBHF}), there is redshift factor $\alpha_n^2$ multiplying
the hopping $XX$ coupling, the on-site mass and chemical potential carry a single $\alpha_n$ factor, and the chiral term is independent of it.

The flat charge takes the lattice form:
\begin{equation}
    Q_\text{flat}=\sum_{n=1}^N\frac{Z_n+(-1)^n}{2a} \ ,
    \label{flat}
\end{equation}
while the weighted charge coupled to the chemical potential in the Hamiltonian (\ref{HBHF}) reads:
\begin{equation}
 Q_{weighted} = \frac{1}{2a}\sum_{n=1}^N\alpha_n(Z_n+(-1)^n) \ .
 \label{QC}
\end{equation}

\if{
Using the Jordan-Wigner transformation (\ref{JW}), which respects this (flat) anti-commutation relation, the qubit Hamiltonian \eqref{BHH} reads:
\begin{align}
    \begin{aligned}
        H=&\frac{1}{4a}\sum_{n=1}^{N-1}(X_nX_{n+1}+Y_nY_{n+1})+\frac{a}{8L^2}\sum_{n=1}^{N-1}nw_n^2(X_nY_{n+1}-Y_nX_{n+1})\\
        &+\frac{m}{2}\sum_{n=1}^N(-1)^nw_n(Z_n+1)-\frac{\mu}{2}\sum_{n=1}^Nw_n(Z_n+(-1)^n) \ , 
    \end{aligned}
    \label{HBH}
\end{align}
where $w_n= \alpha_n^{-1}$, and 
constant terms have been be neglected. 

The Hamiltonian (\ref{HBH}) consists of uniform $XX$ hopping, a small chiral bond term from the spin connection, and a staggered on-site mass (and chemical potential) weighted by the curved-space factor $w_n$.
$w_n$ 
varies on the curvature scale $L$ of the curved background, whereas our lattice spacing $a$ is taken to be much smaller, $a\ll L$.  In this regime a Taylor expansion gives:
\begin{equation}
w_{n+1} - w_n
= w(r_n + a) - w(r_n)
\approx w’(r_n)\,a + \mathcal O(a^2) \ ,
    \end{equation}
but since $w’(r)\sim \mathcal O(1/L)$ and $a\ll L$, the leading difference is suppressed:
\begin{equation}
w_{n+1}-w_n = \mathcal O\!\bigl(a/L\bigr)\,\ll 1,   
\end{equation}
and we can take  
\begin{equation}
w_{n+1} - w_n = 0 \ .  
\label{weight}
\end{equation}

The last term in (\ref{HBH}) is the $-\mu a Q$, where $Q$ is the physical charge:
\begin{equation}
 Q = \frac{1}{2a}\sum_{n=1}^Nw_n(Z_n+(-1)^n) \ ,
 \label{QC}
\end{equation}
and is straightforward to check using (\ref{weight}) that $[H,Q]=0$.
We will also consider the flat charge 
\begin{equation}
    Q_\text{flat}=\sum_{n=1}^N\frac{Z_n+(-1)^n}{2a} \ ,
    \label{flat}
\end{equation}
that commutes with the Hamiltonian ($[H,Q_\text{flat}]=0$) without approximation, as 
can be confirmed by the relations
\begin{align}
    [Z_n+Z_{n+1},X_nX_{n+1}+Y_nY_{n+1}]&=0,\nonumber\\
    [Z_n+Z_{n+1},Y_nX_{n+1}-X_nY_{n+1}]&=0 \ .
\end{align}

The axial charge (\ref{QC_axial}) reads:
\begin{equation}
 Q_5 = \frac{1}{4a}\sum_{n=1}^{N-1} w_n(X_nY_{n+1}-Y_nX_{n+1}) \ ,
 \label{QCaxial}
 \end{equation}
 and using (\ref{weight}) we get (\ref{NC}).

}\fi
 
The continuum limit is obtained by taking the limit
$a\rightarrow 0, N \rightarrow \infty$, such that 
the outermost site
$r_N \;=\; r_h + Na \rightarrow \infty$. In the limit, the horizon $0<r_h<\infty$, the AdS radius $L$, the mass $m$ and the chemical potential $\mu$ are fixed.  Any lattice sum maps as:
\begin{equation}
 \sum_{n=0}^{N}aF(r_n)  \rightarrow \int_{r_h}^{\infty}F(r)\,dr \ .
\end{equation}
It is straightforward to check that the qubit Hamiltonian (\ref{HBHF}) is mapped in the continuum limit to 
the Hamiltonian (\ref{BHHF}) \footnote{If we would have used the unscaled variables $\psi_n$, we would have obtained a Hamiltonian  that differs by its on site and link coefficients,thus away from the $a\!\ll\!L$ regime, it would have exhibited
$\mathcal O(a/L)$ differences in finite-size spectra, local densities, and short-range correlators. 
All these vanish in the continuum limit $\frac{a}{L}\to 0$.}.

A plane–wave ansatz $\psi_n\propto e^{i k n a}$ turns the finite difference in $r$ into the usual $\sin$–dispersion.  One finds
\begin{equation}
\psi_{n+1}-2\psi_n+\psi_{n-1}
=-4\,\sin^2\!\Bigl(\frac{k\,a}{2}\Bigr)\,\psi_n \ ,
    \end{equation}
and including the redshift (\ref{redshiftL} gives the lattice dispersion relation:
\begin{equation}
\varepsilon_n(k)
=\alpha_n\;\sqrt{\,m^2 + \frac{4\,\alpha_n^2}{a^2}\,\sin^2\!\Bigl(\tfrac{k\,a}{2}\Bigr)\,} \ 
\label{dra}
\end{equation}
In the continuum limit $a\to 0$, $\sin(\frac{ka}{2})\to \frac{ka}{2}$, so
$\displaystyle\frac{4}{a^2}\sin^2(\tfrac{k\,a}{2})\to k^2$, and $\alpha_n\to \alpha$, reducing (\ref{dra}) to (\ref{dr}).

Finally, let us make two comments.

\noindent
{\it Boundary clock:} 
When going to the boundary we need to normalize the boundary clock so that the physical time is 
$t_{\rm bdy}$, hence at the boundary $\alpha_n\to1$. This means that 
as seen by a boundary observer, there is an effective redshift factor
\begin{equation}
\alpha_n^{(\rm eff)} \;=\;\frac{\alpha_n}{\alpha_N}
\;=\;\sqrt{\frac{r_n^2 - r_h^2}{r_N^2 - r_h^2}} \ ,   
\end{equation}
and $\alpha_N^{(\rm eff)}=1$.
The effective mass at site $n$ is:
\begin{equation}
m^{(\rm eff)}_n = \alpha_n^{(\rm eff)} m \ .   
\label{meffective}
\end{equation}

\noindent
{\it Redshift effect:}
There are two ways to compare the effect of the horizon size on the redshift factor.
If we compare two geometries at the same continuum radius $r$, then (\ref{redshift})
decreases when $r_h$ grows, which is the usual picture of stronger gravitational redshift (local clocks run slower) as the black hole gets bigger.
On the other hand, if we compare at fixed lattice index $n$, then because
in our discretization $r_n=r_h + n a$, then increasing $r_h$ also moves that lattice site farther out (larger $r_n$), and $\alpha_n$ increases with $r_h$.  However, this is a
comparison between different physical radii.

\if{
On the lattice of size $a$ we replace $\partial_x\psi(x)$ by the symmetric finite-difference
\begin{equation}
\partial_x\psi\;\to\;\frac{\psi_{j+1}-\psi_{j-1}}{2a} \longrightarrow
i\,\frac{\sin(k_n a)}{a}\,\tilde\psi_n \ ,
\end{equation}
where $\psi_j\equiv\psi(x_j)$, $x_j = j\,a$, and $n$
refers to the $n$th plane‐wave mode.
$i\,\frac{\sin(ka)}{a}$
is the lattice analogue of the continuum factor $ik$.
Since we are on $AdS_2$, the spatial coordinate is not the 
an ordinary flat direction, but rather a hyperbolic one, hence, we replace 
$sin$ by $sinh$. 
This, together with the inclusion of the redshift factor (\ref{redshift}),
gives the dispersion relation:
\begin{equation}
\varepsilon(k)
= \sqrt{\,m^2 \;+\;\Bigl[\tfrac{2}{a}\,\sinh\!\bigl(\tfrac{\alpha\,a\,k}{2}\bigr)\Bigr]^2\,} \ ,  
\label{dra}
\end{equation}
which in the continuum limit gives (\ref{dr}).
For $k\,a\ll1$, even a large redshift factor $\alpha$ gives:
\begin{equation}
\sinh\!\Bigl(\tfrac{\alpha\,a\,k}{2}\Bigr)
\approx \tfrac{\alpha\,a\,k}{2} \ ,    
\end{equation}
which is simply a steeper linear dispersion—every mode, blueshifted by a factor $\alpha$.
Once $\alpha\,a\,k/2\gtrsim 1$, the $sinh$ crosses over to
\begin{equation}
\sinh\!\bigl(\tfrac{\alpha\,a\,k}{2}\bigr)
\sim\tfrac12\exp\!\bigl(\tfrac{\alpha\,a\,k}{2}\bigr) \ ,    
\end{equation}
and the band energy (\ref{dra}) 
grows exponentially fast in $k$, rather than flattening out near the Brillouin‐zone edge as a $\sin(ka)$ dispersion does.

}\fi
\subsection{Ground State}

\subsubsection{Energy}

The ground state energy of the chain 
\begin{equation}
E_{0} =\sum_{(n,k):\,\varepsilon_n(k)\le\mu}\varepsilon_n(k)   
\end{equation}
takes the general form:
\begin{equation}
E_{0}(L,r_{h},a;m,\mu)
=\frac{r_{h}}{a}\,\mathcal E(ma,\mu a)
\;+\;\frac{1}{L}\,G\!\Bigl(\tfrac{r_{h}}{L},mL,\mu L\Bigr)
\;+\;\frac{1}{r_{h}}\,H\!\bigl(mr_{h},\mu r_{h}\bigr)
\;+\;\cdots, 
\label{EL}
\end{equation}
where we expand in the three small parameters $a\ll 1, \frac{1}{L}\ll1, \frac{1}{r_h}\ll1$, 
and $\cdots$ stands for sub-leading corrections. These include terms of the form $\frac{a}{r_{h}},(\frac{a}{r_{h}})^2...$ beyond the leading $\tfrac{r_{h}}{a}$ one,
finite-size corrections in powers of $\frac{1}{L^2}$ and  $\frac{1}{r_h^2}$, and mixed corrections that depend combinations of $\frac{a}{L}$ and $\frac{a}{r_{h}}$, or higher-order functions of the dimensionless arguments $(ma,\mu a, mL,\mu L, mr_{h},\mu r_{h})$.
In a full perturbative expansion one systematically generates these terms by expanding to higher orders in $a, \frac{1}{L}$, and $\frac{1}{r_{h}}$.  
As $a\to 0$, we get a divergent bulk piece:
\begin{equation}
\frac{r_{h}}{a}\,\mathcal E(m\,a,\mu\,a)
\;\longrightarrow\;
\frac{r_{h}}{a}\,\mathcal E(0,0) = O(a^{-1}) \ ,   
\label{bulk}
\end{equation}
which is the usual UV divergence of the vacuum energy that should be subtracted/renormalized. The bulk term counts modes up to the cutoff 
and scales with $r_h$.

The other terms remain finite and constitute the renormalized ground state energy. The $AdS_2$ shift captures how placing the fermion in a curved $AdS_2$ geometry modifies the continuum zero-point energy relative to flat space:
\begin{equation}
\frac{1}{L}\,G\!\bigl(\tfrac{r_{h}}{L},mL,\mu L\bigr)
\;=\;O(L^{-1}) \ .   
\end{equation}
The Casimir‐like part across the interval $[r_{h},\infty)$ 
captures the effect of having a horizon at $r_h$:
\begin{equation}
 \frac{1}{r_{h}}\,H\!\bigl(mr_{h},\mu r_{h}\bigr)
\;=\;O(r_{h}^{-1}) \ .   
\end{equation}

Larger $r_h$ increases the redshift (\ref{redshiftL}), which raises every one-particle energy $\varepsilon(k)$, hence
summing over all occupied modes results in the increase with $r_h$ of the total ground state energy.
As $mL$ increases, the fermion becomes heavier compared to the $AdS_2$ curvature scale
and contributes less to the vacuum energy. Thus, $G$ is a decreasing function of $mL$.
The bulk and horizon pieces are unchanged by $mL$, since they depend on $ma$ and $mr_h$, respectively.

It is possible to have an explicit expression for the ground state energy by filling every single-particle mode with energy below the chemical potential. The one-particle dispersion is:
\begin{equation}
\label{eq:one-particle-dispersion}
E_n(k)\;=\;\alpha_n\,\sqrt{m^2+\alpha_n^2k^2}\;-\;\mu \ ,   
\end{equation}
and the local Fermi momentum reads:
\begin{equation}
k_{F,n}
=\frac{1}{\alpha_n^2}\,\sqrt{\mu^2-\alpha_n^2m^2}
\;\Theta(\mu-\alpha_n m) \ .  
\label{KF}
\end{equation}
The contribution of site $n$ to the ground state energy is then
\begin{equation}
E_{0,n}
=\int_{-k_{F,n}}^{k_{F,n}}\!\frac{dk}{2\pi}\,\bigl(E_n(k)\bigr)
=\frac{1}{2\pi}\!\left[
m^2\ln\!\Big(\frac{\alpha k_F}{m}+\sqrt{1+\frac{\alpha^2k_F^2}{m^2}}\Big)
-\mu k_F\right] \ ,
\end{equation}
and the total ground state energy is the sum over all sites,
$E_0
=\sum_{n=0}^{N}E_{0,n}$,
which can be expanded to give the three terms above.
As expected, it also satisfies 
$E_0(m,\mu)=E_0(-m,-\mu)$ (\ref{gse}).

\subsubsection{Charge}

Consider the local charge distribution.
On the lattice the flat charge (\ref{flat}) is the sum of a site‐by‐site charge density\footnote{The weighted
charge (\ref{QC}) is $Q_{\text{phys}}=\sum_i q_i^{\text{phys}}$ with $q_i^{\text{phys}}=\alpha_i\,q_i$.}:
\begin{equation}
Q =\sum_{i=i}^{N}q_{i} \ ,    
\end{equation}
where $q_{i}=\langle\psi_{i}^{\dagger}\psi_{i}\rangle$ is the ground‐state occupation at site $i$. Recall that $r_i=r_h+ia$. 
By the same separation of scales that gave (\ref{EL}), the local density $q_{i}$ splits into three pieces.
The bulk (UV‐extensive) piece is:
\begin{equation}
q_{i}^{(Bulk)} \;=\;q(m a,\;\mu a),\quad 1\ll i\ll N
 \ .    
 \label{BT}
\end{equation}
It is constant across most of the lattice, that is
away from the horizon and boundary each site carries essentially the same flat‐space type filling fraction.
The $AdS$ term is the effect of the background curvature on $q_{i}$: 
\begin{equation}
q_{i}^{(AdS)}
\;=\;\frac{1}{L}\;
g\!\Bigl(\tfrac{r_{h}}{L},\,mL,\,\mu L;\;\tfrac{i\,a}{L}\Bigr) \ .
\end{equation}
The horizon correction reads:
\begin{equation}
q_{i}^{(Horizon)}
\;=\;\frac{1}{r_{h}}\;
h\bigl(m\,r_{h},\,\mu\,r_{h};\;i\bigr) \ .    
\end{equation}
Within a few sites of the horizon $ia$, the boundary condition at the black hole horizon perturbs the occupancy,
which decays as we move into the bulk.

Putting these together,
\begin{equation}
q_{i}
=q(ma,\mu a)
+\frac{1}{L}g\!\bigl(\tfrac{r_{h}}{L},mL,\mu L;\tfrac{i\,a}{L}\bigr)
+\frac{1}{r_{h}}h\!\bigl(mr_{h},\mu r_{h};i\bigr)
+\cdots \ .
\label{finalq}
\end{equation}
In the deep bulk $i \gg 1$, the constant $q$ dominates.
On the scale of the $AdS_2$ radius $i\,a\sim L$, the $g$-term imprints an $O(\frac{1}{L})$ modulation of the filling.
Near the horizon $i\sim 1$, the $h$-term produces an $O(\frac{1}{r_{h}})$ deviation that decays into the bulk.
Thus, the ground‐state charge is essentially flat across the lattice, with small, localized ripples at the horizon and a curvature‐driven drift across the full system.

We can derive an explicit expression for the charge as:
\begin{equation}
q_i
=\int_{-k_{F,i}}^{\,k_{F,i}}\frac{dk}{2\pi}
=\frac{k_{F,i}}{\pi} \ ,    
\end{equation}
and using (\ref{KF}):
\begin{equation}
Q
= \sum_{i=0}^N q_i
= \sum_{n=0}^N
\frac{1}{\pi\,\alpha_n^2}
\sqrt{\mu^2 - \alpha_n^2m^2}
\;\Theta(\mu-\alpha_n m) \ ,
\label{Qexact}
\end{equation}
which in the continuum takes the form:
\begin{equation}
Q(r)
= \frac{1}{\pi\alpha^2}
\sqrt{\mu^2 - \alpha^2 m^2}
\;\Theta\!\Bigl(\mu-m \alpha\Bigr) \ .
\end{equation}
Only sites $n$ for which $\alpha_n m<\mu$. i.e. redshifted mass below the chemical potential, carry nonzero charge, and $q_n$ decreases as the redshift grows, and vanishes once $\alpha_n m\ge\mu$.
As expected, (\ref{Qexact}) satisfies
$Q(m,\mu)=Q(-m,-\mu)$ (\ref{chargesymmetry}).

\subsubsection{Entanglement Entropy}

The ground state entanglement depends on the UV cutoff $a$, the gap $\Delta\sim|\alpha_{\min}m-\mu|$, and the redshift profile $\alpha_n$ (\ref{redshiftL}). For $|\mu|<\alpha_{\min}|m|$, the state is gapped and obeys an area law; at $|\mu|=\alpha_{\min}|m|$ it becomes gapless and the entanglement of a block of length $\ell$ scales logarithmically with central charge $c=1$ \cite{CalabreseCardy2004,EisertCramerPlenio2010}. At fixed $\mu$ below threshold, increasing $m$ reduces entanglement; at fixed $m$ below threshold, increasing $|\mu|$ raises entanglement as the gap closes.

\subsection{Energy Gap}
\label{GAP}

In the limit of infinite $N$, the band minimum at site $n$ occurs at $k=0$,
$\varepsilon_{n,\min}=\alpha_n|m|$, and the energy gap is:
\begin{equation}
\Delta =\min_n\big(\alpha_n|m|-\mu\big)\;=\;\alpha_{\min}|m|-\mu \ ,    
\end{equation}
where $\alpha_{\min}=\min_n\alpha_n$.
More precisely, the global single particle gap is the smallest local gap:
\begin{equation}
\Delta\;=\;\min_{r}\,\Delta(r)\;=\;\min_{r}\, \big|\alpha(r)\,m-\mu\big| \ .
\end{equation}
Because, with the boundary normalization, $\alpha(r)$ varies continuously from $0$ at the horizon up to $1$ at the boundary, the set
$\{\alpha(r)m\}$ fills the interval $[0,|m|]$. Therefore, the global gap is the distance from $|\mu|$ to that interval:
\begin{equation}
\Delta(m,\mu)
=\operatorname{dist}\!\big(|\mu|,\,[0,|m|]\big)
=\begin{cases}
0 & |\mu|\le |m|,\\
|\mu|-|m| & |\mu|>|m|.
\end{cases}
\end{equation}
Equivalently, the gapless region in the continuum is the closed cone
$|\mu|\le |m|$ bounded by the lines $\mu=\pm m$,
and the system is gapped outside that cone. The two straight boundaries $\mu=\pm m$ are the first places where a local mode closes, at the boundary where $\alpha=1$.

At finite $N$, the open boundary conditions:
\begin{equation}
k_n = \frac{\pi n}{(N+1)\,a},~~~k=1,2,\cdots \ .   
\label{mode}
\end{equation}
Assuming a slowly varying redshift in the bulk window, $\alpha_n\simeq\alpha$, and using (\ref{dra}), the finite $N$ gap is
\begin{equation}
\Delta_N=\min_{n\ge1}\left[\ \alpha\,\sqrt{\,m^2+\frac{4\alpha^2}{a^2}\sin^2\!\frac{k_n a}{2}\,}\ -\ \mu\ \right] \ .    
\end{equation}
Using the smallest nonzero mode in (\ref{mode}),
$k_1 = \frac{\pi}{(N+1)\,a}$, and 
expanding for large $N$ we get:
\begin{equation}
\label{eq:Delta_N}
\Delta_N=\alpha m-\mu+\frac{\alpha^3}{2m}\,\Big[\frac{\pi}{(N+1)a}\Big]^2+O\left(\frac{1}{N^4}\right) \ .
\end{equation}
Thus, the leading finite-size correction is
$O(1/N^2)$, and it increases the gap \footnote{See \cite{P} for a similar effect in the solution of the transverse‐field Ising chain.}
Note that, on a finite open chain, the boundary conditions allow for exponentially localized edge states that lie just below the bulk gap \cite{Kitaev_2001}, whose energy splitting from the continuum reads:
\begin{equation}
E_{edge} \sim e^{-l/\xi_{\rm eff}},~~\xi_{\rm eff}\;\approx\;\frac{r_h}{L\,(m-\mu)},~~l=O(a) \ ,    
\end{equation}
which reduces the gap.

\if{
Let us now derive an explicit for the gap when $|mL|$ or $|\mu L|$ is large.
On the lattice of spacing $a$, momentum is quantized in units of
\begin{equation}
k_n \;=\;\tfrac{n\pi}{a\,N}\,,\quad n=0,1,2,\dots \ ,    
\end{equation}
so the two lowest levels are
\begin{equation}
\varepsilon_0 = \sqrt{m^2 + (0)^2}
\;=\;|m|\,,
\quad
\varepsilon_1 = \sqrt{m^2 + (\alpha\,\tfrac{\pi}{aN})^2} \ ,
\end{equation}
where $\alpha$ is the redshift factor.
The gap is:
\begin{equation}
\Delta E \;=\;\varepsilon_1 - \varepsilon_0
=\;\sqrt{m^2 + \bigl(\alpha\,\tfrac{\pi}{aN}\bigr)^2}\;-\;|m| \ .    
\end{equation}
Expanding in the regime $|m| \gg \frac{\alpha\pi}{aN}$, the leading $1/N$ expression reads:
\begin{equation}
 \Delta E
\;\simeq\;
\frac{L\,|m|}{a\,N}\;\frac{1}{\sqrt{1 + 2\,r_h/(aN)}} \ ,
\quad
\bigl(|m/\mu|\gg1\bigr)   \ ,
\end{equation}
and similarly 
$|m|\leftrightarrow|\mu|$.
Note that we identified the $AdS_2$ box size as $L\approx aN$, and used $\alpha\le1$, hence 
\begin{equation}
mL \;\gg\;1 
\quad\Longrightarrow\quad
m \;\gg\;\frac{\alpha}{aN} \ .
\end{equation}

}\fi

\subsection{First Excited State}

\subsubsection{Energy}
\label{E1}
Unlike the ground‐state energy $E_0$, which is a smooth integral over all modes up to the Fermi momentum and therefore yields regular level sets in the $(m,\mu)$ plane, the first‐excited‐state energy:
\begin{equation}
E_1 \;=\; \sum_{k\le k_F}\varepsilon(k)\;+\;\bigl[\varepsilon(k_1)-\varepsilon(k_F)\bigr] \ ,
\end{equation}
involves a discrete jump from the highest occupied mode at $k_F$ to the next available mode $k_1$.  
Since $k_F$ depends implicitly on $\mu$ via $\varepsilon(k_F)=\mu$, a tiny change in $\mu$ can shift which discrete $k_n$ is the last filled mode.
The first excited energy then picks up $\varepsilon(k_F + \Delta k)$ instead of $\varepsilon(k_F)$, where $\Delta k = \pi/((N+1)a)$ for the open chain\footnote{More precisely, with $r_h\neq 0$ we should use level indices $j$ instead of
plane wave $k$, and view $\Delta k$ as a heuristic for the uniform limit.}. As a result, the contours of constant $E_1$ wiggle whenever $\mu$ crosses one of those discrete level thresholds.

The energy of the first excited states also exhibits a dispersion‐curvature sensitivity,
since it depends on the local second derivative $\varepsilon''(k)$ at the band edge, and small non-linearities in $\varepsilon(k)$ around $k_F$ show up in the shift 
$\varepsilon(k_1)-\varepsilon(k_F)$. This 
distorts the level sets away from the straight $m=\pm\mu$ lines of $E_0$.
Lastly, there are red‐shift amplification, since $\varepsilon(k)$ depends on the redshift factor, whose
variations 
magnify the non-linearities in the band structure.  The stronger the redshift, i.e. larger $r_h/L$, the more pronounced the warping of the $E_1$ contours relative to those of $E_0$.

\subsubsection{Charge}
\label{charge1}
At zero temperature the ground state fills all modes up to $k_F$, and the local
flat charge at site $n$ \footnote{An analogous discussion follows for the local weighted charge 
with a redshift weight $w_n$.}:
\begin{equation}
Q_n^{(0)}
=\sum_{j:\,k_j\le k_F}\bigl|\psi_{k_j}(n)\bigr|^2 \ ,
\end{equation}
where $\psi_{k}(n)$ is the normalized real‐space wavefunction of the mode $k$.

The first excited state is obtained by removing the fermion in the highest filled mode $k_{j_F}=k_F$ and putting it into the next mode $k_{j_F+1}$.  Hence, its local flat charge is
\begin{equation}
Q_n^{(1)}
=\sum_{j:\,k_j< k_F}\bigl|\psi_{k_j}(n)\bigr|^2
\;+\;\bigl|\psi_{k_{j_F+1}}(n)\bigr|^2
\;=\;
Q_n^{(0)}
\;-\;\bigl|\psi_{k_F}(n)\bigr|^2
\;+\;\bigl|\psi_{k_{F+1}}(n)\bigr|^2 \ .
\end{equation}
Thus, 
the change in the expectation value of the flat charge is
\begin{equation}
\delta Q_n
\;=\;
Q_n^{(1)} - Q_n^{(0)}
\;=\;
\bigl|\psi_{k_{F+1}}(n)\bigr|^2
\;-\;
\bigl|\psi_{k_F}(n)\bigr|^2 \ . 
\end{equation}
We still have net neutrality, $\sum_n\delta Q_n=1-1=0$, as the excitation carries no net charge, and just redistributes it. As to the spatial structure, $\psi_{k_F}(n)$ and $\psi_{k_{F+1}}(n)$ differ, so $\delta Q_n$ oscillates across the chain.
The dependence on $m,\mu,L,r_h,a$ enter through the mode wavefunctions $\psi_k(n)$ via the dispersion $\varepsilon(k)$ (\ref{dra}). 

\subsubsection{Entanglement Entropy}

The ground‐state half‐chain entanglement entropy is
\begin{equation}
S_{EE}^{(0)} \;=\; -\sum_{\ell=1}^{N/2}\bigl[\nu_\ell\ln\nu_\ell + (1-\nu_\ell)\ln(1-\nu_\ell)\bigr] \ ,
\end{equation}
where $\{\nu_\ell\}$ are the eigenvalues of the correlation matrix $C_{ij}=\langle c_i^\dagger c_j\rangle$ restricted to the left half, and
$(c_j,c^{\dagger}_j)$ are the fermionic annihilation/creation operators on lattice site $j$. In the first excited state, we swap one occupied mode at $k_F$ for the next one at $k_{F+1}$. Thus, the bulk of the spectrum is unchanged, 
and all the correlator eigenvalues $\nu_\ell$ associated with modes below the Fermi level remain identical to the ground state and contribute the same amount to $S_{EE}^{(0)}$.
The difference comes from the swap of the two modes:
\begin{equation}
\Delta S_{EE} \;=\; S_{EE}^{(1)} - S_{EE}^{(0)} \sim O\left(\frac{1}{N}\right) \ ,
\label{DS}
\end{equation}
and it generically increases the entropy, since adding a quasiparticle across the cut tends to boost entanglement.
Since the excitation only changes one momentum mode near $k_F$, and neighboring quantized $k$ values differ by
$\sim\frac{\pi}{N}$, thus the correlation spectrum and $S_{EE}$ shift only by $O(\frac{1}{N})$.

\subsection{Summary: From Critical to Gapped Regime}

In our staggered-fermion $AdS_2$ black hole chain, the critical regime is where at least one local mode remains gapless, so correlations span the entire half-chain and $S_{EE}$ is large. 
In the continuum, the global single-particle gap is $\Delta=\min_r|\alpha(r)m-\mu|$. Because $\alpha(r)\in[0,1]$, the set $\{\alpha(r)m\}$ spans $[0,|m|]$, so the gapless region is the closed cone $|\mu|\le |m|$ bounded by
$\mu=\pm m$.

At lattice site $n$, the dispersion’s minimum energy is:
\begin{equation}
\varepsilon_{n,\min}
=\alpha_n\,|m| \ ,    
\end{equation}
and with chemical potential $\mu$ the local gap reads:
\begin{equation}
\Delta_n
=\bigl|\alpha_n\,m - \mu\bigr| \ ,   
\end{equation}
where $\alpha_n$ (\ref{redshiftL})  is the redshift at that site.
The system as a whole remains critical as long as
$\min_n\Delta_n = 0$,
i.e. at least one $\Delta_n$ vanishes and there’s a gapless mode.  Since $\alpha_n\le1$, with the maximum at the boundary site, $\alpha_N=1$, the first closing always occurs at $n=N$, giving the critical (gapless) lines:
\begin{equation}
\mu = +m
\quad\text{or}\quad
\mu = -\,m \ .    
\end{equation}
Inside that $X$-shaped region in the $(m,\mu)$ plane one has gapless excitations and long-range correlations. Outside it the global gap $\min_n\Delta_n>0$, and the chain is gapped.

Once $\mu$ crosses the line $\mu=\pm m$, even the smallest local gap at the boundary becomes positive, $\Delta_N=|m-\mu|>0$, and all other $\Delta_n\ge\Delta_N$. Hence, 
$\min_n\Delta_n>0$ and the chain enters the gapped phase where 
the correlation length $\xi\sim \frac{1}{min_n\Delta_n}$ becomes finite, and entanglement across the midpoint decays exponentially with system size.
In the gapped phase $S_{EE}$ saturates to an $O(1)$ value (area law), rather than the $O(\ln N)$ (or larger) behavior in the critical region.
The redshift profile $\{\alpha_n\}$ set by the horizon radius $r_h$ controls how sharply the transition happens. When $r_h=0$,  $\alpha_n = \frac{n a}{L}$, so all sites gap out almost simultaneously as $m$ crosses the threshold, giving a sharp transition.
Strong redshift near the horizon, $\alpha_n\ll 1$ for small $n$, means that interior sites remain effectively gapless until very small $m$, so the gapping of the entire chain is smeared out over a range of $m$, producing a broader crossover.

In summary, moving from critical to gapped in our model is the process of lifting the last zero of 
$\Delta_n$, first at the boundary, then throughout the bulk, thereby turning long-range entanglement into a finite-correlation, area-law regime.

\section{\label{sec:simulation}Quantum Simulations}
In all our quantum simulations we will set $a=1$, unless explicitly stated. Also, the number of qubits 
$N$ is even, hence $\sum_{i=1}^N (-1)^n = 0$. The radius $L$ of $AdS_2$ sets the length scale.
We perform the large $N$ simulations using Matrix Product State (MPS) representations.

\subsection{Ground State}

In the following we study the properties of the ground state of the Hamiltonian (\ref{HBHF}).
\subsubsection{Energy}

\begin{figure}[H]
    \centering
    \includegraphics[width=0.49\linewidth]{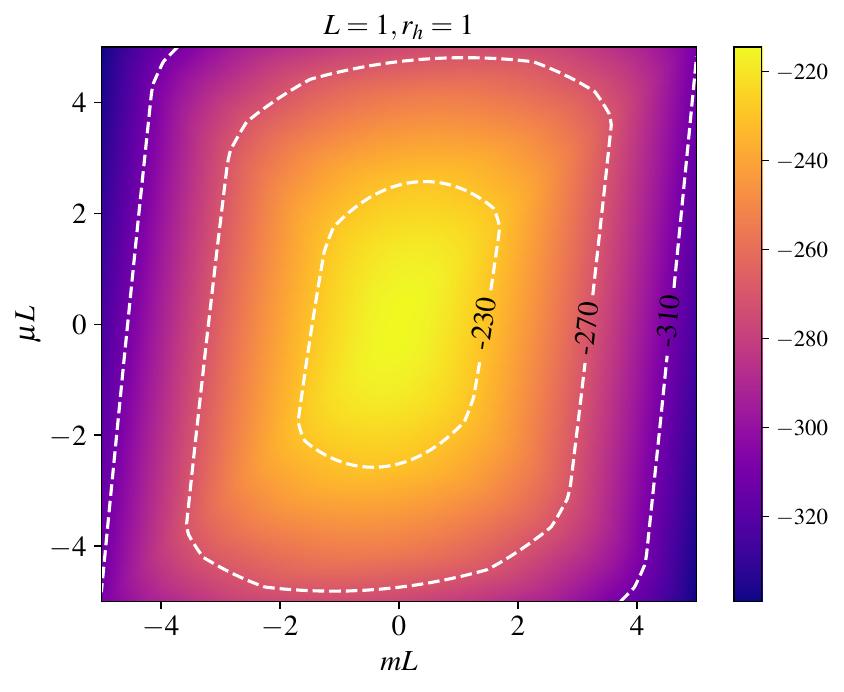}
    \includegraphics[width=0.49\linewidth]{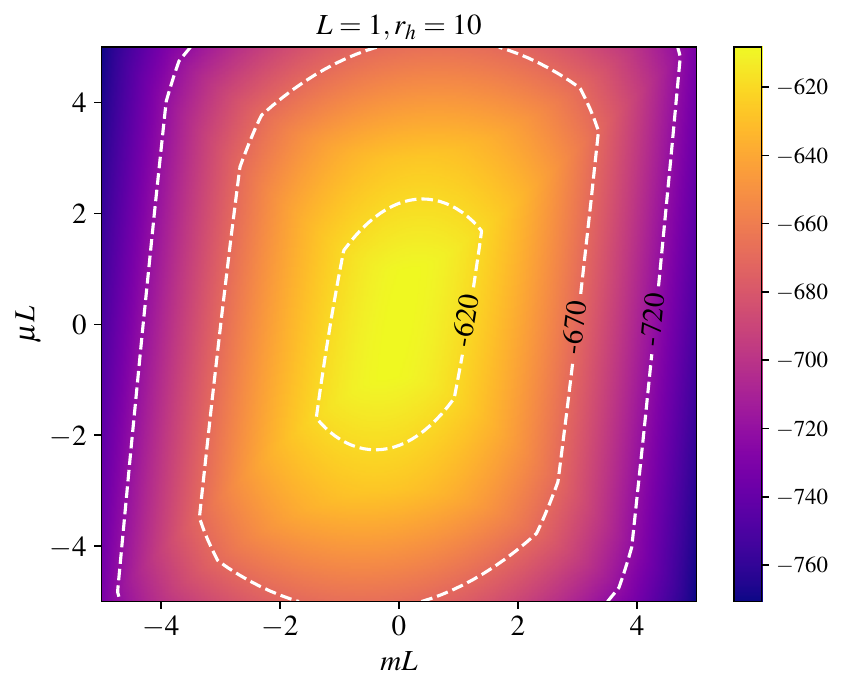}
    \caption{The ground state energy
    for $N=12$ qubits, with horizon radius $r_{h} =1$ (left)
   $r_{h} =10$ (right). There is a symmetry: $m\rightarrow -m,~\mu \rightarrow -\mu$.  The energy becomes more negative as the horizon radius grows, and as $mL$ and $\mu L$ increase.}
    \label{fig:E0_heatmap_BH_N12}
\end{figure}

In Fig.~\ref{fig:E0_heatmap_BH_N12} we present the heatmap of the ground state energy for $N=12$ qubits and different values of the horizon radius $r_h$. The energy heatmap exhibits 
the combined symmetry:
\begin{equation}
(m,\mu)\leftrightarrow (-m,-\mu) \ .   
\label{Sym}
\end{equation}
This is the charge conjugation symmetry (\ref{HC}), (\ref{spec}) and  (\ref{gse}).
When the horizon is larger (compare left vs.\ right panels), the entire energy surface shifts towards more negative values, i.e the ground state energy decreases with $r_h$.
Physically, this is the effect of the gravitational redshift, larger $r_h$ increases the redshift factor.
As the fermion mass $|mL|$ grows, the energy becomes more negative, because heavy fermions have shorter Compton wavelengths, cutting off long-wavelength vacuum fluctuations and reducing both bulk and curvature/horizon contributions.
Also, since we work at zero temperature, every single‐particle level with $\varepsilon(k)\le\mu$ is filled, and in our qubit chain filling an extra mode contributes a negative amount to the renormalized ground state energy.  Thus, as $\mu L$ grows, we include more and more modes, making the total ground state more negative.

Consider now the dependence of the ground state energy on the number of qubits $N$.
We will work in the regimes of large $\mu$ or large $m$, where one of the last two terms in (\ref{HBHF}) dominates the Hamiltonian. The ground state is a product state and 
the lowest energy can be approximated by:
\begin{equation}
    E_0\approx
    \begin{cases}
        -\frac{|m|}{2}\sum_{n=1}^N \alpha_n& |m/\mu|\gg1,\\
        -\frac{|\mu|}{2}\sum_{n=1}^N \alpha_n& |\mu/m|\gg1 \ .
    \end{cases}
    \label{app}
\end{equation}
Define $S_N(\beta)=\sum_{n=1}^N\sqrt{n(n+\beta)}$, then
\begin{eqnarray}
    \sum_{n=1}^N \alpha_n=\frac{1}{L}S_N\left(2r_h\right) \ .
\label{SN}    
\end{eqnarray} 
There is no closed analytical formula for (\ref{SN}). When $\beta$ is small, it can be expanded into generalized harmonic numbers as 
\begin{equation}
    S_N(\beta)=\frac{N(N+1)}{2}+\frac{\beta}{2}N-\frac{\beta^2}{8}H^{(1)}_N+\frac{\beta^3}{16}H^{(2)}_{N}+\cdots\;,
    \label{series}
\end{equation}
where $H^{(r)}_N:=\sum_{n=1}^Nn^{-r}$.  \if{
For fixed $\beta$ and large $N$, $S_N(\beta)\approx 2\ln N$ for large $N$, while when $\beta = \ell N$, where $\ell>0$ is fixed \ki{to be updted}:
\begin{equation}
S_N(\ell N)
=\frac{2}{\sqrt{\ell}}\,
\operatorname{sinh}^{-1}\!\bigl(\ell^{-1/2}\bigr)
\;+\;\frac{13}{24\sqrt{\ell}}\,N^{-1/2}
\;+\;O\bigl(N^{-1}\bigr) \ .
\label{series}
\end{equation}
}\fi

\begin{figure}[H]
    \centering
    \includegraphics[width=0.32\linewidth]{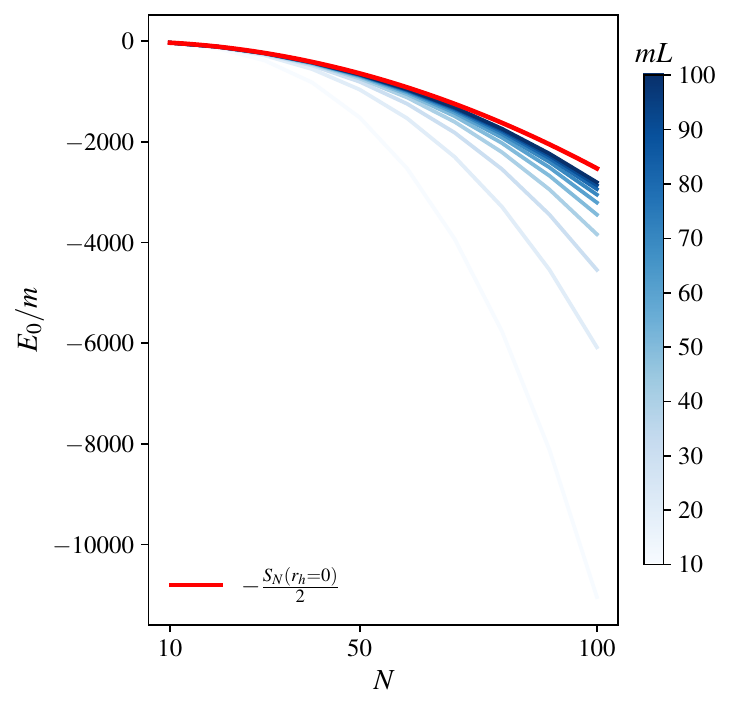}
    \includegraphics[width=0.32\linewidth]{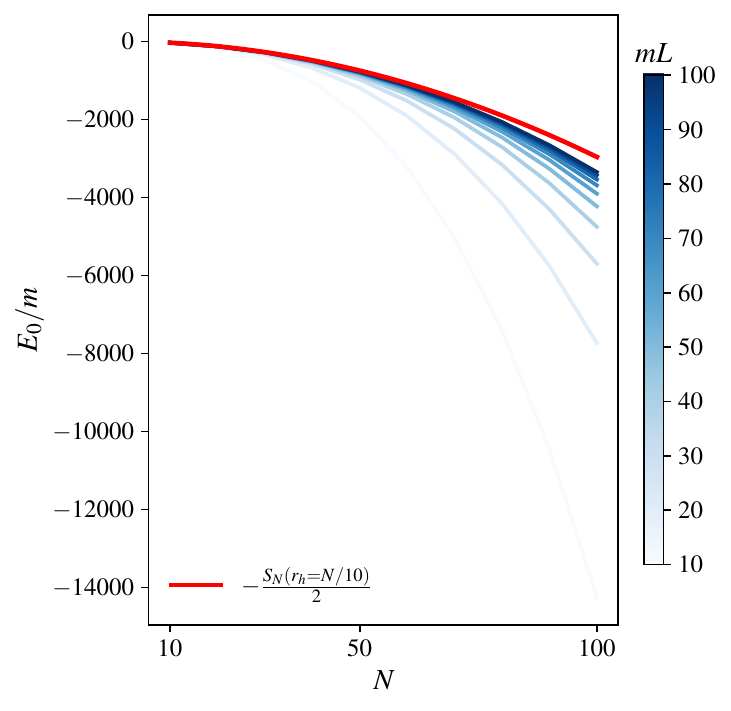}
    \includegraphics[width=0.32\linewidth]{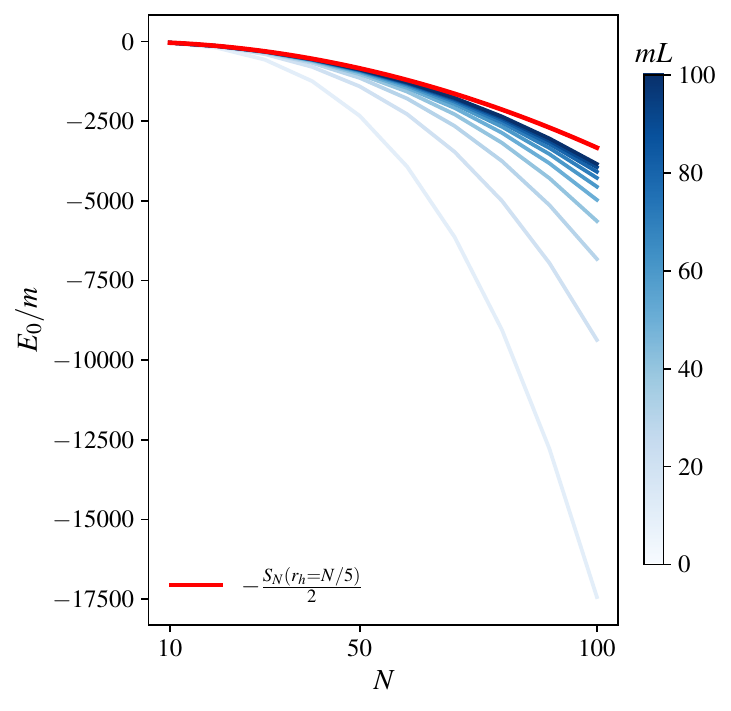}
    \caption{The ground state energy 
per mass $E_0/m$ as a function of the system size $N$, at $\mu=0, L=1$, for four choices of the horizon radius $r_h\in\{0,\;N/10,\;N/5\}$ (from left to right panels).}
    \label{fig:E0_m_100}
\end{figure}

In Fig.~\ref{fig:E0_m_100} we plot the ground state energy 
per mass $E_0/m$ as a function of the system size $N$, at $\mu=0$.
The three panels corresponding (from left to right) to horizon radii
$r_h = 0,\quad r_h = \tfrac{N}{10},\quad r_h = \tfrac N5$.
The colored curves track  different dimensionless fermion mass $mL$.
As $mL$ increases, the curves flatten. Indeed, as we increase $m$,
we decrease the Compton wavelength of the fermion $\lambda_c \sim 1/m$,
so that $\lambda_c\ll L_{box}\sim a N$, and the particle’s wavefunction is localized on scales much smaller than the box.
Since the finite size of the lattice is effectively invisible to fermions, the ground state energy per mass settles to a constant as we increase $N$.
At small $N$, varying $r_h$ has a pronounced effect on $E_0/m$, where
larger $r_h$ lowers the energy due to stronger redshift.
As $N$ increases, all curves for different $r_h$ converge to the same value, indicating that in the continuum limit the black hole redshift effects become subleading at fixed mass.

We can also verify analytically the curves in Fig.~\ref{fig:E0_m_100}: on the left panel we have a summation of the arithmetic series, while the middle and the right panels can be matched at large mass and large $N$ to (\ref{series}).

\begin{figure}[H]
    \centering
    \includegraphics[width=0.32\linewidth]{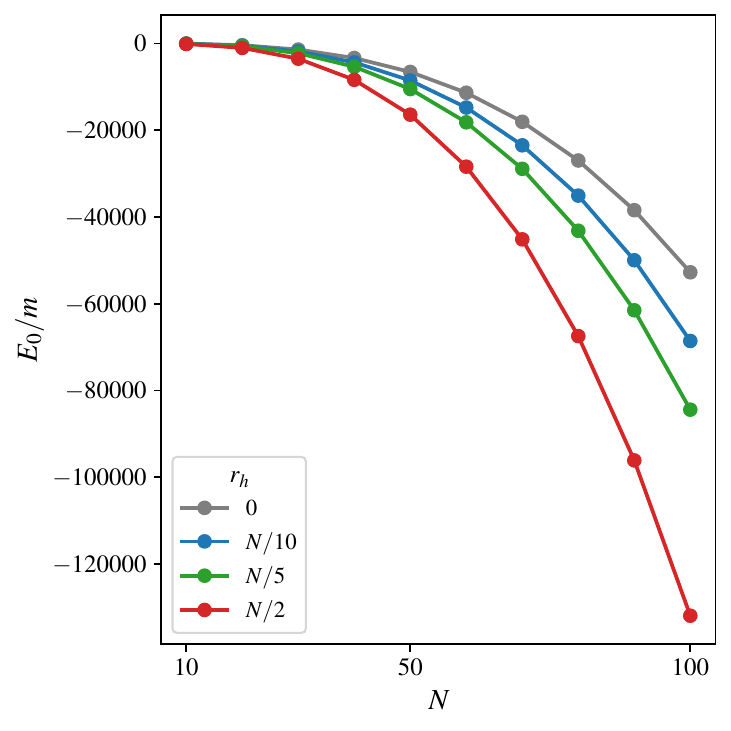}
    \includegraphics[width=0.32\linewidth]{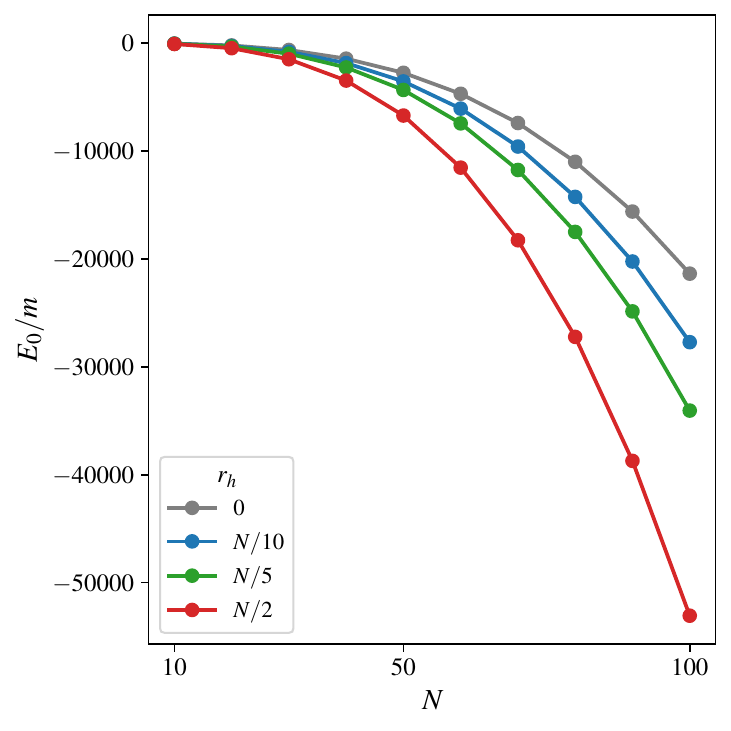}
    \includegraphics[width=0.32\linewidth]{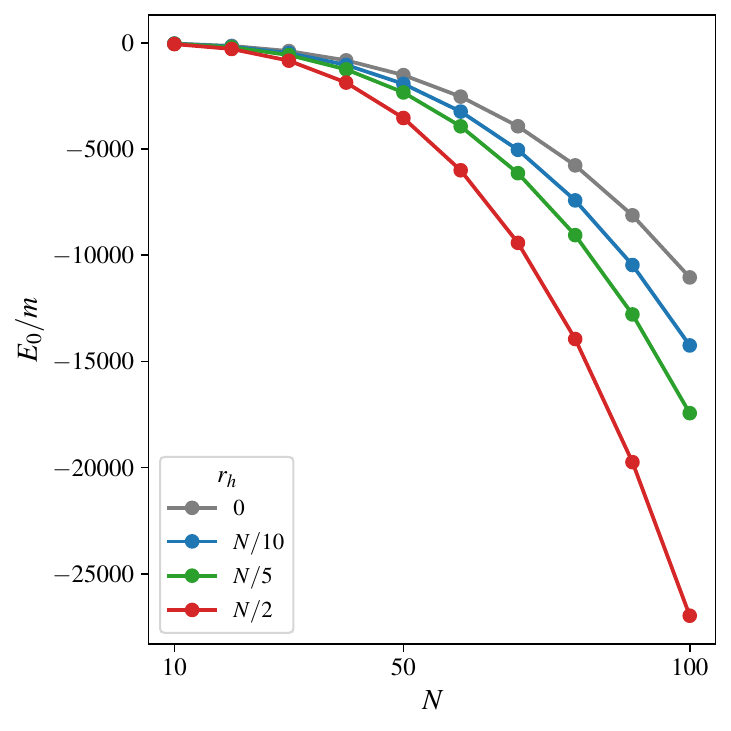}
    \caption{The ground state energy per mass,
$\frac{E_0}{m}$, as a function of $N$, at $\mu=0$, for three different fermion masses $mL=2,5,10$ and $L=1$ (from left to right panel), and for three choices of the horizon radius $r_h\in\{0,\;N/10,\;N/5,\;N/2\}$.}
    \label{fig:three_r0_m}
\end{figure}

Fig.~\ref{fig:three_r0_m} shows how the ground‐state energy per mass,
$\frac{E_0}{m}$,
scales with the total number of lattice sites $N$ (the box size), for three different fermion masses and for several choices of the $AdS_2$ horizon radius $r_h$. 
At small $N$, we see that increasing $r_h$ pulls the energy downward (stronger gravitational redshift). As $N$ grows, all of these curves converge to the same asymptote, meaning that in the continuum limit (large box) the black hole redshift becomes a subleading effect at fixed fermion mass.
For lighter fermions ($mL=2$), the curves are steeper with $N$ since the finite box supports many low-energy modes, hence $\frac{E_0}{m}$ grows (in absolute value) roughly linearly in $N$.
As $mL$ increases ($mL=5,10$), the curves flatten out sooner, since a heavy fermion’s Compton wavelength $\lambda_c\sim \frac{1}{m}$ becomes much smaller than the box, so the lattice’s finite size is no longer seen by the particle and $\frac{E_0}{m}$ saturates.

More precisely,  we see a decreasing increment $\Delta(E_0/m)$ per $\Delta N$ as $mL$ grows. 
To see the zero slope flattening for $mL=10$, we need to increase $N$ by an order of magnitude.
When $r_h=0$ there is no gravitational redshift, so the only thing setting the ground‐state energy is the finite box of length $L_{\rm box}\sim aN$.  That means at small $N$ we see the linear-in-$N$ growth for light fermions, since more modes fit in as you enlarge the box, and a flattening for heavy fermions once $\frac{1}{m}\ll L_{\rm box}$.
In summary, Fig.~\ref{fig:three_r0_m} illustrates how both finite-size (small $N$) and curvature/redshift ($r_h\neq0$) effects interplay, and that in the large-$N$ (continuum) limit these gravitational corrections become negligible, especially for heavier fermions.

\subsubsection{Charge}
\if{
\begin{figure}[H]
    \centering
    \includegraphics[width=0.49\linewidth]{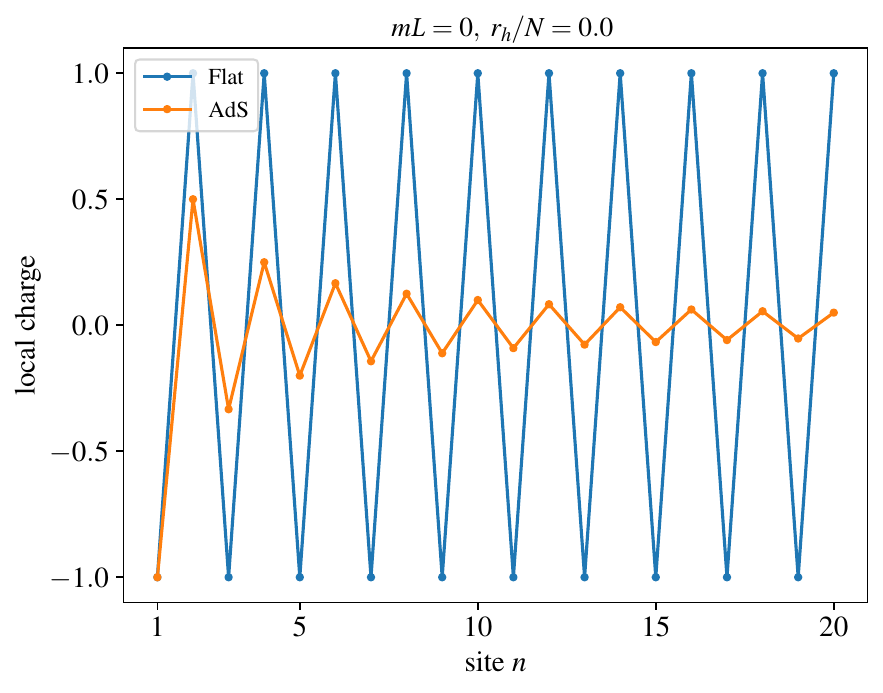}
    \includegraphics[width=0.49\linewidth]{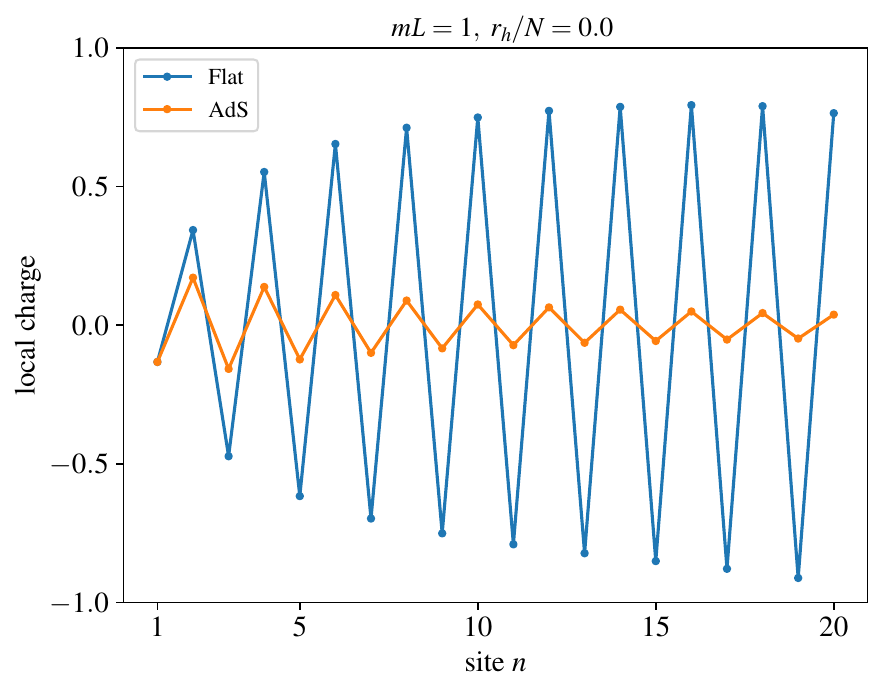}
    \includegraphics[width=0.49\linewidth]{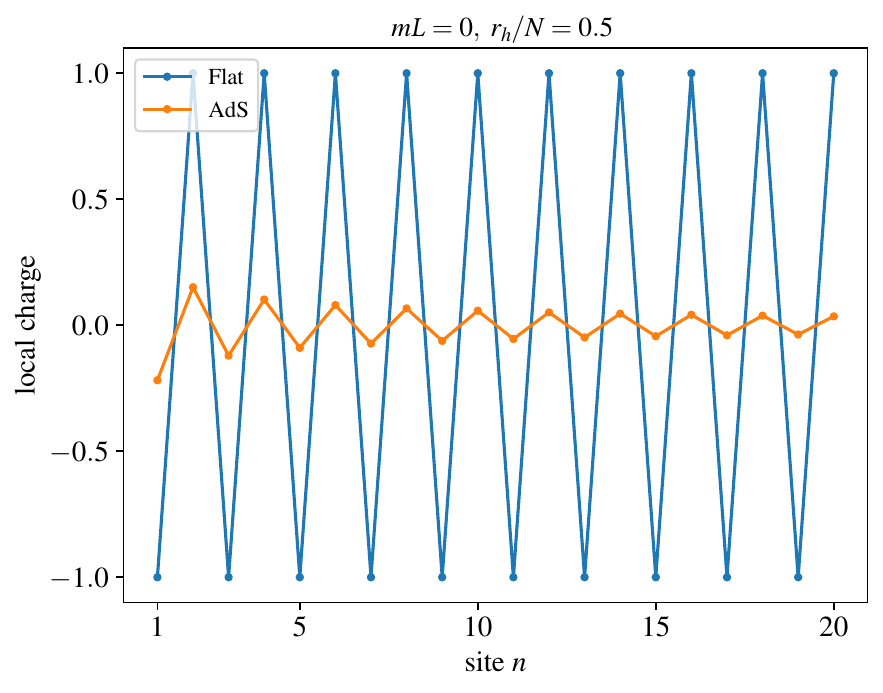}
    \includegraphics[width=0.49\linewidth]{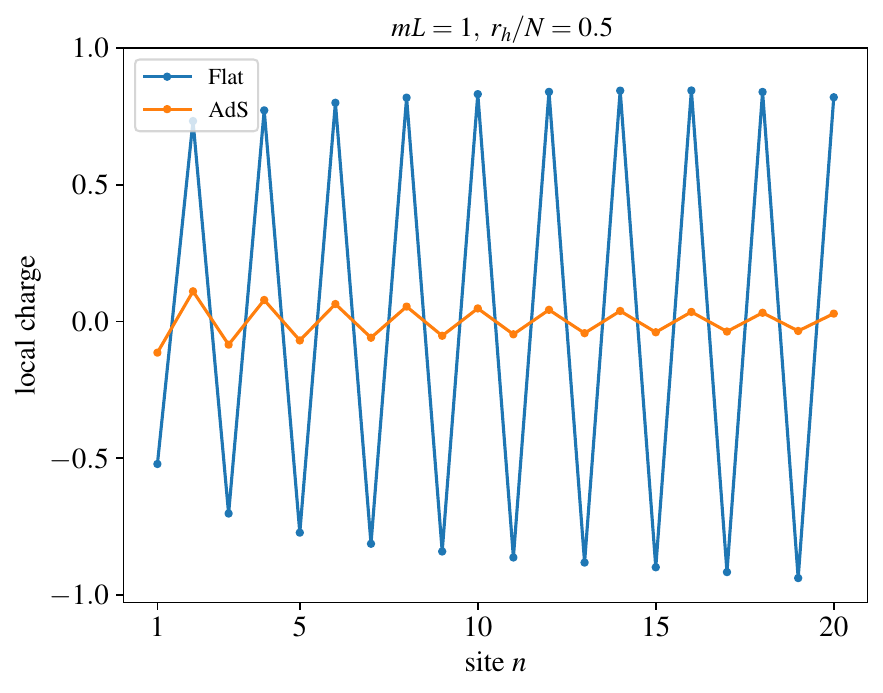}
    \caption{Local charge evaluated by $Q_{\text{flat}}(n)=\frac{Z_n+(-1)^n}{2}$ and $Q_\text{AdS}(n)=w_nQ_\text{flat}(n)$. $\mu=0,N=20$. \ki{$m,L,r_h$}}
    \label{localcharge}
\end{figure}
}\fi

Consider the site‐by‐site local charge profiles. 
We will consider both the flat charge (\ref{flat}) and the weighted charge (\ref{QC}). 
As discussed in Section 3.3, the ground state weighted charge is essentially flat across the lattice, with small, localized ripples at the horizon and a curvature‐driven drift across the full system.
More precisely, in the deep bulk $i \gg 1$, a constant amplitude dominates, 
while on the scale of the $AdS_2$ radius $i\,a\sim L$ there is a $O(\frac{1}{L})$ modulation of the filling,
and near the horizon $i\sim 1$, there is an $O(\frac{1}{r_{h}})$ deviation that decays into the bulk.

Consider next the effect of the mass on the local charge density.
Heavier fermions are less influenced by the horizon 
since their Compton wavelength $\lambda_{c}\sim \frac{1}{m}$ sets how far quantum modes probe the region near $r_{h}$.  
The horizon‐induced charge ripple is governed by:
\begin{equation}
\delta q_{i}^{\rm (Horizon)}
=\frac{1}{r_{h}}\,h\!\bigl(m r_{h},\,\mu r_{h};(i-1)\bigr) \ , 
\label{horizoneffect}
\end{equation}
where $h(x,y;0)$ is the peak amplitude at the site right against the horizon.  
Light fermions have $\frac{m}{r_{h}}\ll 1$, hence $h(m\,r_{h},\mu\,r_{h};0)$ is $O(1)$, and the local‐charge deviation at the horizon is of order $\frac{1}{r_{h}}$.
Heavy fermions have $m\,r_{h}\gg 1$, the modes are exponentially suppressed over distances of order $\frac{1}{m}$, so that $h(m\,r_{h},\mu\,r_{h};0)\sim e^{-\,m\,r_{h}}\ll 1$.  The horizon ripple amplitude becomes $\sim e^{-m r_{h}}/r_{h}$, hence negligible compared to the bulk.
When we set $r_{h}=0$, the horizon contribution 
$\frac{1}{r_{h}}\,h(mr_{h},\mu r_{h})$
is ill‐defined at $r_{h}=0$, but physically it disappears since there is no horizon.

Similarly, the curvature‐driven modulation of the site charge
\begin{equation}
\delta q_i^{\rm (AdS)} \;=\; \frac{1}{L}\;g\!\Bigl(\tfrac{r_h}{L},\,mL,\,\mu L;\,\tfrac{i\,a}{L}\Bigr) \ ,   
\label{AdSeffect}
\end{equation}
is controlled by the dimensionless mass parameter $mL$.  
Hence for light field $mL\ll 1$,
the Compton wavelength is large compared to the $AdS_2$ radius and
\begin{equation}
g\bigl(r_h/L,\;mL,\;\mu L;\;x\bigr)\;\sim\;O(1)
\quad\Longrightarrow\quad
\delta q_i^{(AdS)}\sim \frac{1}{L} \ ,  
\end{equation}
leading to $O(\frac{1}{L})$ tilt across the entire lattice.
Heavy fermions  $mL\gg1$ have a small Compton wavelength and low‐energy modes do not probe the curvature deeply.
Thus,
\begin{equation}
  g\bigl(r_h/L,\;mL,\;\mu L;\;x\bigr)\;\propto\;e^{-\,mL} \ ,  
\end{equation}
and the curvature‐induced drift is exponentially suppressed:
$\delta q_i^{AdS}\;\sim\;\frac{e^{-\,mL}}{L}\;\ll\;\frac{1}{L}$.

\begin{figure}[H]
    \centering
    \includegraphics[width=0.49\linewidth]{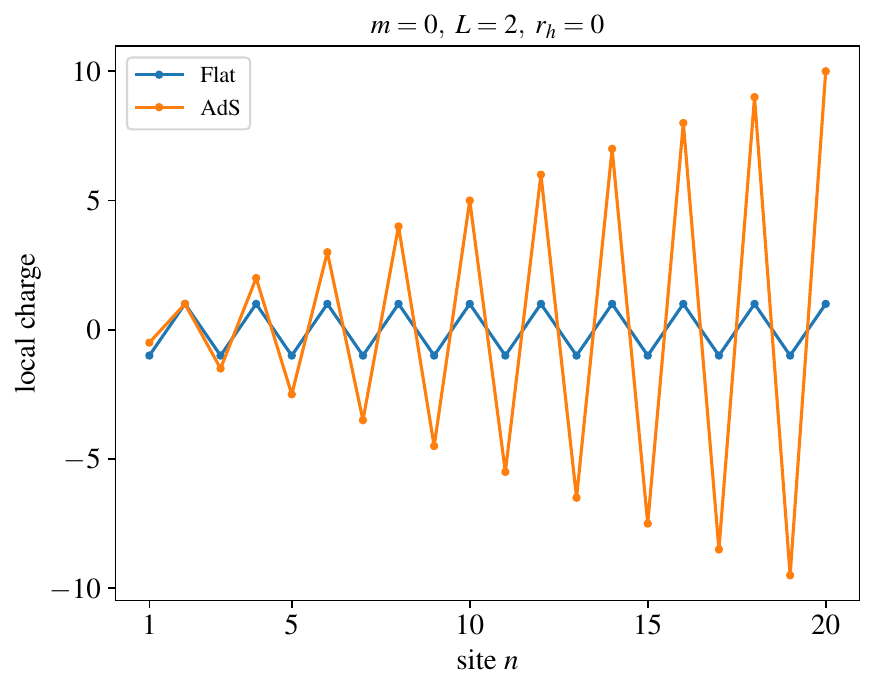}    \includegraphics[width=0.49\linewidth]{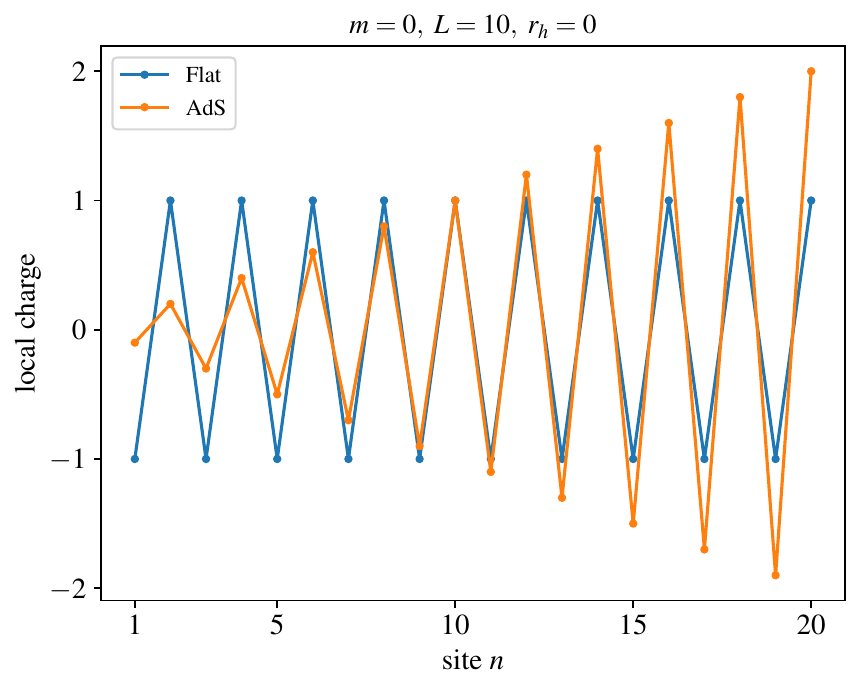}
    \caption{The flat and weighted charges for pure $AdS_2$ ($r_h=0$), 
two choices of $AdS_2$ radius $L$ (2 vs.\ 10), and mass $m=0$. The weighted charge is affected by the 
$AdS$ curvature and differs from the flat charge (left panel). This effect decreases as the radius $L$ increases, $O(\frac{1}{L})$, (right).}
    \label{m0}
\end{figure}

In Fig. \ref{m0}, we plot the flat and weighted charges for pure $AdS_2$ ($r_h=0$), 
two choices of $AdS_2$ radius $L$ (2 vs.\ 10), and mass $m=0$. The weighted charge is affected by the 
$AdS$ curvature and differs from the flat charge (left panel). This effect decreases as the radius $L$ increases,  $O(\frac{1}{L})$, (right) (\ref{AdSeffect}). In general, the local charge is negative for odd sites and positive for even sites.  
These odd-even sites oscillations of the local charge are a consequence of the 
$(-1)^n$ term. 
In the continuum, these oscillations average away over distances $\gg a$,
and the charge density has no built-in oscillations as seen between the odd-even sites. Physically, low-momentum observables live on length scales large compared to the lattice spacing, so any $\pi/a$ oscillatory piece is non-universal and disappears in the infrared physics.

\begin{figure}[H]
    \centering
    \includegraphics[width=0.49\linewidth]{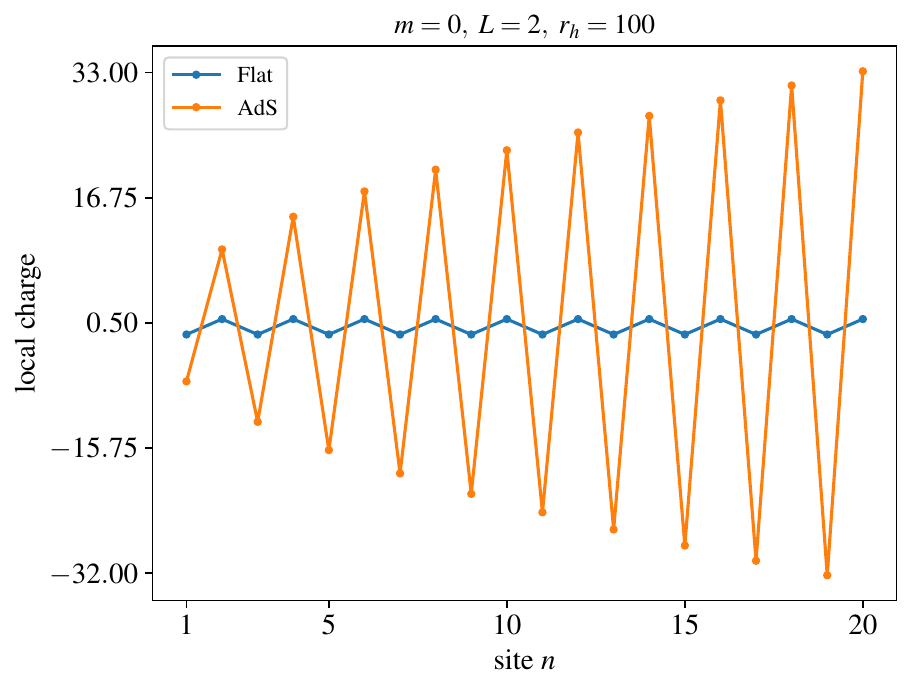}
    \includegraphics[width=0.49\linewidth]{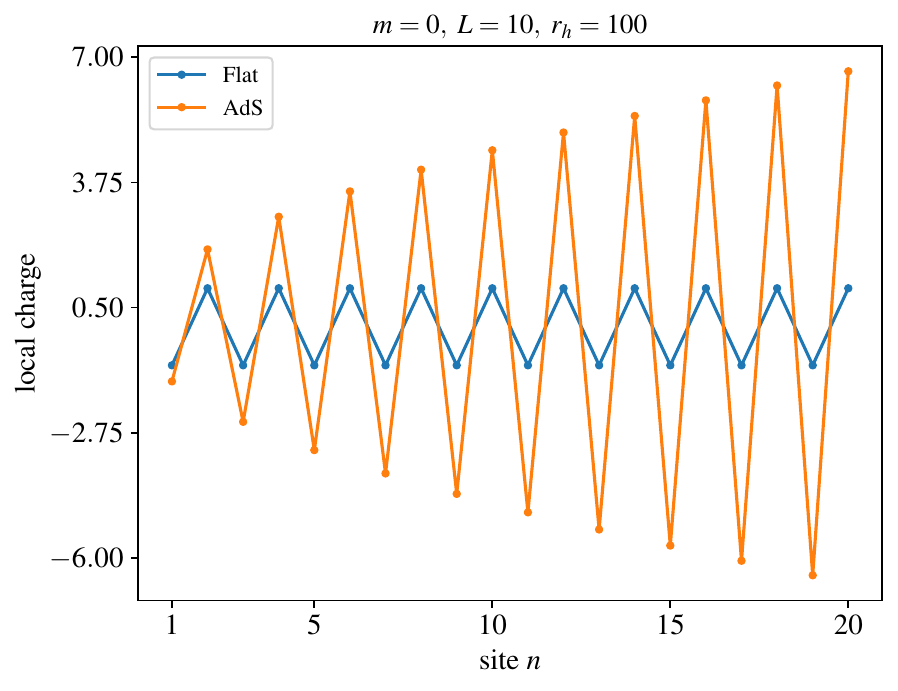}
    \caption{The flat and weighted charges for $AdS_2$ black hole
with a large horizon radius ($r_h=100$), 
two choices of $AdS_2$ radius $L$ (2 vs.\ 10), and mass $m=0$. The weighted charge is affected by 
the horizon (site $1$ and its neighborhood). The 
$AdS$ curvature affects the bulk site, which is large in the left panel since $\frac{L}{r_h}$ is small,
and decreases as we increase the radius $L$ (right).
 }
    \label{r100}
\end{figure}

In Fig. \ref{r100}, we plot the flat and weighted charges for $AdS_2$ black hole
with a large horizon radius ($r_h=100$), 
two choices of $AdS_2$ radius $L$ (2 vs.\ 10), and mass $m=0$. The weighted charge is affected by the 
the horizon (site $1$ and its neighborhood): the near-horizon ripple scales as $O(1/r_h)$ and decays into the bulk
(\ref{horizoneffect}). The 
$AdS$ curvature affects the bulk site, which is large in the left panel since $\frac{L}{r_h}$ is large,
and decreases as we increase the radius $L$,$O(\frac{1}{L})$, (right).

\begin{figure}[H]
    \centering
    \includegraphics[width=0.49\linewidth]{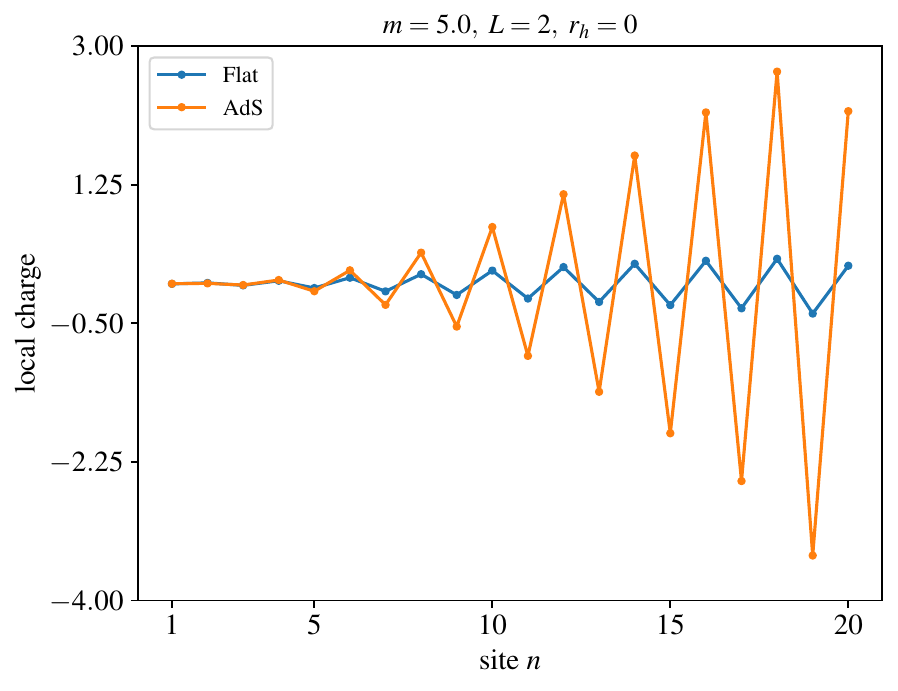}
    \includegraphics[width=0.49\linewidth]{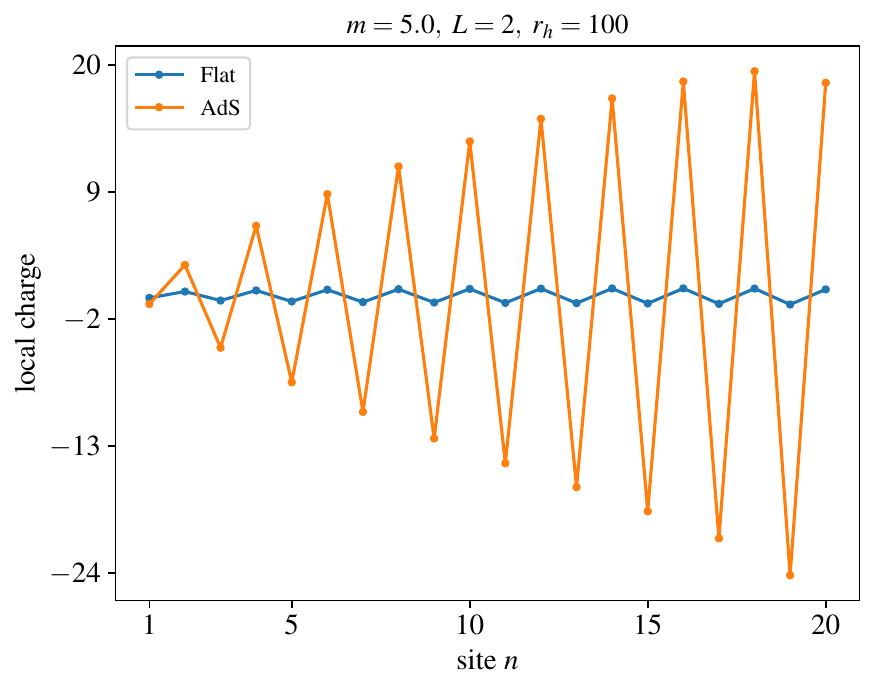}
    \caption{Raising the fermion mass to $m=5$ shortens the fermion’s Compton wavelength,
so that it cannot resolve the length scales set by $L$ or $r_h$. 
Here, $mL=10$ and $m r_h\in\{0,500\}$, 
so both $e^{-mL}$ and $e^{-m r_h}$ are tiny. Thus,
the effects of the $AdS_2$ curvature and the horizon red shift are suppressed, and the fermionic
system tends to exhibit uniform filling.}
\label{m5}
\end{figure}

Raising the fermion mass $m$ shortens the fermion’s Compton wavelength,
so that it cannot resolve the length scales set by $L$ or $r_h$. Here, $mL=10$ and $m r_h\in\{0,500\}$, 
so both $e^{-mL}$ and $e^{-m r_h}$ are tiny, as discussed above.  Thus,
the effects of the $AdS_2$ curvature and the horizon redshift are suppressed, and the fermionic
system tends to exhibit uniform filling.
Note that the absolute vertical offset of $q_i$  
includes the homogeneous bulk term $q(ma,\mu a)$ (\ref{BT}), which varies with $m$ (and $\mu$).  Fig. \ref{m5} highlights the site-to-site variation (the $n$-dependence), which is strongly suppressed for heavy $m$. Therefore, direct comparisons of the absolute vertical axis range with earlier figures, e.g. Fig. \ref{m0}, are not meaningful.

\if{

\begin{figure}[H]
    \centering
    \includegraphics[width=0.49\linewidth]{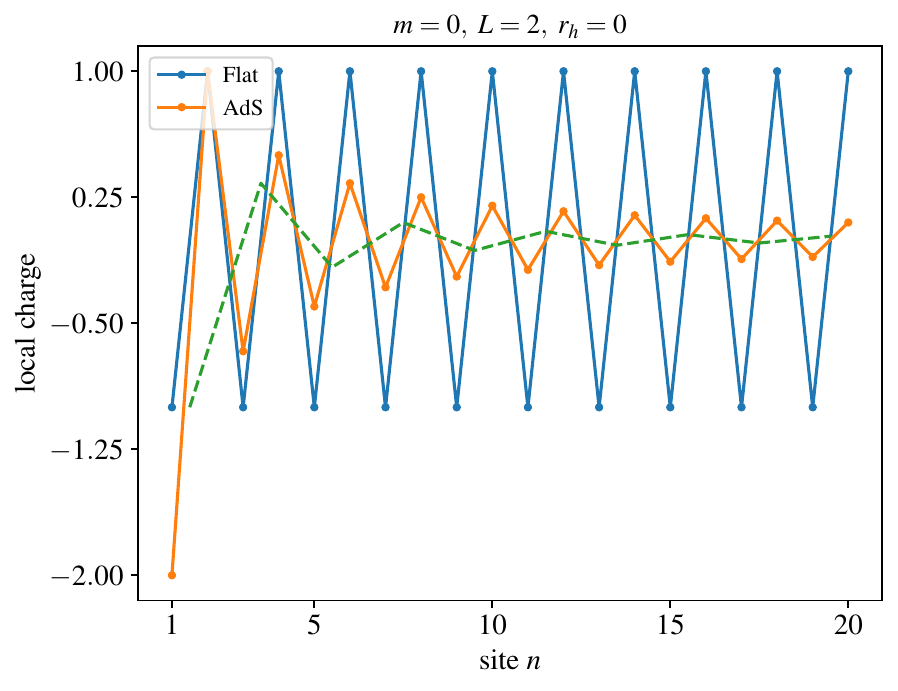}
    \includegraphics[width=0.49\linewidth]{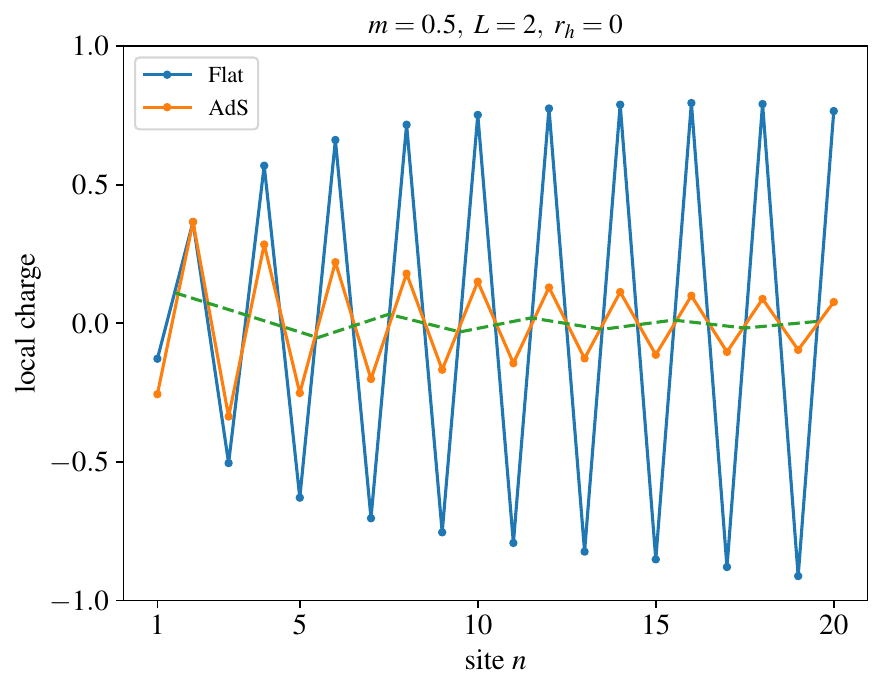}
    \includegraphics[width=0.49\linewidth]{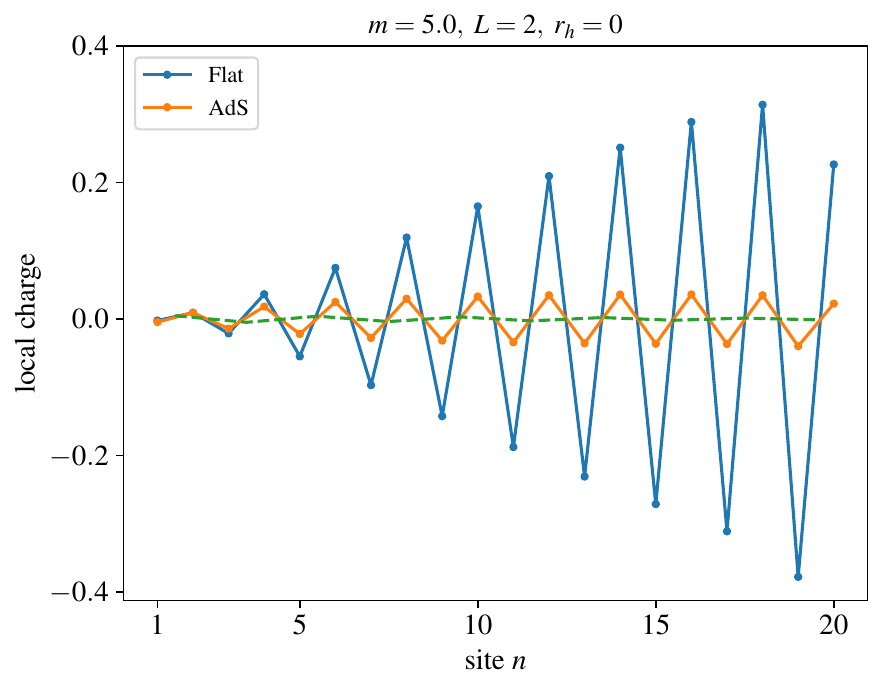}
    \includegraphics[width=0.49\linewidth]{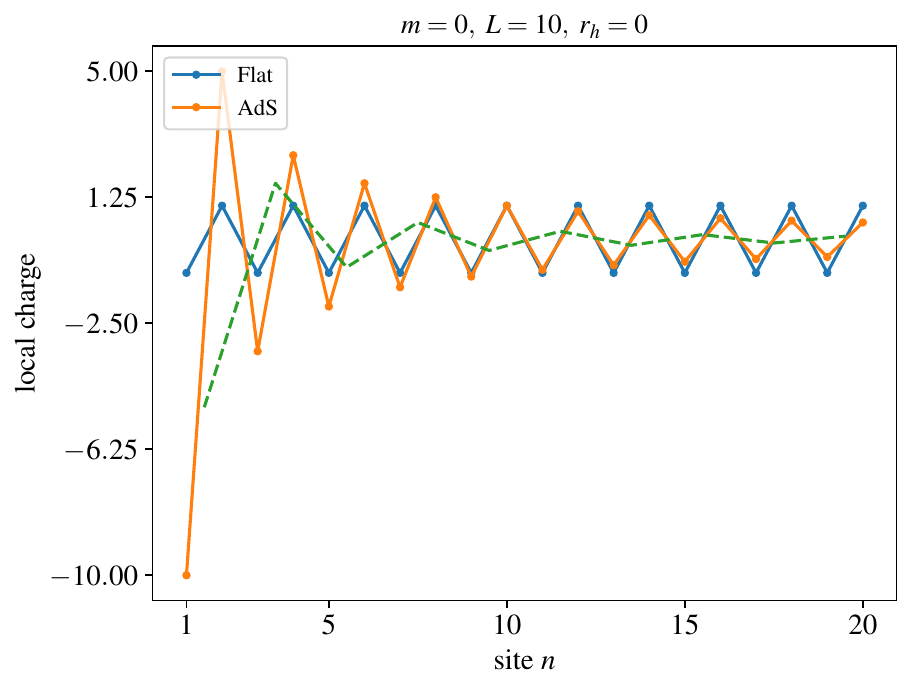}
    \includegraphics[width=0.49\linewidth]{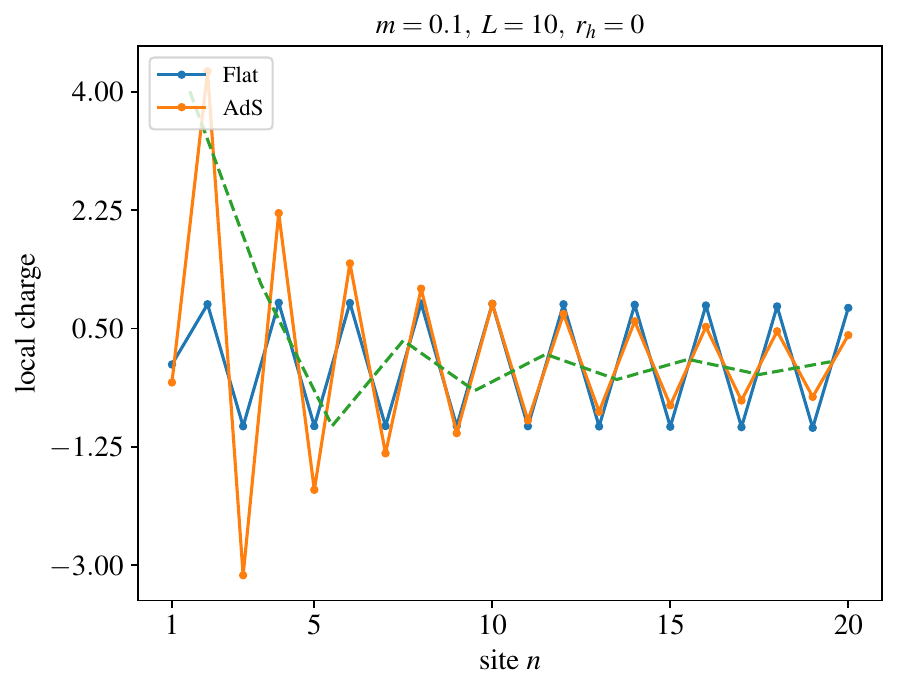}
    \includegraphics[width=0.49\linewidth]{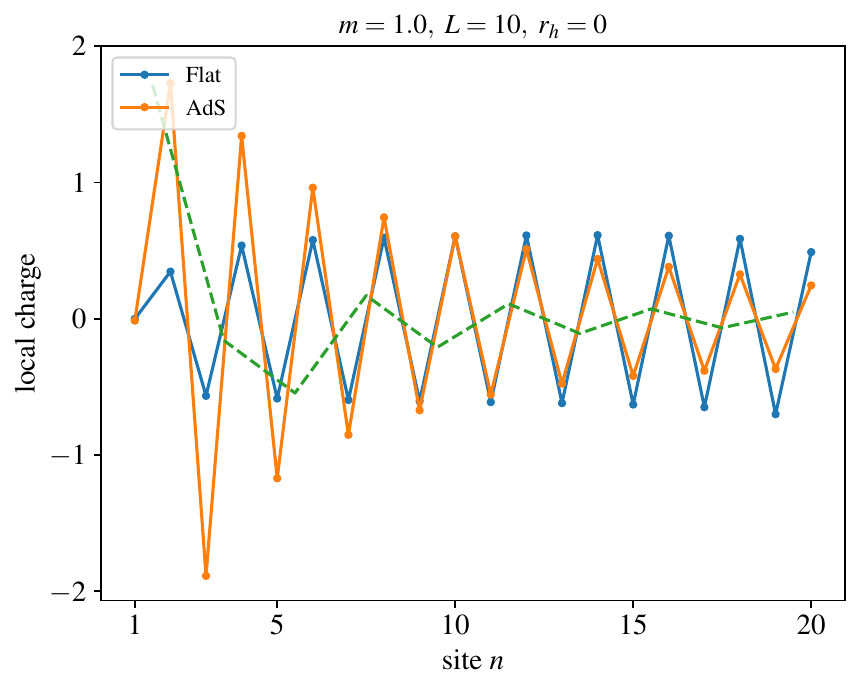}
    \caption{For $r_h=0$. The dashed line is physical charge $Q_{AdS,n}+Q_{AdS, n+1}$ in AdS. }
    \label{fig:enter-label}
\end{figure}

\begin{figure}[H]
    \centering
    \includegraphics[width=0.49\linewidth]{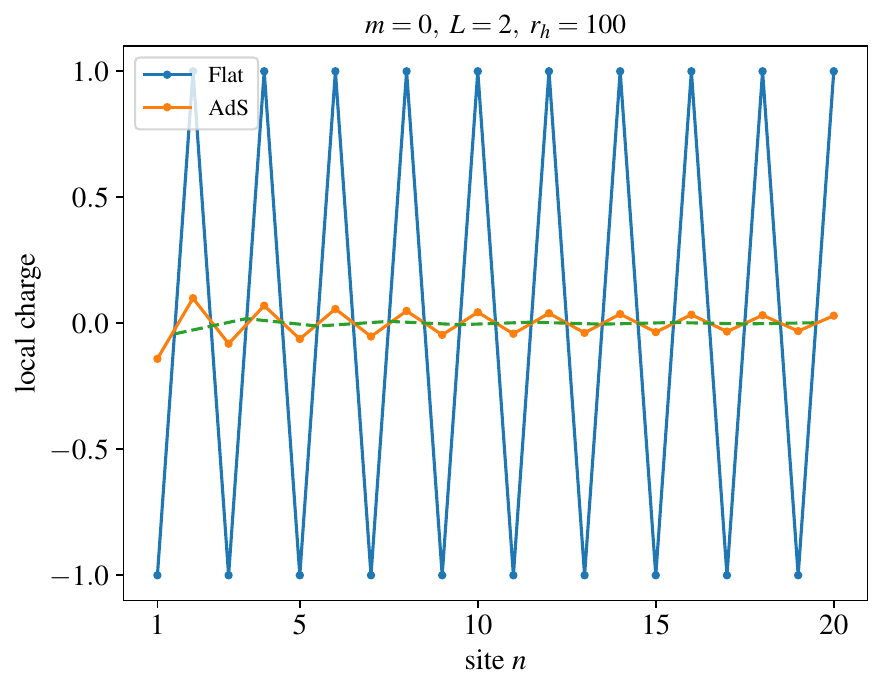}
    \includegraphics[width=0.49\linewidth]{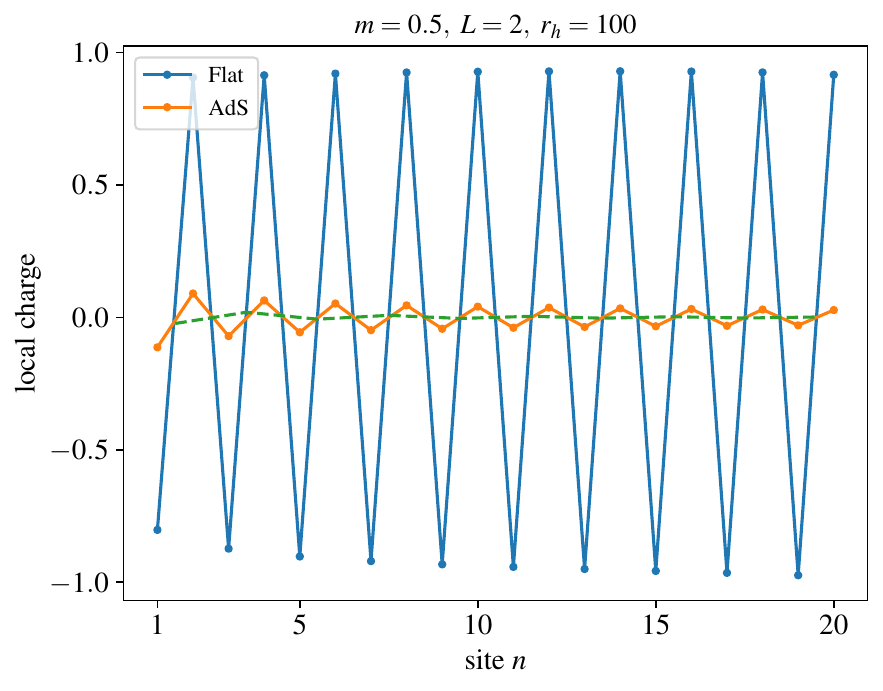}
    \includegraphics[width=0.49\linewidth]{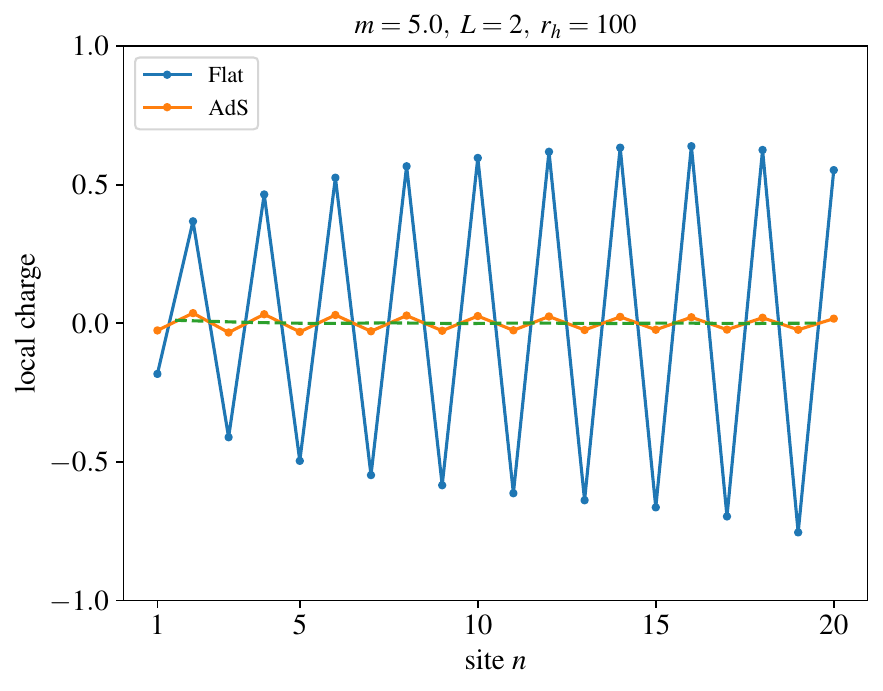}
    \includegraphics[width=0.49\linewidth]{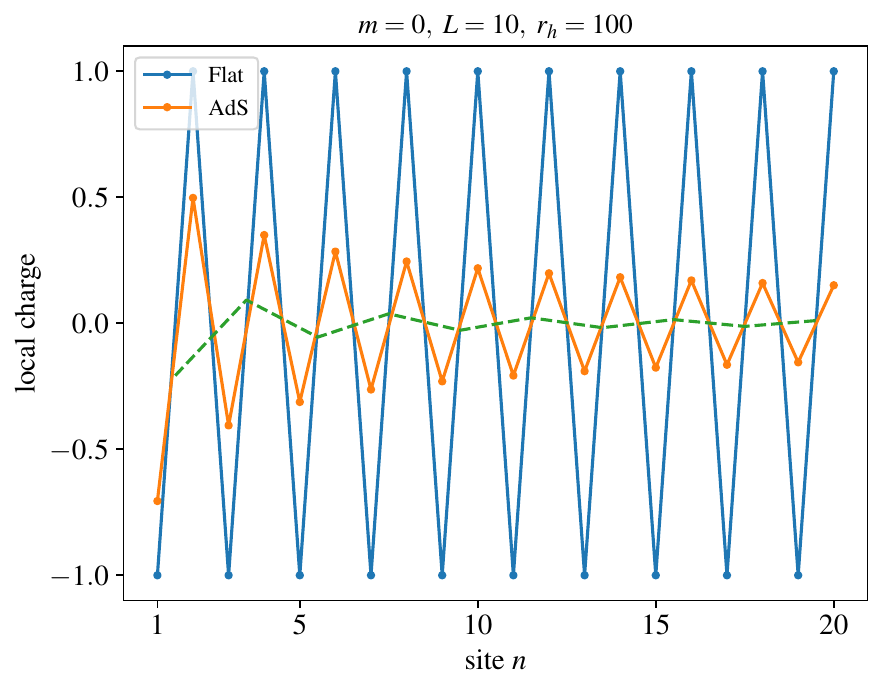}
    \includegraphics[width=0.49\linewidth]{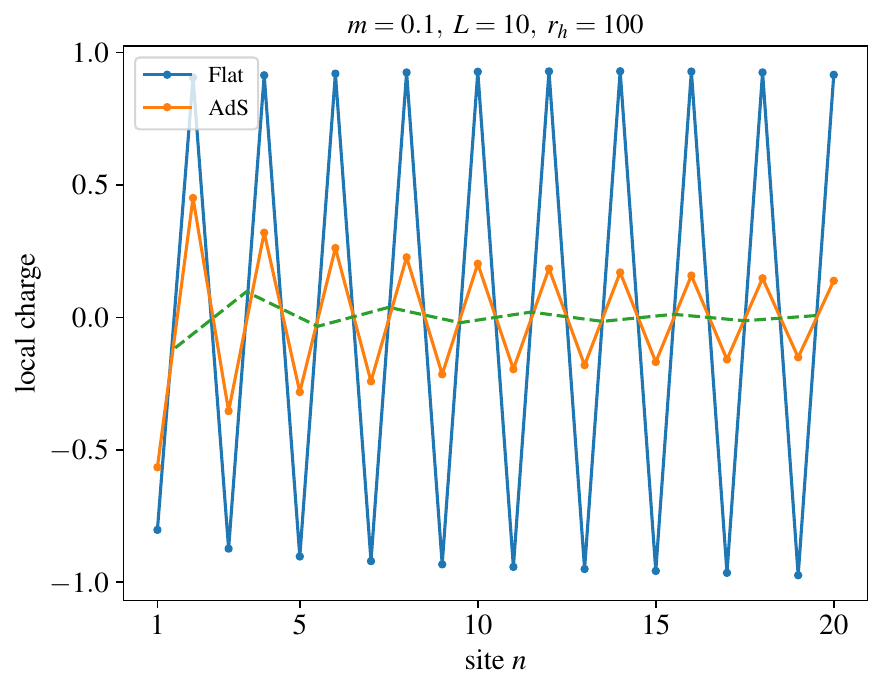}
    \includegraphics[width=0.49\linewidth]{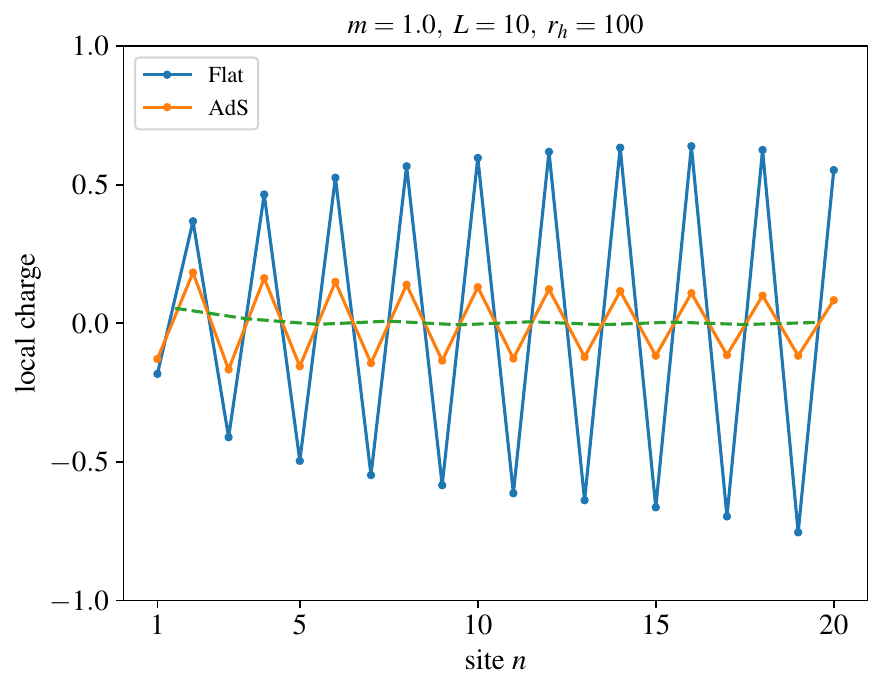}
    \caption{For $r_h=100$. The dashed line is physical charge $Q_{AdS,n}+Q_{AdS, n+1}$ in AdS. }
    \label{fig:enter-label}
\end{figure}

\begin{figure}[H]
    \centering
    \includegraphics[width=0.49\linewidth]{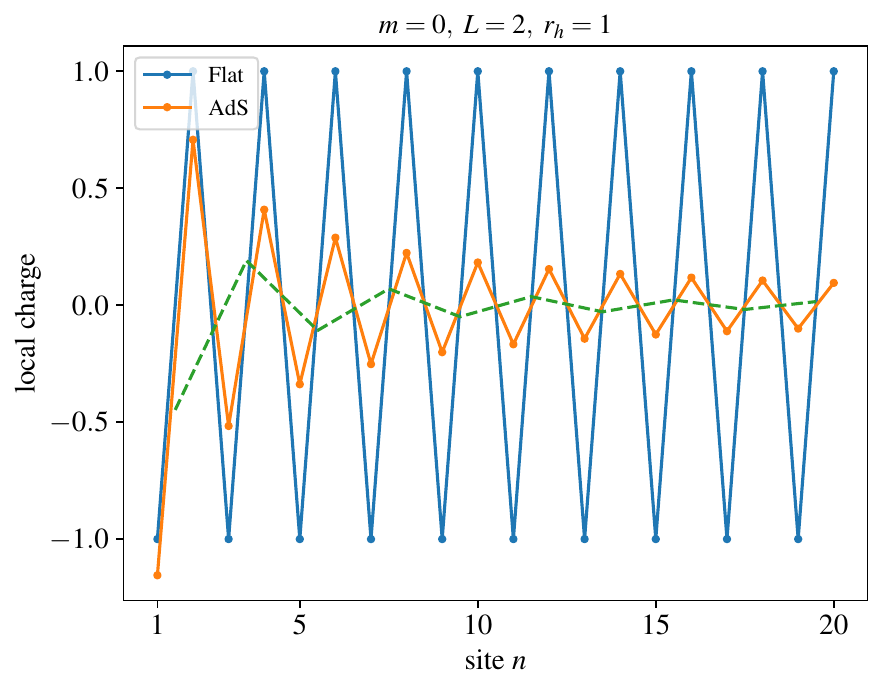}
    \includegraphics[width=0.49\linewidth]{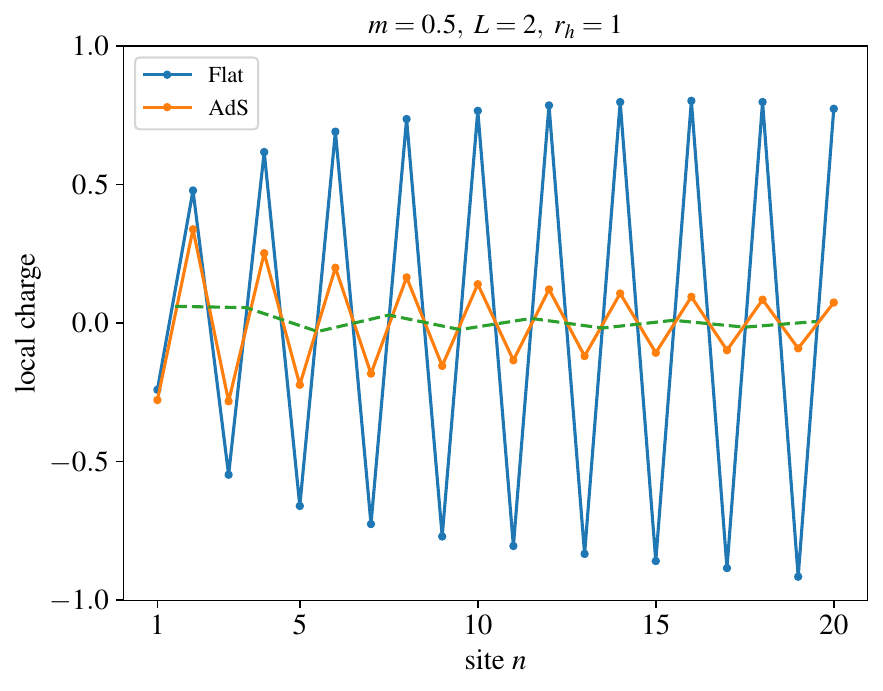}
    \includegraphics[width=0.49\linewidth]{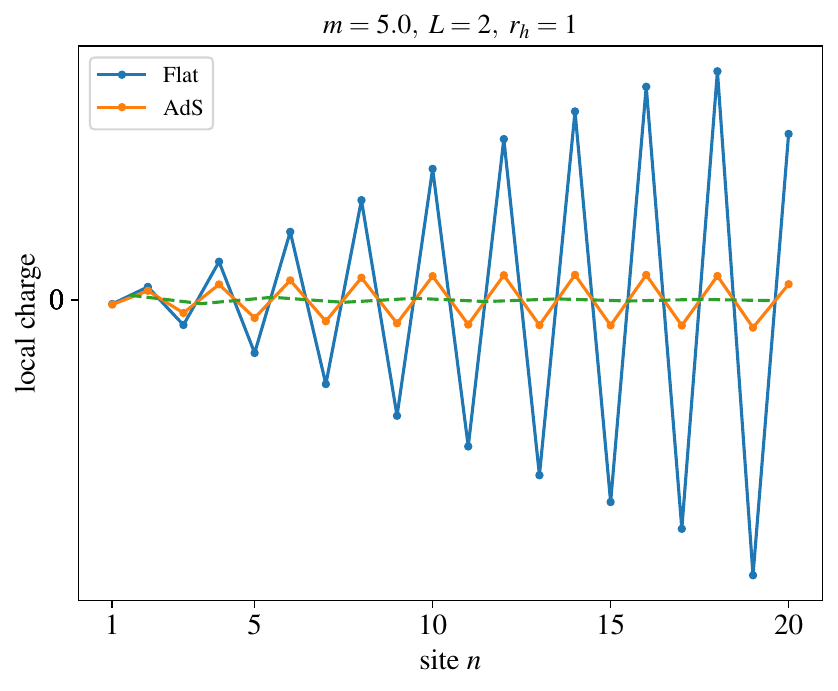}
    \includegraphics[width=0.49\linewidth]{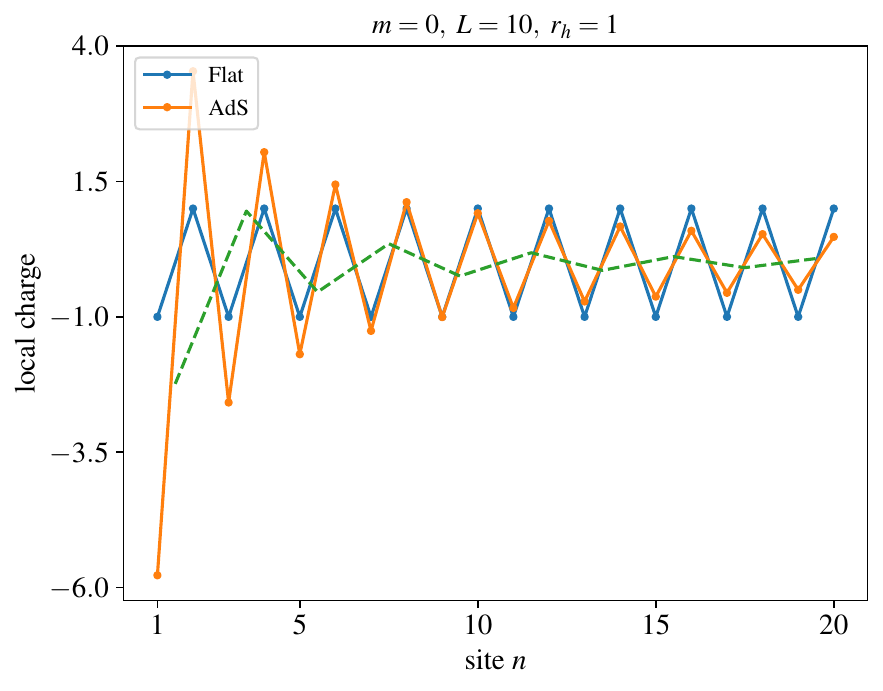}
    \includegraphics[width=0.49\linewidth]{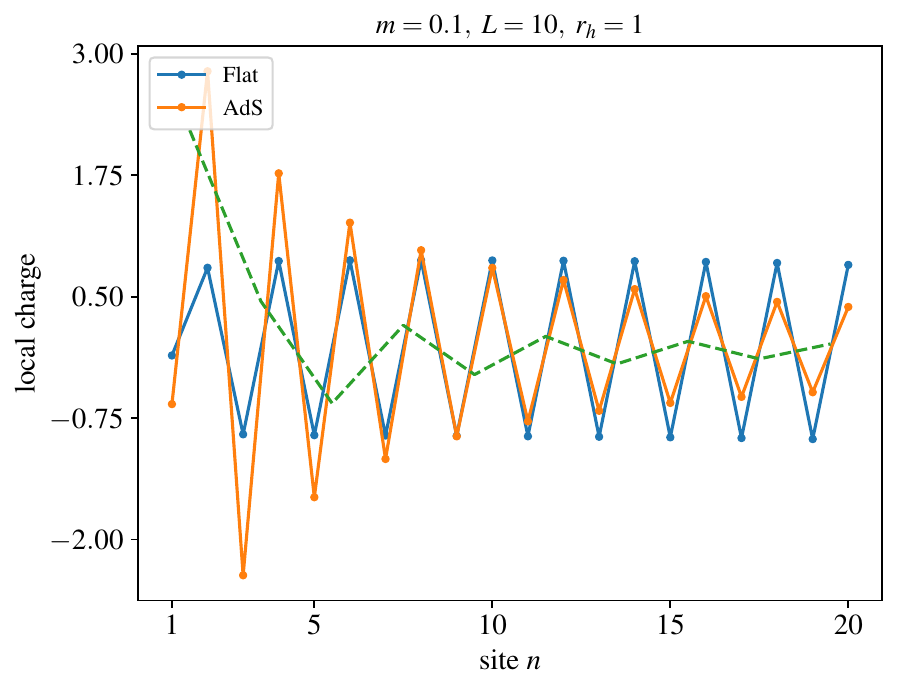}
    \includegraphics[width=0.49\linewidth]{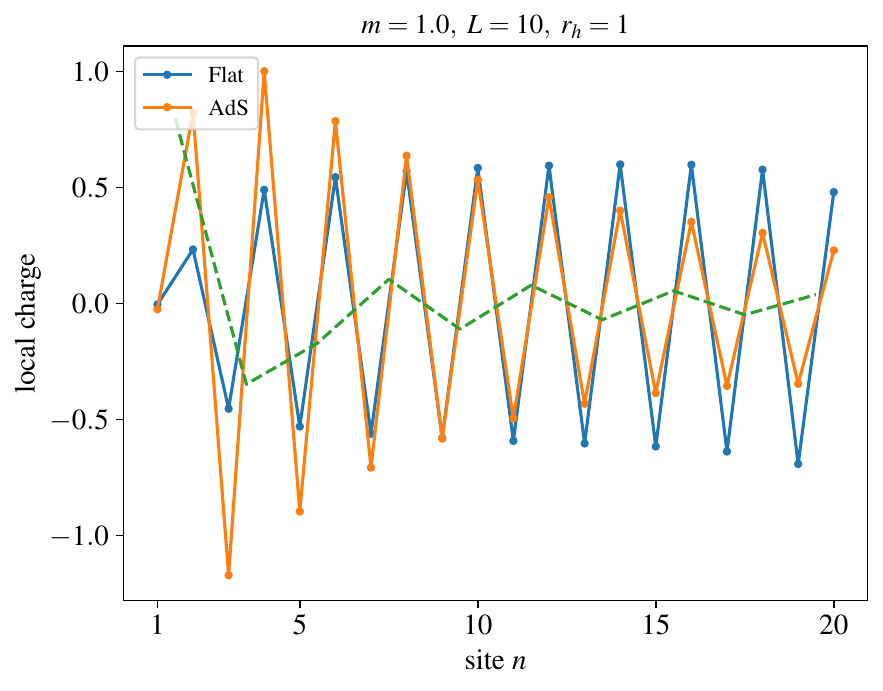}
    \caption{For $r_h=1$. The dashed line is the physical charge $Q_{AdS,n}+Q_{AdS, n+1}$ in AdS. }
    \label{fig:enter-label}
\end{figure}

}\fi

\begin{figure}[H]
    \centering
    \includegraphics[width=0.49\linewidth]{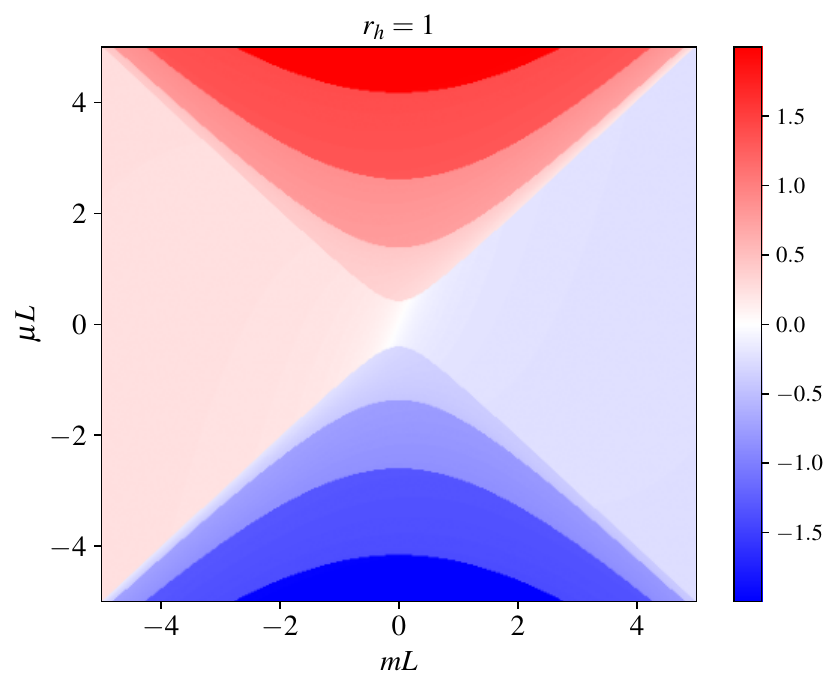}
    \includegraphics[width=0.49\linewidth]{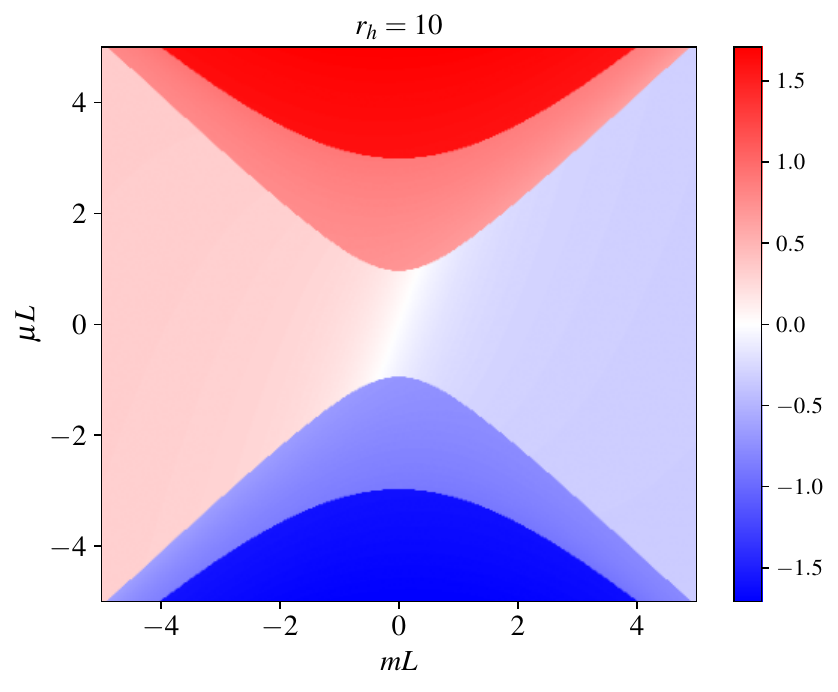}
    \caption{The expectation value of the weighted charge density $\langle Q_{weighted}\rangle/N$ for $N=12$ and $L=1$ in the ground state. The structure reflects the transition from the vacuum polarization regime ($|\mu|<|m|$) to the filled‐sea regime ($|\mu|>|m|$), and how both curvature and redshift (\ref{redshift}) modify those expectation values.
     We see the symmetry: $m\rightarrow -m,~\mu \rightarrow -\mu,~ Q_{weighted}\rightarrow -Q_{weighted}$.}
    \label{fig:gs_physical_charge}
\end{figure}

In Fig.~\ref{fig:gs_physical_charge} we plot the 
global charge density heatmap, i.e. 
the average weighted charge per site, $\langle Q_{weighted}\rangle/N$, as a function of $(mL,\mu L)$, for two horizon radii 
$r_h=1$ (left) and $r_h=10$. (right). Color scale: Blues indicate net negative charge density, reds net positive.
The key features of the heatmap are, a pronounced dip (blue) for $|\mu|<|m|$ (gapless cone in the continuum; near-zero gaps at finite $N$). As $r_h$ grows, the entire color range shrinks (peak-to-trough amplitude falls), reflecting that stronger redshift (\ref{redshift}) suppresses net polarization. There is a symmetry: $m\rightarrow -m,~\mu \rightarrow -\mu,~ Q_{weighted}\rightarrow -Q_{weighted}$.

\begin{figure}[H]
    \centering
\includegraphics[width=0.49\linewidth]{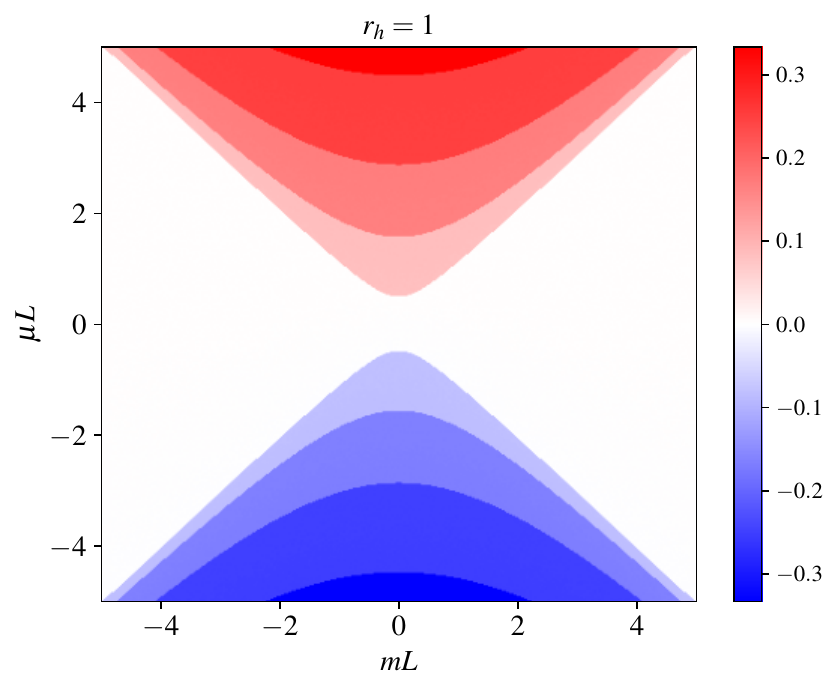}
    \includegraphics[width=0.49\linewidth]{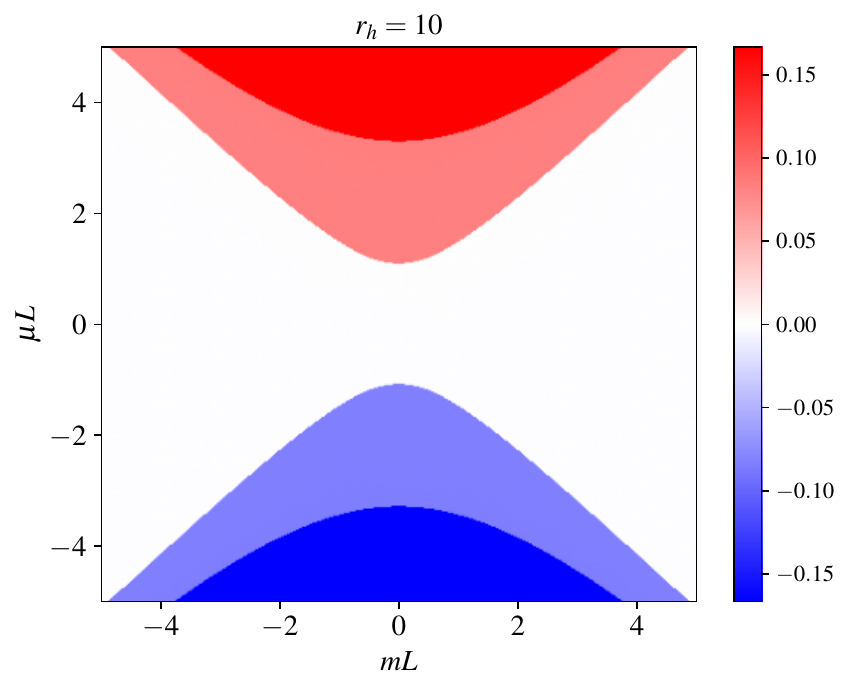}
    \caption{The expectation value
of the site-occupation operator $\langle Q_{flat}\rangle/N$ for $N=12$ and $L=1$ in the ground state.
The structure reflects the transition from the vacuum polarization regime ($|\mu|<|m|$) to the filled‐sea regime ($|\mu|>|m|$), and how both curvature and redshift (\ref{redshift}) modify those expectation values.
We see the symmetry: $m\rightarrow -m,~\mu \rightarrow -\mu,~ Q_{flat}\rightarrow -Q_{flat}$.
}
    \label{fig:BH_gs_heatmap_charge}
\end{figure}

In Fig.~\ref{fig:BH_gs_heatmap_charge} we plot the heatmap of the expectation value
of the site-occupation operator $\langle Q_{flat}\rangle/N$ (\eqref{flat}), i.e. 
the average charge per site, as a function of $(mL,\mu L)$, for two horizon radii 
$r_h=1$ (left) and $r_h=10$. (right). $Q_{flat}$ commutes with the Hamiltonian, 
and Fig.~\ref{fig:BH_gs_heatmap_charge} is the flat-charge analogue of the weighted charge
in Fig. \ref{fig:gs_physical_charge}, showing the $\mu=m$ transition and how a larger black hole horizon shrinks the overall charge response.
In the $\mu<m$ blue region $\langle Q_{flat}\rangle/N < 0$. 
In $|\mu|<|m|$, only sites with $\alpha_n<|\mu|/|m|$ can fill; at finite $N$ this is a small near-horizon set, so the average per-site charge is typically negative.
In the $|\mu|>|m|$ red region $\langle Q_{flat}\rangle/N > 0$. Here, modes fill up to the Fermi level, net charge density builds up.
We see the symmetry: $m\rightarrow -m,~\mu \rightarrow -\mu,~ Q_{flat}\rightarrow -Q_{flat}$.
Increasing $r_h$ at fixed $L$ dampens both the vacuum and filled parts, because the stronger gravitational redshift (\ref{redshift}) flattens out the lattice dispersion and reduces the net polarization.
In the case of a flat space, $Q_{flat}$ corresponds to the charge of the fermions, and one  
obtains similar phase diagrams for several common models, including the massive free fermion and the Schwinger model \cite{PhysRevD.108.L091501}.

\if{

, we present the phase diagram drawn by $\langle Q_\text{flat}\rangle/N$ (see eq.\eqref{flat}), where the expectation value is evaluated by the ground state of the Hamiltonian. Since $Q_\text{flat}$ commutes with the Hamiltonian, it represents a conserved quantity and thus serves as an order parameter for the phase diagram, as illustrated in the figures. Specifically, $\langle Q_\text{flat}\rangle$ measures the imbalance between the number of 0s and 1s ($2aQ_\text{flat} = \#0 - \#1$) in the (ground) state $\ket{\psi}$, when it is written as $\ket{\psi}=\sum_{z\in\{0,1\}^{2^N}}c_z\ket{z}$. When the space is flat, it corresponds to the charge of the fermions, and one can confirm similar diagrams for several common models, including the massive free fermion and the Schwinger model \cite{PhysRevD.108.L091501}. The figures suggest that $\langle Q_\text{flat}\rangle$ plays as a background-independent order parameter for spin-chains. 
}\fi

\if{
In Fig.~\ref{fig:BH_gs_local_charge}, we present the ground state expectation value of the local charge, $Q_{\text{flat},n} = \frac{Z_n + (-1)^n}{2}$, for $N = 20$ and $\mu = 0$. When $m = 0$, the local charge is either $-1$ (for odd $n$) or $1$ (for even $n$), regardless of the site $n$. However, when $m \neq 0$, the local charge depends on the site $n$, reflecting the geometry of the curved space. Since it is the ground sate of the model with $\mu=0$, the sum of them is 0. 
Note that the odd-even sites oscillations of the local charge are a consequence of the 
$(-1)^n$ term. In the continuum, these oscillations average away over distances $\gg a$,
and the charge density has no built-in oscillations as seen between the odd-even sites. Physically, low-momentum observables live on length scales large compared to the lattice spacing, so any $\pi/a$ oscillatory piece is non-universal and disappears in the infrared physics.

\begin{figure}[H]
    \centering
    \includegraphics[width=0.49\linewidth]{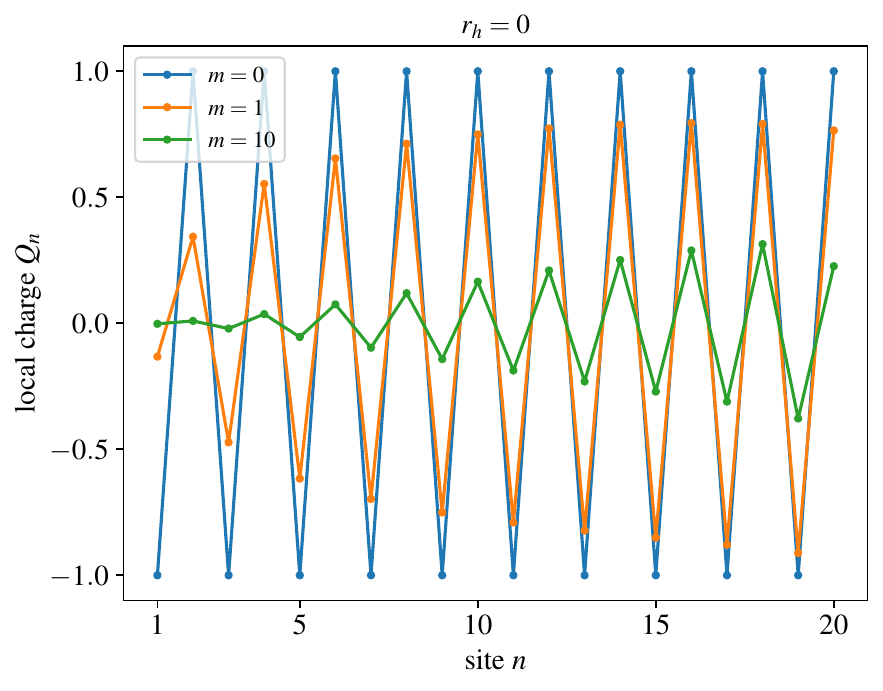}
    \includegraphics[width=0.49\linewidth]{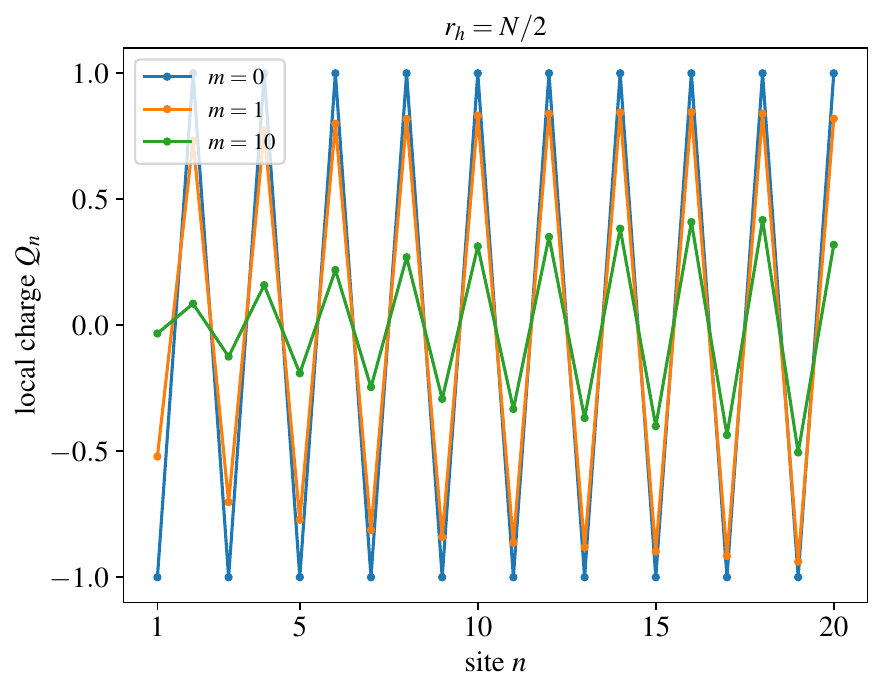}
    \caption{Local charge $Q_{\text{flat},n}=\frac{Z_n+(-1)^n}{2a}$ for various values of $m$ with $N=20$ and $\mu=0$. $m$ is in unit of $L$. From left to right: $r_h/N=0,1/5,1/2$.}
    \label{fig:BH_gs_local_charge}
\end{figure}

}\fi

\subsubsection{Entanglement Entropy}

\begin{figure}[H]
    \centering    \includegraphics[width=0.49\linewidth]{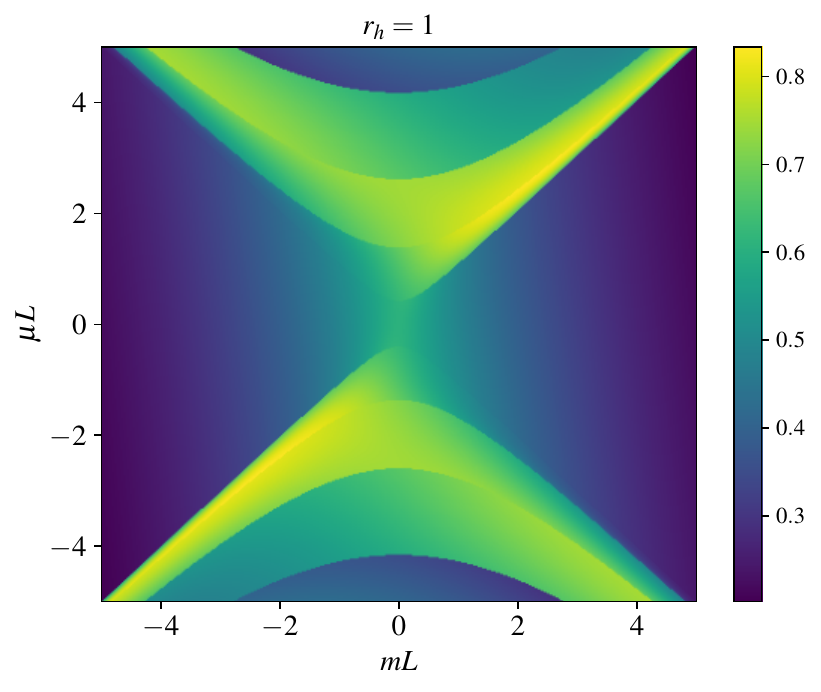}
    \includegraphics[width=0.49\linewidth]{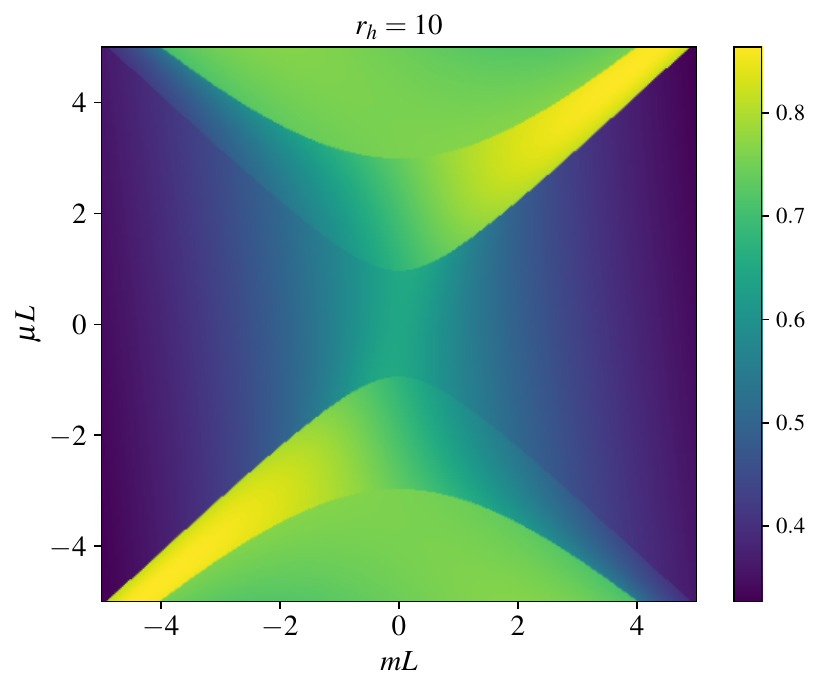}
    \caption{The ground state entanglement entropy for $N=12$ and $L=1$, with horizon radius $r_{h} =1$ (left)
   $r_{h} =10$ (right). There is a symmetry: $m\rightarrow -m,~\mu \rightarrow -\mu$. 
  As we increase $r_h$ we see the effect of the stronger gravitational redshift.
   }
    \label{fig:gs_ee_heatmap}
\end{figure}

We explore the entanglement entropy $S_{EE}(\ell)$ between $A=[1,\cdots, \ell]$ and $B=[\ell+1,\cdots,N]$. In Fig.~\ref{fig:gs_ee_heatmap} we present the heatmaps of the entanglement entropy when $\ell=\tfrac{N}{2}$. 
We see the symmetry $(m,\mu)\to(-m,-\mu)$ (point-reflection symmetry about the origin).
In the continuum gapped region ($|\mu|>|m|$), the expected entropy is low because the single-particle spectrum remains unfilled and the ground state is nearly a product state.
In the continuum gapless region ($|\mu|< |m|$), the 
entropy is expected to rise as the Fermi sea forms and long-range correlations span the bi-partition.
Interestingly, we see a different structure at finite $N$, where the system is always gapped. 
Although the finite $N$ spectrum has a nonzero level spacing (finite-size gap), increasing $\mu$ changes the set of occupied extended modes (therefore the charge changes as seen in Figs.~\ref{fig:gs_physical_charge} and \ref{fig:BH_gs_heatmap_charge}); each time $\mu$ crosses a level with support on both halves, the correlation eigenvalues move toward $1/2$ and the half-chain entanglement grows stepwise. Thus $S_{EE}$ increases with $\mu$, peaking at intermediate fillings and diminishing near empty/full limits.
The right panel ($r_h=10$), is shifted upward relative to ($r_h=1$), reflecting the stronger gravitational redshift (\ref{redshift}).

\begin{figure}[H]
    \centering
    \includegraphics[width=0.49\linewidth]{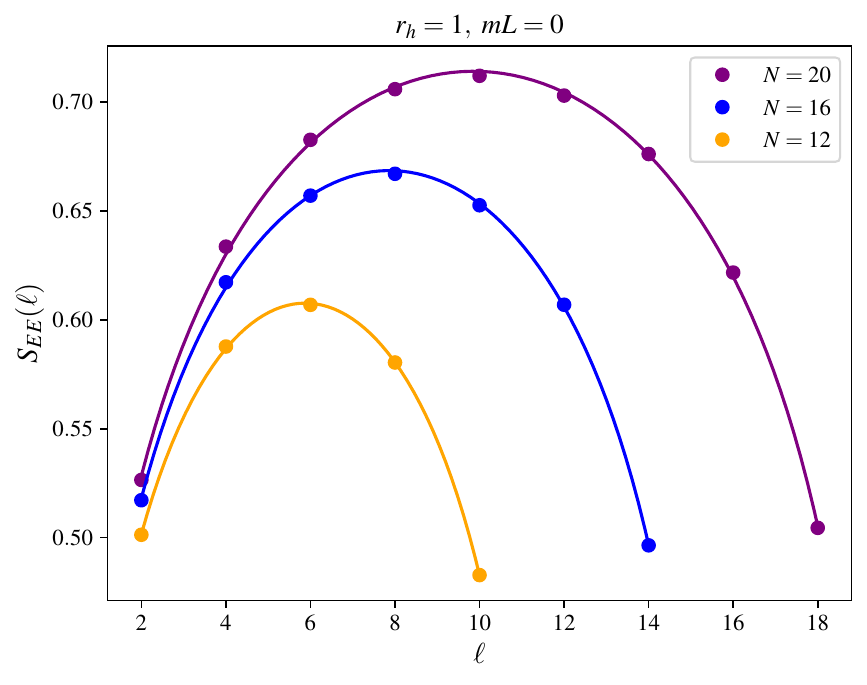}
    \includegraphics[width=0.49\linewidth]{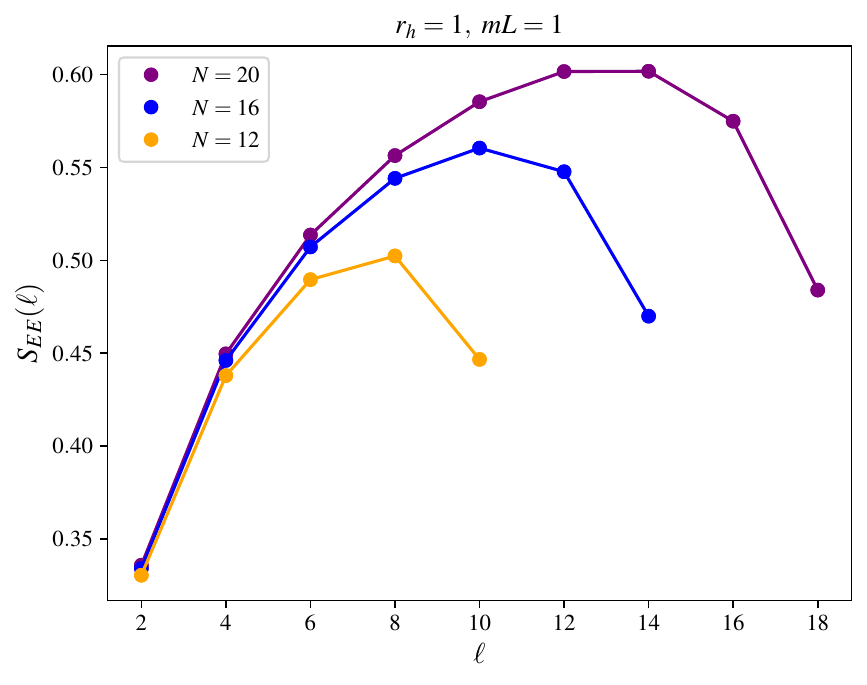}
    \caption{The entanglement entropy $S_{EE}(\ell)$ with $mL=0$ (left), mL=1 (right), for $r_h=1,\mu=0$. For both plots, $L=1$ is used.}
    \label{fig:gs_ee_fit}
\end{figure}

Fig.~\ref{fig:gs_ee_fit} shows the entanglement entropy $S_{EE}(\ell)$ vs. the
subsystem size $\ell$, at $\mu=0$ and $r_h=1$.
This figure has two panels, each showing $S_{EE}(\ell)$ for three total system sizes $N=12,16,20$. 
In the left panel ($mL=0$), the entanglement curve is not symmetric about the midpoint $\ell=N/2$,
because at $m=0$, the AdS background geometry breaks the parity symmetry of the lattice Hamiltonian. This highlights a key difference from the flat background case. All three curves peak at $\ell\approx N/2$, and their height grows slowly with $N$.
In the right panel ($mL=1$), the symmetry is broken more explicitly, $S_{EE}(\ell)\neq S_{EE}(N-\ell)$, and a finite mass biases the ground state toward one Neél ordering over the other. Thus, cutting off the favored end of the chain yields slightly higher entanglement than cutting off the opposite end.
The three curves still cluster around $\ell=N/2$, but now the peak is skewed and the overall profile is subtly asymmetric. Past $\ell=N/2$, we lose bonds at the cut and $S_{EE}$ falls, mirroring the rise before the midpoint.

The structure that we see in Fig.~\ref{fig:gs_ee_fit}
aligns perfectly with what we would expect on physical grounds for a gapped one-dimensional
fermion chain with and without a sublattice symmetry.
In a gapped theory ($m>0$), connected two-point correlators (in units of $a$) fall off as:
\begin{equation}
\langle\mathcal O_i \,\mathcal O_j\rangle_{\!c}\;\sim\;e^{-|i-j|/\xi},
\quad
\xi\sim\frac{1}{m} \,.    
\end{equation}
Physically, $\xi$ is the size of the region over which degrees of freedom remain significantly entangled or correlated.
In the regime $N\lesssim\xi$, every cut through the chain sits inside a region where correlations are still building up, hence enlarging $N$ adds more correlated sites on each side of the cut, and the entanglement entropy $S_{EE}(\ell)$ at its peak (near $\ell=N/2$) grows with $N$.
In the regime $N\gg\xi$ we have the area-law saturation,
where the two halves of the chain are only correlated across a boundary region of width $\sim\xi$.
Any sites beyond distance $\xi$ from the cut contribute essentially zero additional entanglement,
and further increasing $N$ no longer increases the peak entropy appreciably, and it flattens out to an 
$O(1)$.

\begin{figure}[H]
    \centering
    \includegraphics[width=0.49\linewidth]{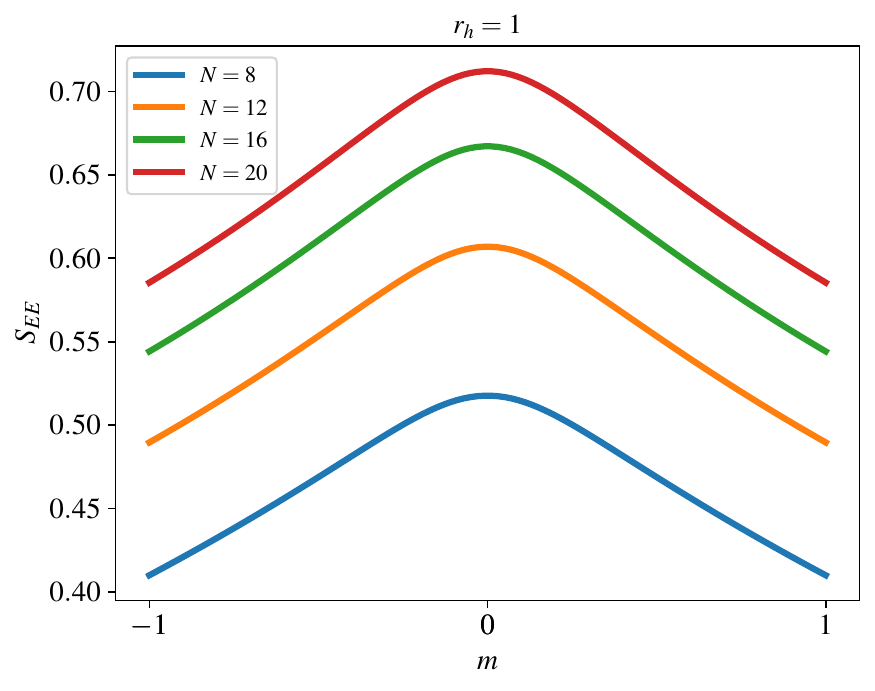}
    \centering
    \includegraphics[width=0.49\linewidth]{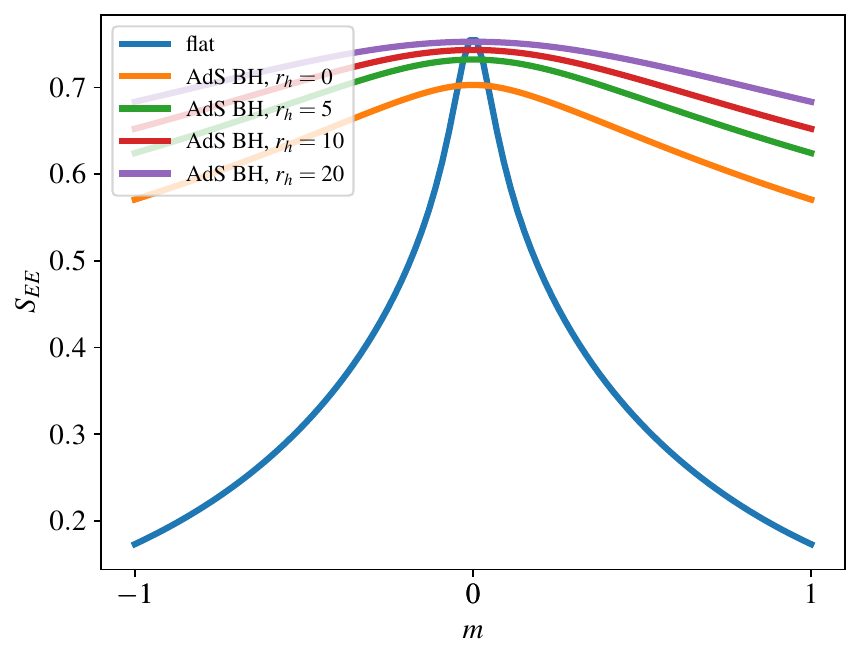}
    \caption{Entanglement entropy $S_{EE}(\ell=\tfrac{N}{2})$ for various system sizes $N\in \{12,16,20\}$ (left) and for various horizon radii $r_h$ for $N=20$ (right). \emph{flat} means $S_{EE}$ of the massive free fermions in the flat background. For both plots, $L=1$ is used.}
    \label{fig:ee_m0}
\end{figure}

Fig.~\ref{fig:ee_m0} examines how the half‐chain entanglement $S_{EE}(\ell=\frac{N}{2})$
varies as we dial the fermion mass $mL$ at zero chemical potential and $r_h=1$. There are two panels:
In the left panel we plot $S_{EE}(N/2)$ versus $mL$ for system sizes $N=8,12,16,20$. 
All curves peak sharply at $mL=0$, reflecting maximal entanglement when the theory is massless (gapless).
As we increase $N$, the peak grows taller and narrower: larger chains support more entanglement near criticality but still collapse to low entropy once $mL\gtrsim1$.
In the right panel we fix $N=20$ and consider the flat space case, as well as vary the horizon radius 
$r_h/L \in \{0,5,10,20 \}$.
All the curves share the same massless peak, but as $r_h$ increases, the sides of the peak become less steep.  

The physical interpretation of these observations is as follows.
At $m=0$, the chain is critical and half-chain cuts capture long-range correlations leading to the highest entropy.
In the flat chain, a finite mass opens a gap, 
where every link feels the same mass $m$, so there’s a single sharp threshold $mL\sim O(1)$, where correlations suddenly decay and $S_{EE}$ collapses. This shows up as a steep cliff.
In the curved chain, the mass is effectively different at each site leading to a cascade  of local gappings, and a smeared, gentler overall decline.
Thus, the horizon softens the entanglement transition, making the $AdS_2$ black hole curves flatter than the flat space one.
That mirrors what we saw in Fig.~\ref{fig:gs_ee_heatmap}. In Fig.~\ref{fig:ee_m0}, the flat-space curve plunges abruptly at a single mass scale, whereas the $AdS_2$ black hole curves spread that drop over a wider mass window. In Fig. \ref{fig:gs_ee_heatmap} this showed up as the contours of constant $S_{EE}$ being more tightly packed (steep gradient) on the left and more spread out (gentler gradient) on the right.

\begin{figure}[H]
    \centering
    \includegraphics[width=0.49\linewidth]{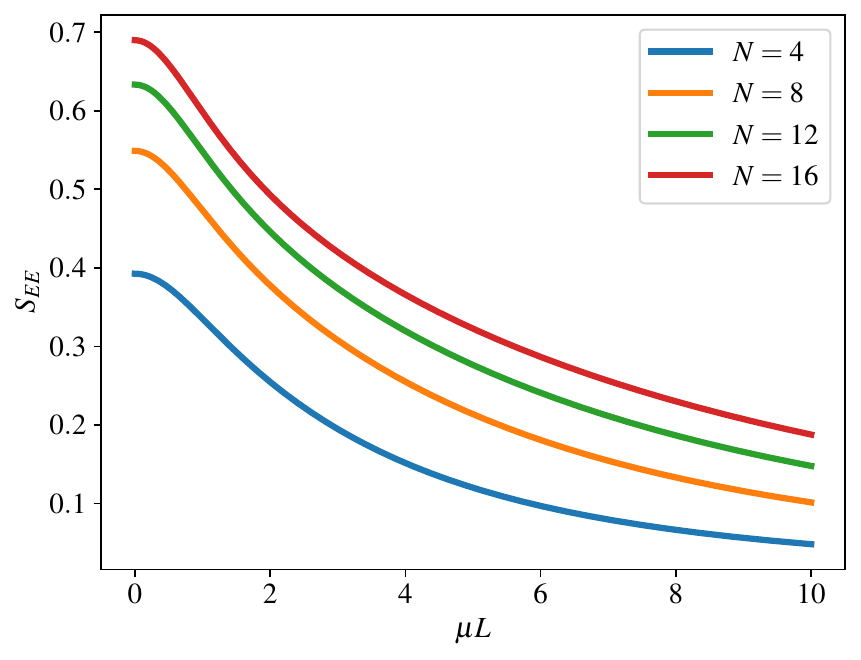}\includegraphics[width=0.49\linewidth]{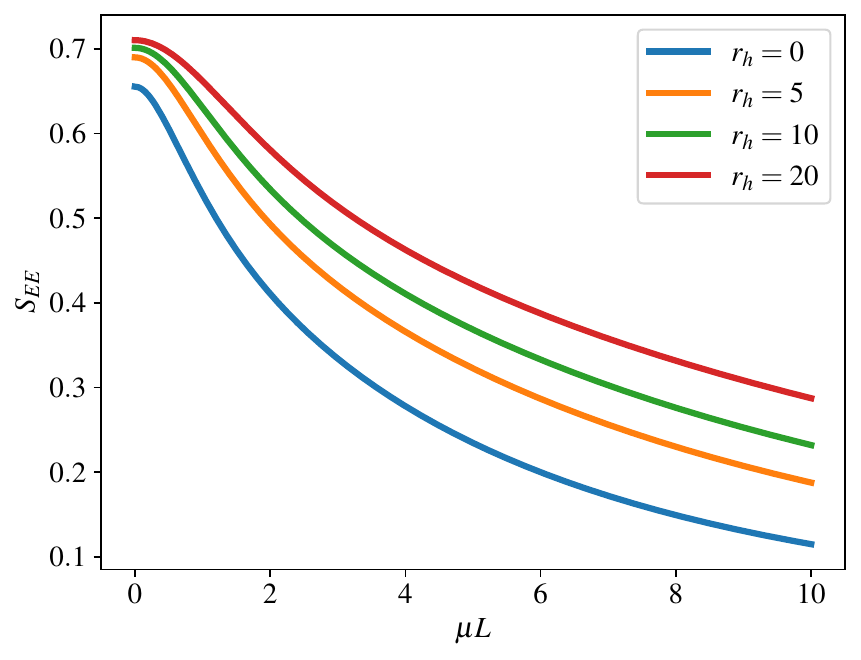}

    \caption{$N$- and $r_h$-dependence of the entanglement entropy along with $mL=\mu L$ for $r_h=5$ (left), and $N=16$ (right). The entanglement entropy is a monotonically decreasing function of $mL$ and $\mu L$. For both plots, $L=1$ is used.}
    \label{fig:gs_EE_m_is_mu}
\end{figure}

Fig. \ref{fig:gs_EE_m_is_mu} shows $S_{\rm EE}(\frac{N}{2})$ along $m=\mu$. In the left panel
we fix $r_h=5$ and vary $N \in \{4,8,12,16\}$. In the right panel we fix $N=16$ and vary $\frac{r_h}{L}\in \{0,5,10,20\}$. 
In both panels $S_{\rm EE}$ decreases monotonically with $\mu L$. Physically, increasing $m$ (and $\mu$) only adds diagonal (on-site) energy terms, which favors more classical, product-state behavior and suppresses quantum correlations. Because the model is invariant under $(m,\mu)\to(-m,-\mu)$, the same decreasing behavior holds if we continue the plot into negative $\mu L$.  The gravitational redshift (which is stronger for larger $r_h$) has an effect on the speed of the monotonic decay of entanglement with $m$.
In summary, Fig.~\ref{fig:gs_EE_m_is_mu}  confirms that locking the mass and chemical potential together drives the system toward a more classical regime as $mL$ becomes large, and that this effect is robust against horizon size.

\subsection{Energy Gap}

\begin{figure}
    \centering
    \includegraphics[width=0.49\linewidth]{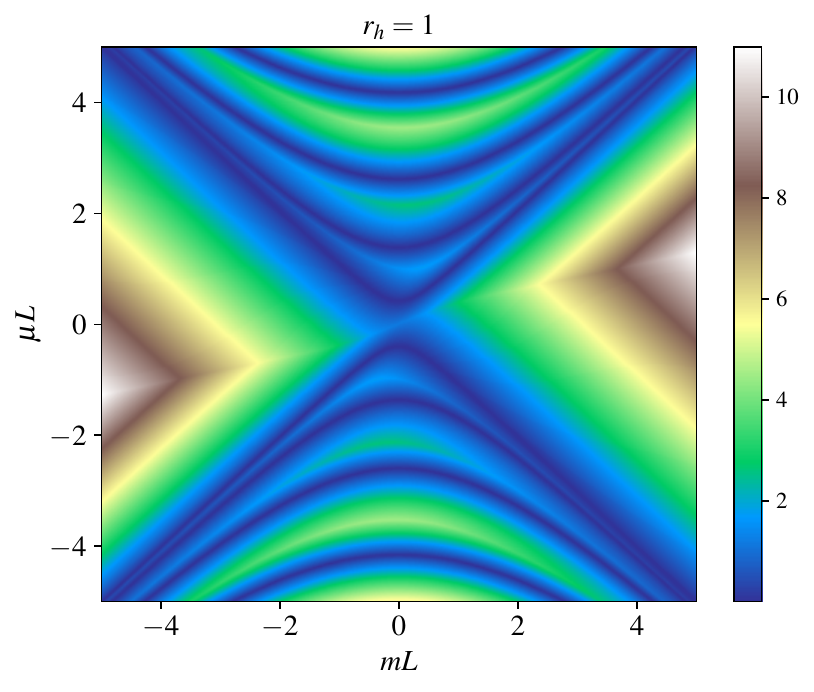}
    \includegraphics[width=0.49\linewidth]{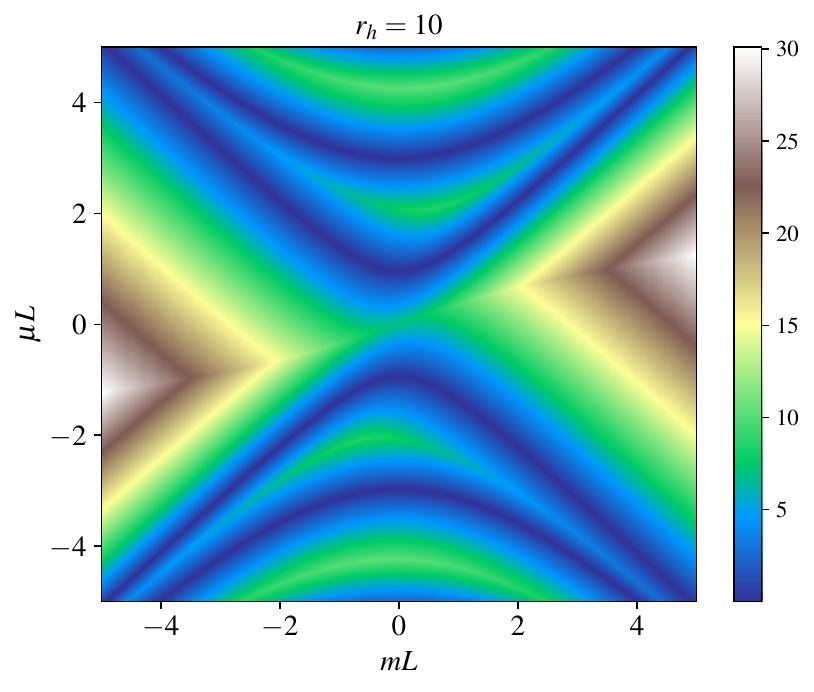}
    \caption{The energy gap for $N=12$, with horizon radius $r_{h} =1$ (left)
   $r_{h} =10$ (right). We see an $X$-shaped valley along $\mu\approx\pm m$, and the
gap vanishes when $\mu=+m$ and $\mu=-m$ signaling the transition from a fully gapped vacuum into a gapless Fermi-sea phase. Note that because the figure is computed at finite $N$ with open-chain quantization $k_j=\pi j/((N+1)a)$ and a discrete set of redshifts $\{\alpha_n\}$, the gap has a positive finite-size floor $O((\pi/(N+1))^2)$ near $\mu=\pm m$; in the continuum limit ($N\to\infty$) the valley closes along $|\mu|\le|m|$. There is a symmetry under 
$m\to -m, \mu\to -\mu$ that reflects the charge-conjugation symmetry 
(\ref{spec}).
The effect of the horizon radius $r_h$ is such that larger $r_h$ implies
an overall suppression of the minimum gap. For both plots, $L=1$ is used.}
    \label{Gap}
\end{figure}

Fig. \ref{Gap}  shows the zero-temperature single-particle energy gap as a function of the dimensionless mass $mL$ (horizontal axis) and chemical potential $\mu L$ (vertical), for a chain of $N=12$ sites, in two gravitational backgrounds: Left panel is small black hole $r_h=1$ and the right is a large black hole $r_h=10$.
There are several key features. We see an $X$-shaped valley along $\mu\approx\pm m$, and the
gap vanishes when $\mu=+m$ and $\mu=-m$, signaling the transition from a fully gapped vacuum into a gapless Fermi-sea phase. Everywhere else $\Delta>0$. There is a symmetry under 
$m\to -m, \mu\to -\mu$ that reflects the charge-conjugation symmetry $E(m,\mu)=E(-m,-\mu)$
(\ref{spec}).
The effect of the horizon radius $r_h$ is such that larger $r_h$ implies
an overall suppression of the minimum gap. Physically it means that stronger gravitational redshift (\ref{redshift}) stretches the lattice dispersion (\ref{dra}), reducing the size of the smallest excitation energy across most of the parameter space.

Note that there is a difference between a finite $N$ and the continuum.
For a finite chain, $N=12$ here, exact zeros are guaranteed on the lines $\mu=\pm m$ if we scale
the boundary site to $\alpha_N=1$. Inside the X-shape, $|\mu|<|m|$, the continuum picture predicts $\Delta=0$, whenever some site has $\alpha_n=\frac{|\mu|}{|m|}$. With discrete $\alpha_n$ this is seen as very small but not necessarily exactly zero gaps except at special $(m,\mu)$.

In the following we consider the extreme case where mass is very large. The mass term is
\begin{equation}
    M=\frac{m}{2}\sum_{n=1}^N\alpha_n(-1)^nZ_n \ . 
\end{equation}
It has the Néel state $\ket{\psi_0}=\ket{01\cdots01}$ (for $m>0$) for the ground state and the corresponding energy eigenvalue is 
\begin{equation}
    E_0=-\frac{|m|}{2}\sum_{n=1}^N\alpha_n \ .
\end{equation}
Unlike the flat case, the first excited state of the mass term is non-degenerated:
\begin{equation}
    \ket{\psi_1}=X_1\ket{\psi_0}=\ket{11\cdots01}. 
\end{equation}
Here it is important that $\alpha_i<\alpha_j$ for all $i<j$. The corresponding energy eigenvalue is
\begin{equation}
    E_1=-\frac{|m|}{2}\left(-\alpha_1+\sum_{n=1}^{N-1}\alpha_n\right)
\end{equation}
Therefore the energy gap at a large $mL$ is:
\begin{equation}
\label{eq:gap}
    \Delta E=|m|\alpha_1=\frac{|m|\sqrt{r^2_1-r^2_h}}{L},\quad r_1=a+r_h\;.
\end{equation}
The energy gap for an extremely large $|\mu| L$ can be obtained similarly. So, when $|mL|$ or $|\mu L|$ is large, the energy gap between the lowest and first excited energies at leading $\frac{1}{N}$ is:
\begin{equation}
    \Delta E=
    \begin{cases}
       |m|\alpha_1 & |m/\mu|\gg1,\\
        |\mu |\alpha_1 & |\mu/m|\gg1 \ .
    \end{cases}
\end{equation}
When $r_h\ll aN$, it corresponds to the case without the black hole. When $r_h/a$ is at the order of $O(N)$, its contribution becomes significant.

\begin{figure}[H]
    \centering
    \includegraphics[width=0.49\linewidth]{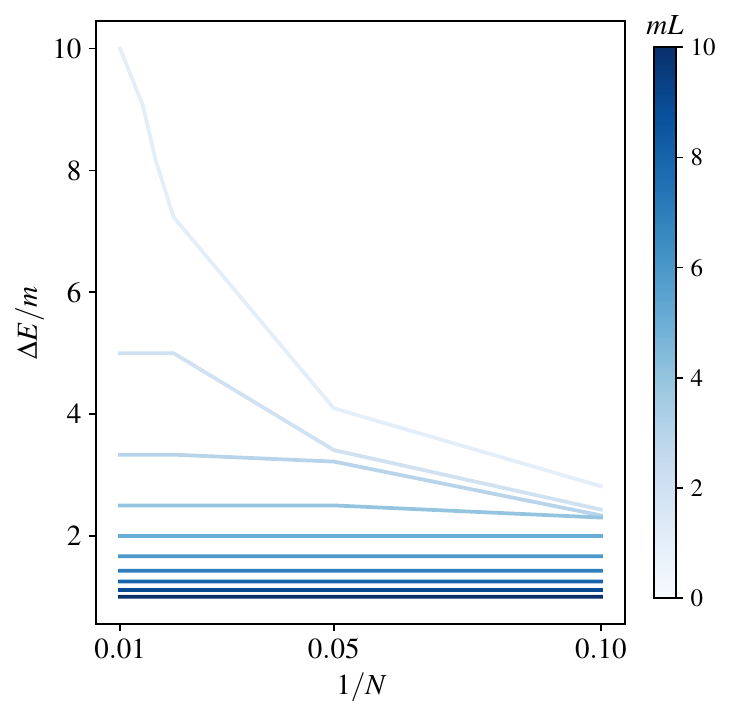}
    \includegraphics[width=0.49\linewidth]{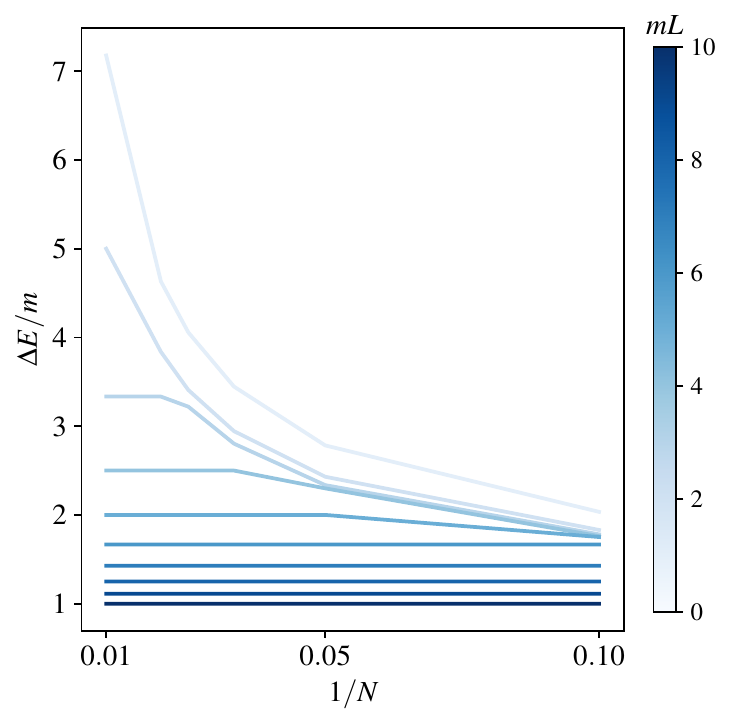}
    \caption{The $N$-dependence of the ratio $\Delta E/m$ when $\mu=0$ and $r_h=\frac{N}{5}$ (left) and $r_h=\frac{N}{10}$ (right) as a function of $1/N$ with $N$ up to $100$. For both plots, $L=1$ is used.}
    \label{fig:AdS_BH_gap_r_halfN}
\end{figure}

In Fig.~\ref{fig:AdS_BH_gap_r_halfN}, we show the case $r_h=\frac{aN}{5}$ (left) and $r_h=\frac{aN}{10}$ (right) as a function of $\frac{1}{N}~(N=10,\cdots,100)$. The other parameters are fixed to $\mu=0$. As expected, $\Delta E/m$ approaches $1/\sqrt{1+\frac{2r_h}{aN}}$ as $m$ increases. In the plot, we take $a=1$ as before. 

\if{
The second-excited state is 
\begin{equation}
    \ket{\psi_2}=X_{N-1}\ket{01\cdots01}
\end{equation}
and the energy eigenvalue is 
\begin{equation}
    E_2=-\frac{|m|}{2}\left(\sum_{n=1}^{N-2}w_n-w_{N-1}+w_{N}\right).
\end{equation}
Thus the gap $\Delta_2=E_2-E_1$ is non-zero (though very small) on a lattice:
\begin{equation}
    \Delta_2=|m|(w_{N-1}-w_N)>0.
\end{equation}
}\fi
More generally, we consider the following excited eigenstate:
\begin{equation}
    \ket{\psi_n}=X_n\ket{\psi_0}, 
\end{equation}
whose corresponding energy eigenvalue is 
\begin{equation}
    E_n=-\frac{|m|}{2}\left(-\alpha_n+\sum_{j\neq n }^N\alpha_n\right)
\end{equation}
and the gap between the lowest energy recovers the local energy dispersion \eqref{eq:one-particle-dispersion}: 
\begin{equation}
    \Delta_n=|m|\alpha_n. 
\end{equation}

\subsection{First Excited State}

\subsubsection{Energy}

\begin{figure}[H]
    \centering
    \includegraphics[width=0.49\linewidth]{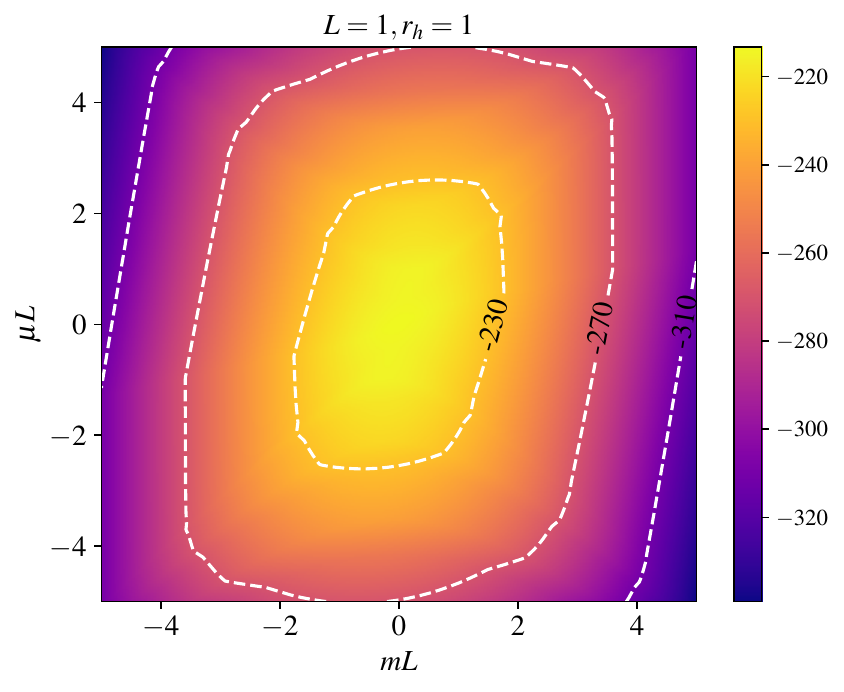}
    \includegraphics[width=0.49\linewidth]{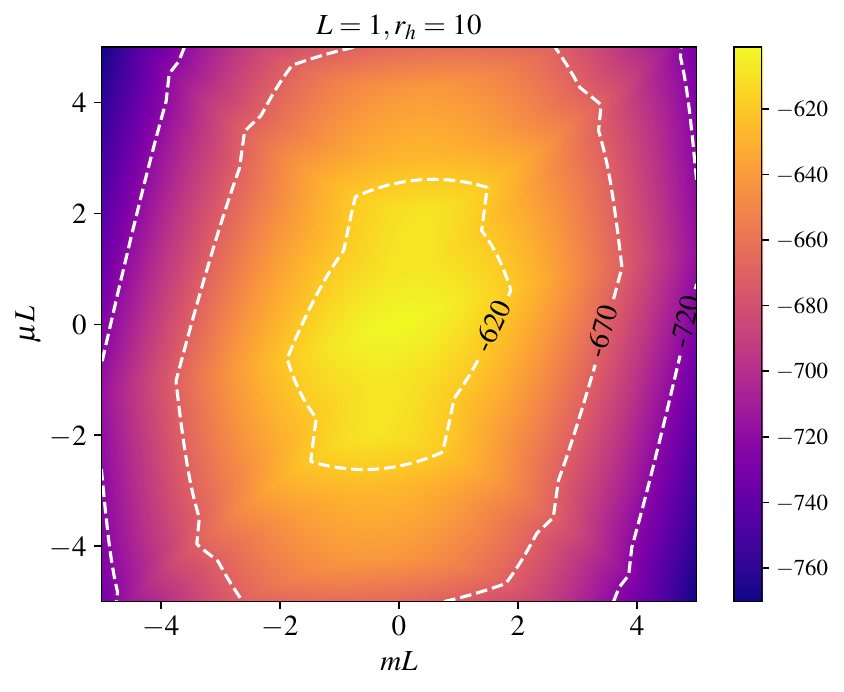}
    \caption{The first excited state energy
    for $N=12$ qubits, with horizon radius $r_{h} =1$ (left)
   $r_{h} =10$ (right). We see the wiggling of the constant $E_1$,
 in contrast to the smoother, straighter contours of the ground‐state energy in Fig.~\ref{fig:E0_heatmap_BH_N12}.
    }
    \label{E1energy}
\end{figure}

In Fig.~\ref{E1energy} we present the heatmaps of the first excited state energy for $N=12$ qubits and different values of the horizon radius $r_h$. The heatmaps exhibit the charge‐conjugation symmetry
$(m,\mu)\to(-m,-\mu)$, and the $X$-shaped valley along $|\mu|=\alpha_{\min}|m|$ (near $\mu=\pm m$ when $\alpha_{\min}\approx 1$).
As the gap closes at $\mu=\pm m$, the first excited state dips lowest (darkest) along these lines, reflecting that the ground and first excited levels become nearly degenerate at the gap-closing transition.
In Section \ref{E1} we outlined the differences between the ground state and the first excited state energies. In particular, as we explained, $E_0$ is a cumulative area under the band up to $\mu$, so its level‐curves follow the simple condition $\mu\approx\pm m$ almost exactly.
$E_1$, however, is the area plus a bump given by the next level. That bump moves around non-smoothly as $m$ and $\mu$ vary, and it is weighted by the local band curvature, which is enhanced by the redshift factor. The result is the wiggling of the constant $E_1$
that we see in Fig.~\ref{E1energy}, in contrast to the smoother, straighter contours of the ground‐state energy in Fig.~\ref{fig:E0_heatmap_BH_N12}.

\subsubsection{Charge}

\begin{figure}[H]
    \centering
    \includegraphics[width=0.49\linewidth]{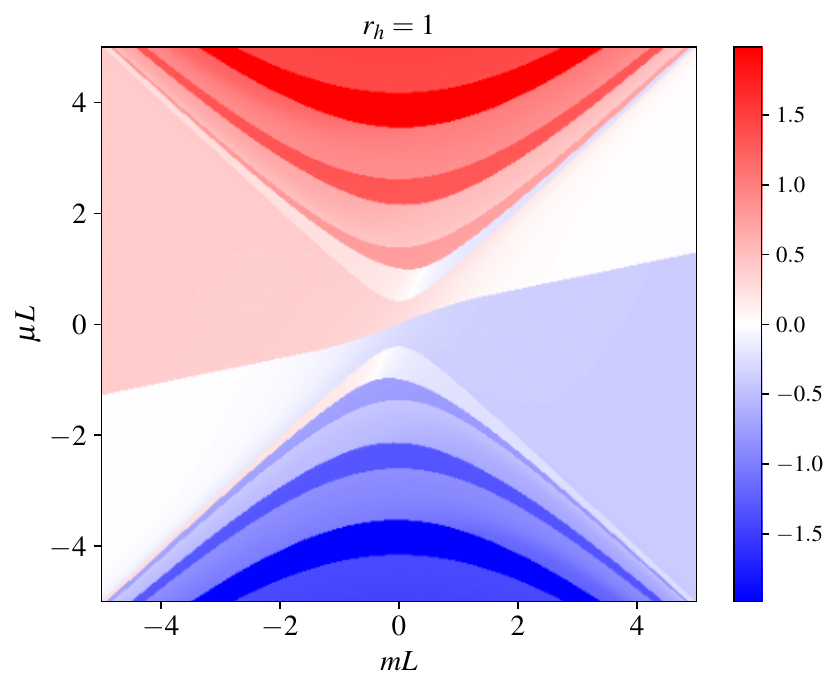}
    \includegraphics[width=0.49\linewidth]{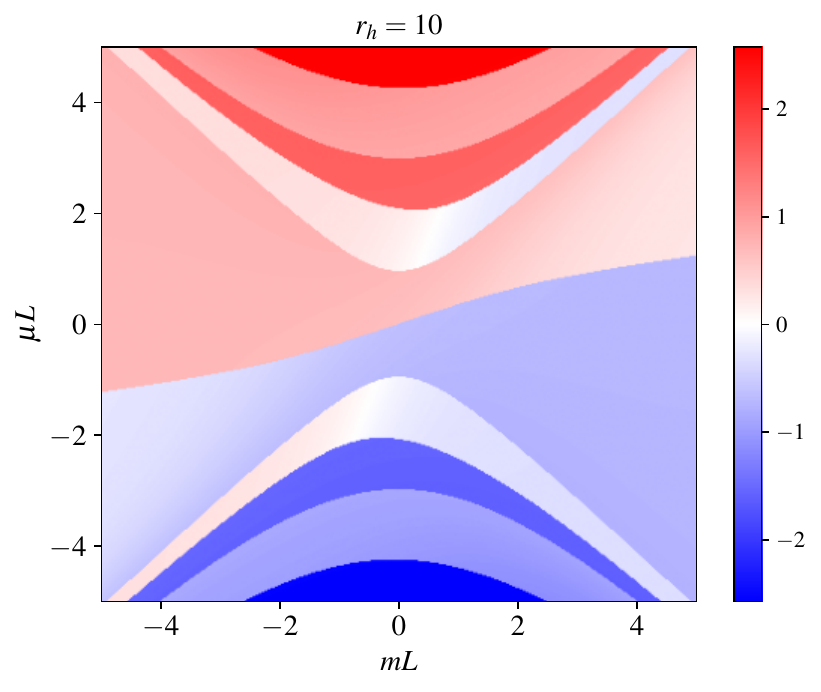}
    \caption{The expectation value of the weighted charge density $\langle Q\rangle/N$ in the first excited state, as a function of the dimensionless mass $mL$ (horizontal axis) and chemical potential $\mu L$ (vertical), for two choices of horizon radius $r_h$ (left: $r_h=1$; right: $r_h=10$). As discussed, compared to the smooth Fermi‐sea profile of the ground state charge, the first excited state shows a localized oscillatory ripple given exactly by replacing one 
    $|\psi_{k_F}(n)|^2$ with $|\psi_{k_{F+1}}(n)|^2$. For both plots, $L=1$ is used.}
    \label{fig:1st_physical_charge}
\end{figure}

In Fig.~\ref{fig:1st_physical_charge} shows 
the heatmaps of the expectation value of weighted charge density in the first excited state,
as a function of the dimensionless mass $mL$ (horizontal axis) and chemical potential $\mu L$ (vertical), for two choices of horizon radius $r_h$ (left: $r_h=1$; right: $r_h=10$).
As in Fig. \ref{fig:gs_physical_charge} we see the  $X$ of sign-change running along the lines $\mu=\pm m$. Below $|\mu|<|m|$ the excited state carries negative net charge (blue tones), and above $|\mu|>|m|$ it carries positive net charge (red tones). There is the same charge-conjugation symmetry under $(m,\mu)\!\to\!(-m,-\mu)$, flipping $\langle Q_{weighted}\rangle\to -\langle Q_{weighted}\rangle$.

As discussed in Section \ref{charge1},
since we removed the highest-filled mode and added the next one, the exchange still carries one unit of charge but can be in a different momentum eigenstate whose spatial profile is non-uniform.  As a result, around $\mu\simeq m$ there is a white band, where the first excited state stays in the same charge sector as the ground state (coming from the $q=0$ sector), so $\langle Q\rangle_{\rm 1st}\approx 0$.
The colored lobes are warped, and their contours wiggle slightly compared to the straight lines of 
Fig.~\ref{fig:1st_physical_charge}
because the extra mode’s charge density $|\psi_{k_{F+1}}(n)|^2$ can oscillate more strongly than the smooth ground state profile. The suppression by the horizon radius $r_h$ follows the same trend (right panel is flatter than left), but the band of zero charge broadens, reflecting that at strong redshift the momentum-quantization and edge-mode effects become comparatively more important.

\begin{figure}[H]
    \centering
\includegraphics[width=0.49\linewidth]{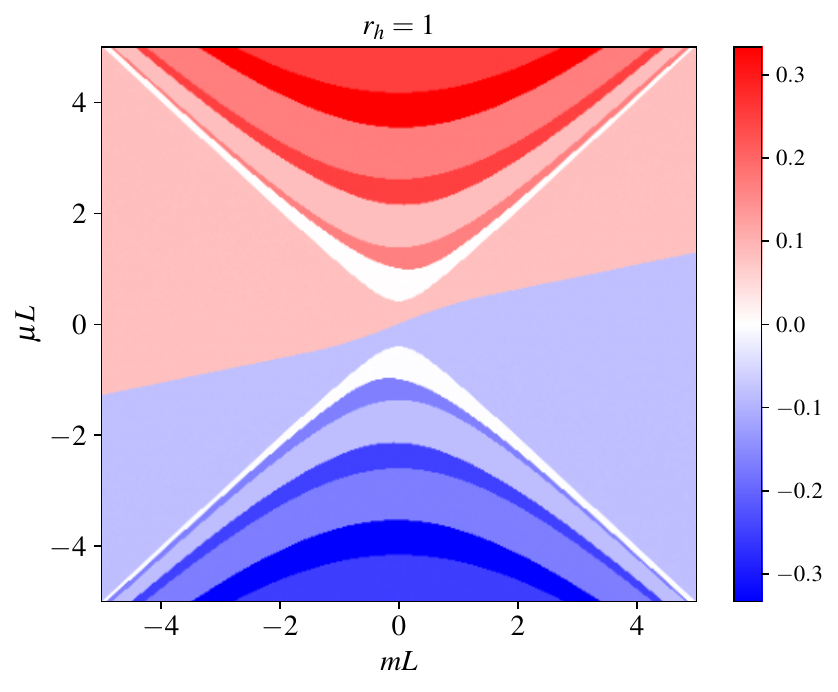}
    \includegraphics[width=0.49\linewidth]{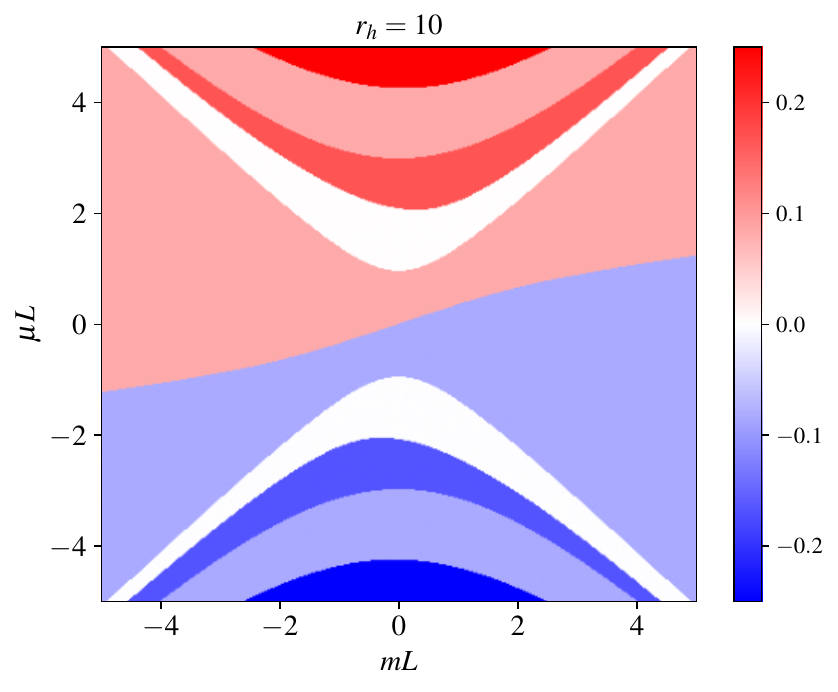}
    \caption{The expectation value
of the site-occupation operator $\langle Q_{flat}\rangle/N$ in the first excited state,
as a function of the dimensionless mass $mL$ (horizontal axis) and chemical potential $\mu L$ (vertical), for two choices of horizon radius $r_h$ (left: $r_h=1$; right: $r_h=10$). Because $Q_{\text{flat}}$ carries no redshift weights $\alpha_n$, any geometry dependence enters only through the wavefunctions of the single-particle modes rather than explicitly through the operator itself. For both plots, $L=1$ is used.
}
    \label{fig:BH_1st_charge}
\end{figure}

\if{

Here we discuss the charge of the first excited state of the Hamiltonian. 
Fig.~\ref{fig:1st_physical_charge} shows 
the heatmap of the physical charge density $\frac{Q_\text{AdS}}{N}$, and 
Fig.~\ref{fig:BH_1st_charge} shows the heatmap of the charge physical charge density $\frac{Q_\text{flat}}{N}$ for $r_h=1$ and $10$ within the system of size $N=10$. The colored regions indicate that the first excited state within each corresponding charged sector $H_{q}$ determines the first excited energy level for the entire system’s Hamiltonian, whereas the white regions originate from the charge-zero sector $q=0$.

}\fi

In Fig. \ref{fig:BH_1st_charge} we plot the expectation value of 
the first excited state flat charge $\langle Q_{\text{flat}}\rangle/N$ across $(mL,\mu L)$,
 for two horizon radii ($r_h=1$ left; $r_h=10$ right). Because $Q_{\text{flat}}$ carries no redshift weights $\alpha_n$, any geometry dependence enters only through the wavefunctions of the single-particle modes rather than explicitly through the operator itself. The color map exhibits an $X$-shaped sign change along the gapless lines $\mu=\pm m$ (charge-conjugation symmetry maps $(m,\mu)\to(-m,-\mu)$ and flips the sign of the plotted quantity). Away from these lines, the magnitude is set by the spatial profile of the mode that defines the excitation: the first excited state is obtained by swapping one filled mode at $k_F$ with the next at $k_F+1$, so its local charge is a ripple governed by
 \begin{equation}
\delta Q_n \;=\; \big|\psi_{k_F+1}(n)\big|^2-\big|\psi_{k_F}(n)\big|^2 \ ,     
 \end{equation}
which integrates to zero but can give sizable local contrasts. Increasing $r_h$ (stronger redshift) smooths these contrasts by reshaping the single-particle wavefunctions, hence the right panel is visually less saturated than the left.

\subsubsection{Entanglement Entropy}

\begin{figure}[H]
    \centering
\includegraphics[width=0.49\linewidth]{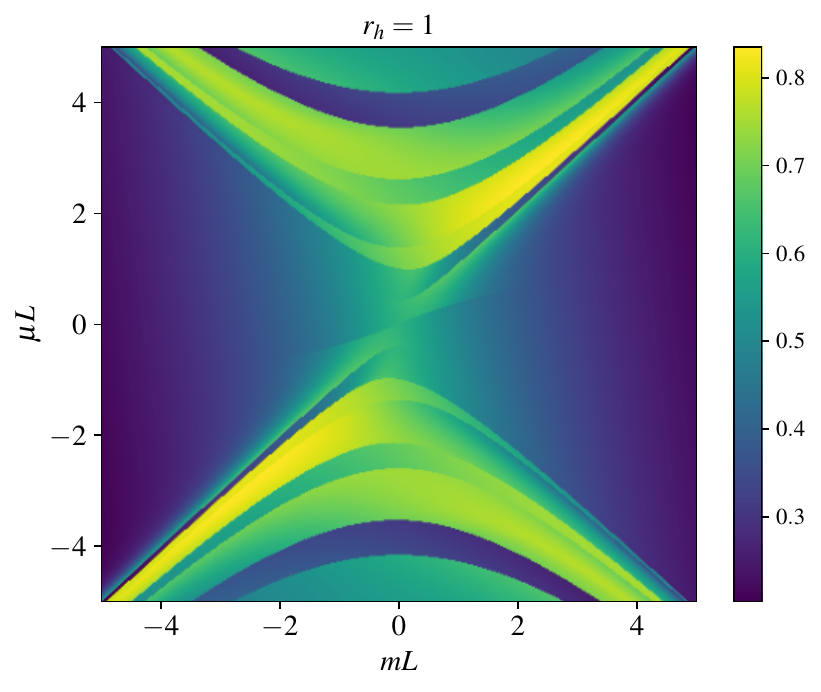}
    \includegraphics[width=0.49\linewidth]{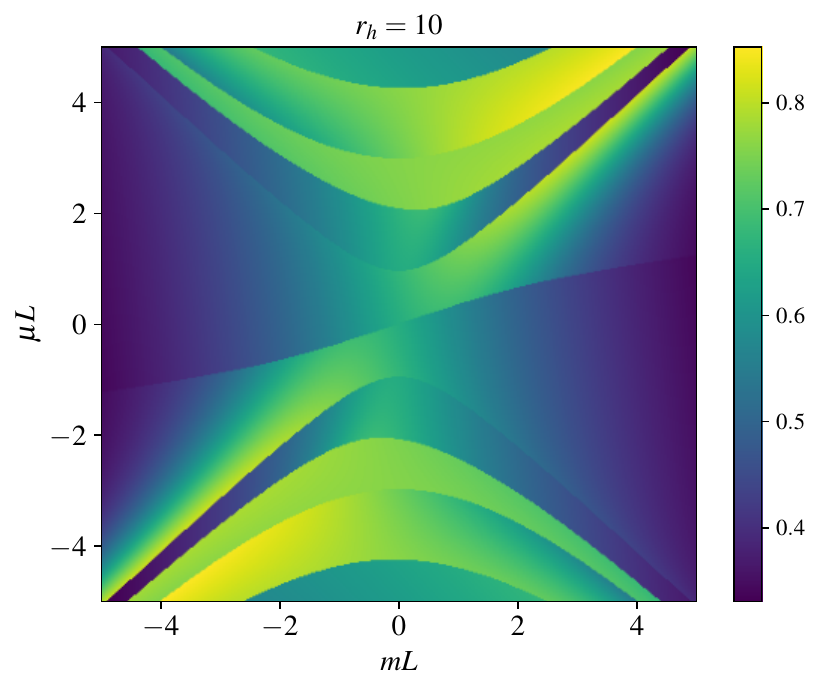}
    \caption{The first excited state half-chain entanglement entropy for $N=12$, $r_h=1$ (left) and $r_h=10$ (right). The $X$-shaped boundary at $\mu\simeq\pm m$ is visible, and the entire surface is uplifted relative to the ground state by $\Delta S=O(\frac{1}{N})$. Larger $r_h$ (stronger redshift) reduces $\Delta E_1$ near its minimum, allowing greater delocalization and hence higher entanglement.
 There is a symmetry: $m\rightarrow -m,~\mu \rightarrow -\mu$. For both plots, $L=1$ is used.}
    \label{EE1}
\end{figure}

In Fig.~\ref{EE1}, we show the half‐chain entanglement of the first excited state for $N=12$. Compared to the ground state entanglement entropy in Fig. (\ref{fig:gs_ee_heatmap}), 
we see the same $X$-shaped rise of entanglement when $\mu$ crosses $\pm m$. However,
everywhere in parameter space it is uniformly higher than 
that of the ground state by a small offset $(\sim 0.05–0.1)$, consistent with the expected $\Delta S=O(1/N)$
increase (\ref{DS}).
The boundary between low and high entanglement regions still follows $\mu\approx m$, and the effect of increasing $r_h$ is the same uplift of the entire surface.

\begin{figure}[H]
\centering
    \includegraphics[width=0.49\linewidth]{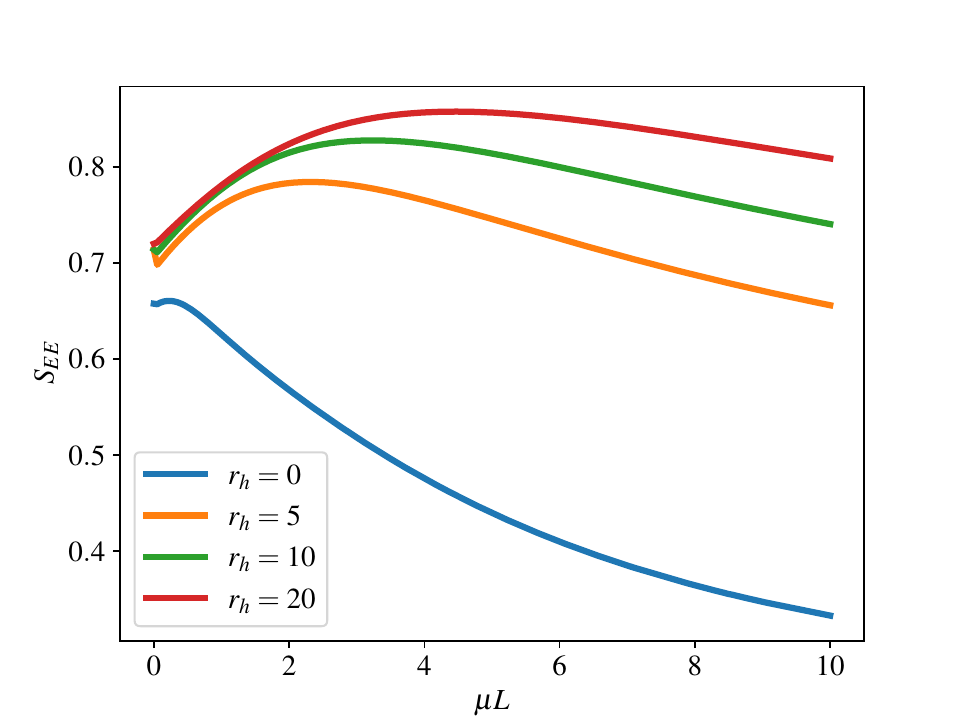}
    \includegraphics[width=0.49\linewidth]{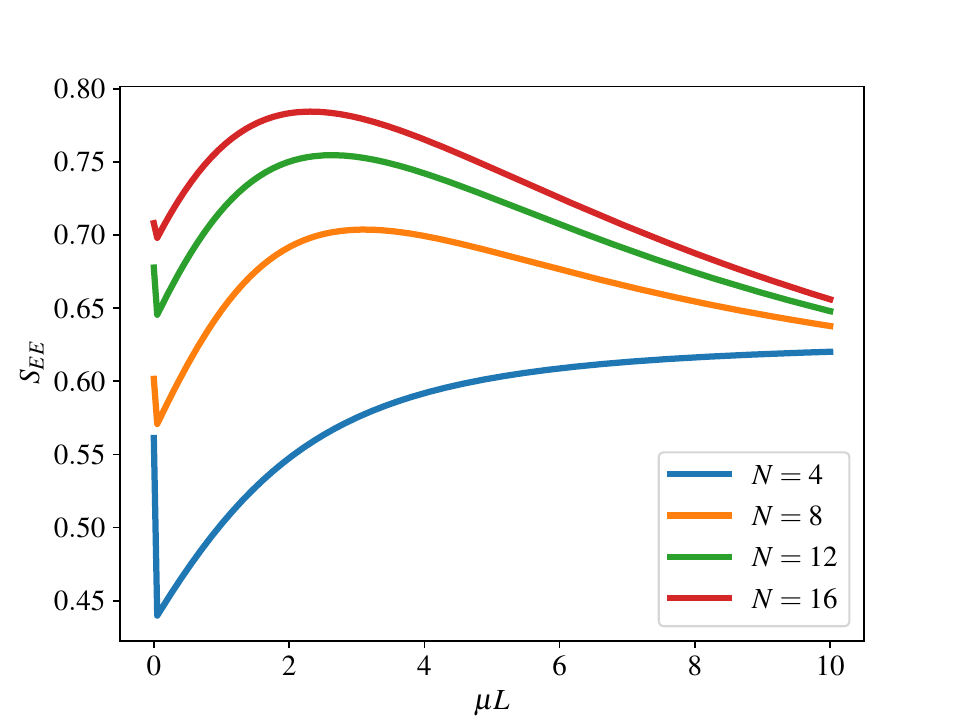}
    \caption{$N$- and $r_h$-dependence of the first excited state entanglement entropy as a function of $\mu=m$. 
    Left panel: $N$ fixed, varying $r_h$; Right panel: $r_h$ fixed, varying $N$. 
  $S_{\rm EE}$ increases with $\mu L$ up to a broad maximum ($\mu L\sim 4$) and then decreases. This reflects a competition between two effects: 
  as m rises from zero, $\Delta E_1$ decreases, enhancing mixing and entanglement; at larger $m$, mass-induced localization dominates and entanglement falls, even though $\Delta E_1$ continues to decrease.
  Thus, unlike the ground state, here the entanglement entropy is not monotonically decreasing function of $m$ and $\mu$. For both plots, $L=1$ is used.
    }
    \label{EEEE}
\end{figure}

In Fig. (\ref{EEEE}) we see the $N$-dependence of the first-excited-state entanglement entropy along the 
line $m=\mu$.  Left panel: $N$ fixed, varying $r_h$; Right panel: $r_h$ fixed, varying $N$. Unlike the ground state, here the entanglement entropy is not monotonically decreasing function of $m$ and $\mu$.
At $\mu=m=0$, the first excited state is the lowest‐lying single‐particle mode above the filled Dirac sea.  Its wavefunction is delocalized across the entire chain, having a comparatively large bipartite entanglement.
The excitation energy at site $n$
reads:
\begin{equation}
E_{1}(n)=\alpha_n\sqrt{m^{2}+\alpha_n^{2}k_{1}^{2}}-\mu  \ ,
\end{equation}
where $k_1\sim\pi/N$ is the lowest nonzero lattice momentum.
The global first positive excitation along $\mu=m$ at the boundary reads:
\begin{equation}
E_{1}\;=\;\sqrt{\,m^{2}+k_{1}^{2}\,}\;-\;m \ ,   
\end{equation}
which decreases with $m$.
The non-monotonic behaviour of $S_{\rm EE}$ comes from the competition between this decreasing gap 
(enhancing mixing) and mass induced localization suppressing the bipartite entanglement.
As we raise $m$ from zero, the first excited wavefunction mixes more strongly with the vacuum fluctuations across the cut—hence its bipartite entanglement increases.
At some larger mass the localization takes over and the entanglement entropy decreases

\if{

Once $m$ grows so that $m\gg\alpha k_1$, we have:
\begin{equation}
 \Delta E_1 \;\approx\; m\Bigl(\sqrt{1 + (\tfrac{\alpha\,k_1}{m})^2} - 1\Bigr)
\;\xrightarrow{m\gg\alpha k_1}\;\frac{\alpha^2\,k_1^2}{2m}\;\sim\;\tfrac{1}{m} \ ,   
\end{equation}
which decreases with $m$.  A larger gap pushes the first excited mode to become spatially localized and its entanglement decreases.
Larger $r_h$ raises the height of that non-monotonic entanglement peak, since a bigger horizon induces a stronger gravitational redshift, which softens the kinetic term for near-horizon modes (smaller $\alpha$), letting the first excitation become more delocalized at its gap minimum.
It lowers the $\Delta E_1$ and thus maximizes the mode’s mixing with vacuum fluctuations, hence boosting its entanglement.
Larger  $r_h$ (stronger redshift) reduces $\Delta E_1$, allowing greater delocalization and hence higher entanglement.

}\fi

\if{
\begin{figure}[H]
    \centering
    \includegraphics[width=0.49\linewidth]{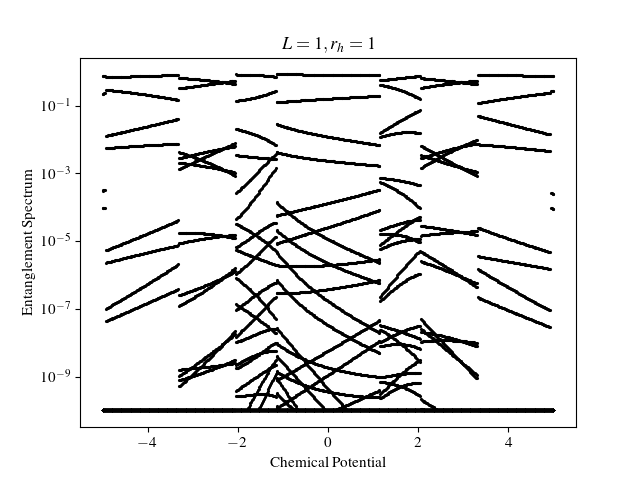}
    \includegraphics[width=0.49\linewidth]{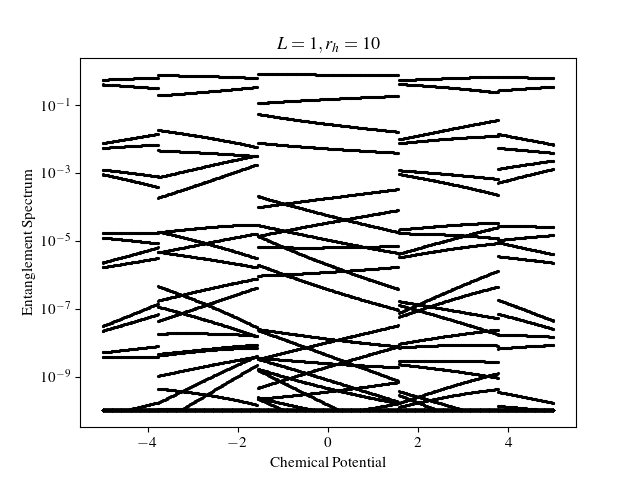}
    \caption{Ground state entanglement spectrum for $N=12$, $m=1$ as a function of $\mu$, $r_h=0.1$ (left)
    and $r_h=10$ (right).}
    \label{fig:enter-label}
\end{figure}
}\fi

\subsection{Charge Sectors}

Since the Hamiltonian commutes with the flat charge (\ref{flat}), 
it can be diagonalized with the basis of $Q_{\text{flat}}$.
The Hilbert space breaks into blocks labeled by the total fermion number $q$, and each block has its own ground and first excited energy surfaces.
The Hamiltonian in this basis takes the form: $H=\bigoplus_{q}H_q\;,$ where $H_q$ means the block-diagonalized Hamiltonian with charge $q$.
For an $N$-qubit system, $q$ takes a value between $-N$ and $N$. Let $E_{n,q}$ be $n$-th eigenvalue of the charge $q$-sector. 

\begin{figure}
    \centering
    \includegraphics[width=0.32\linewidth]{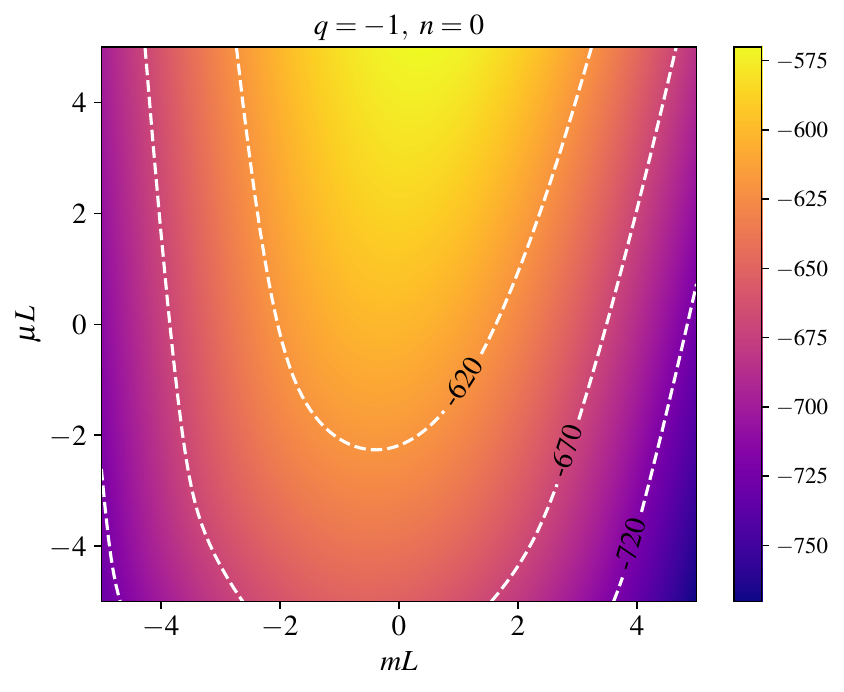}
    \includegraphics[width=0.32\linewidth]{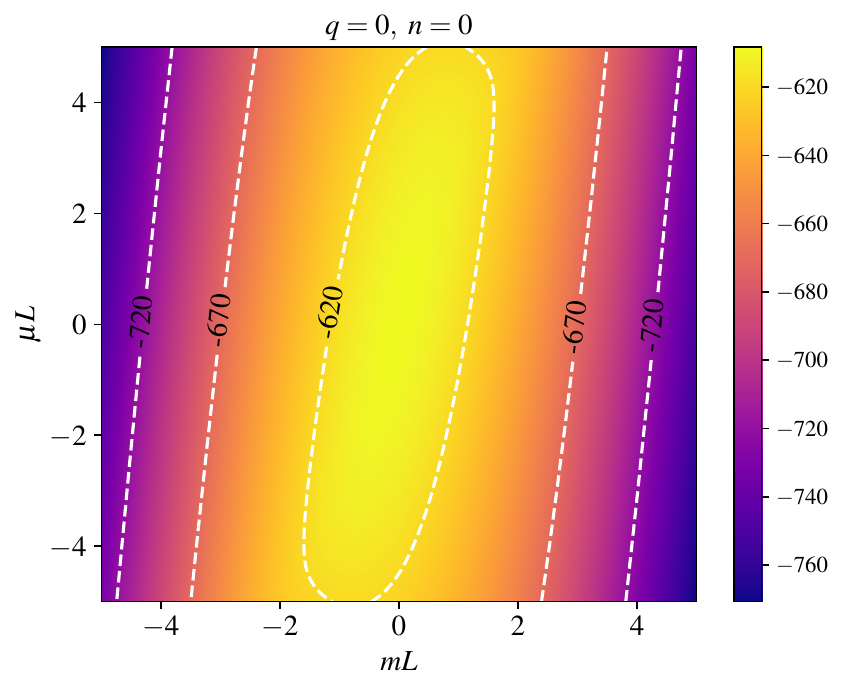}
    \includegraphics[width=0.32\linewidth]{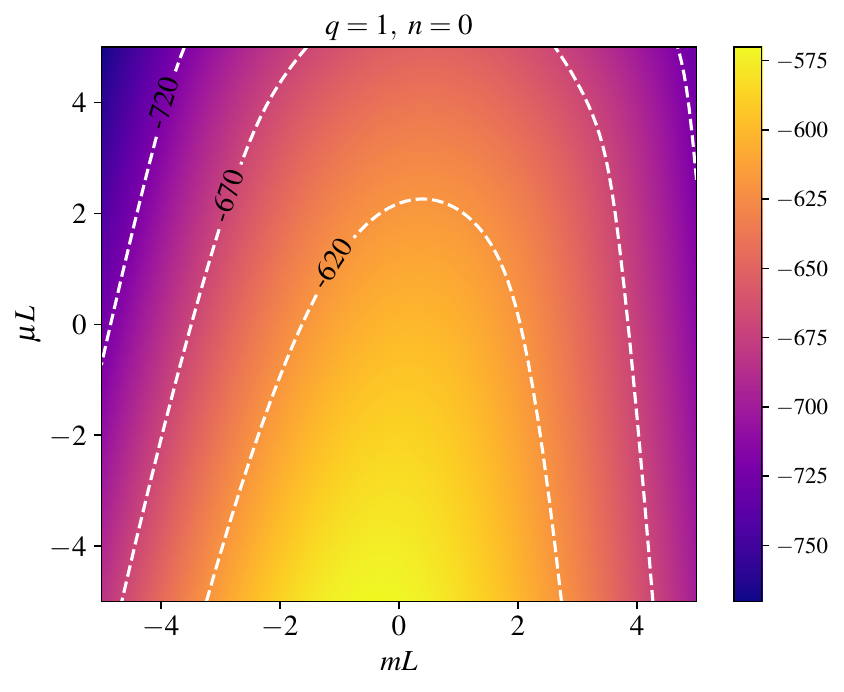}
    \includegraphics[width=0.32\linewidth]{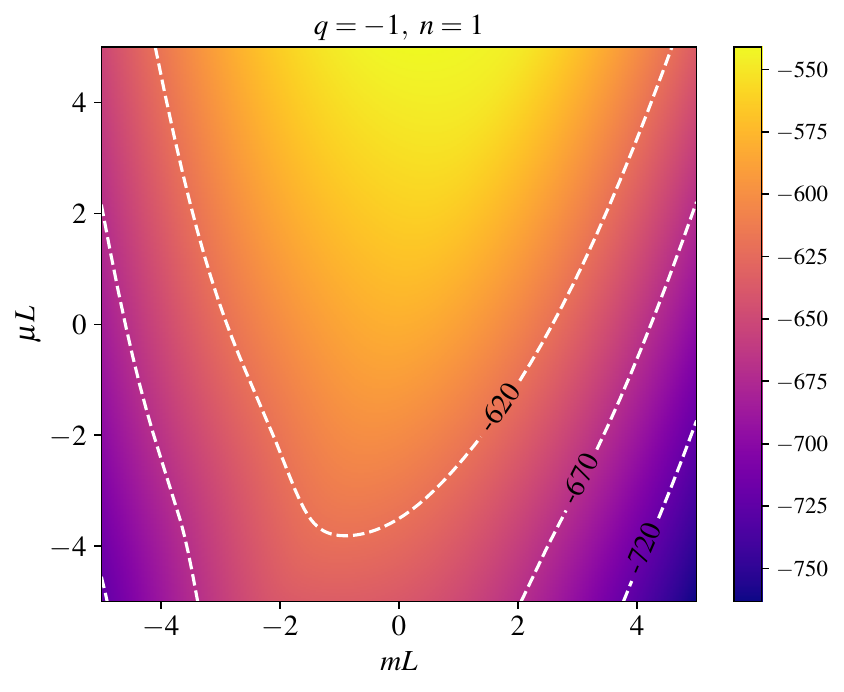}
    \includegraphics[width=0.32\linewidth]{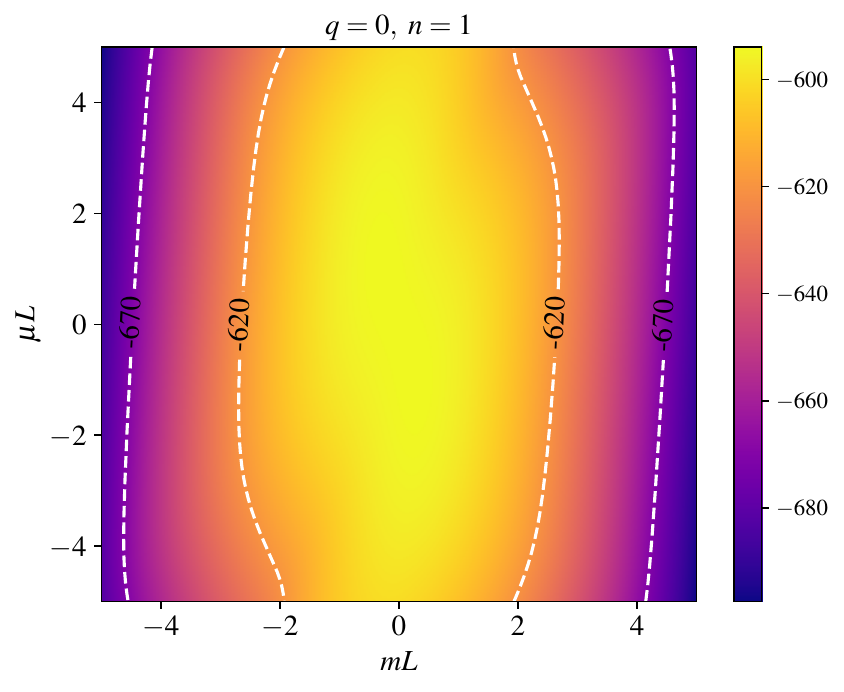}
    \includegraphics[width=0.32\linewidth]{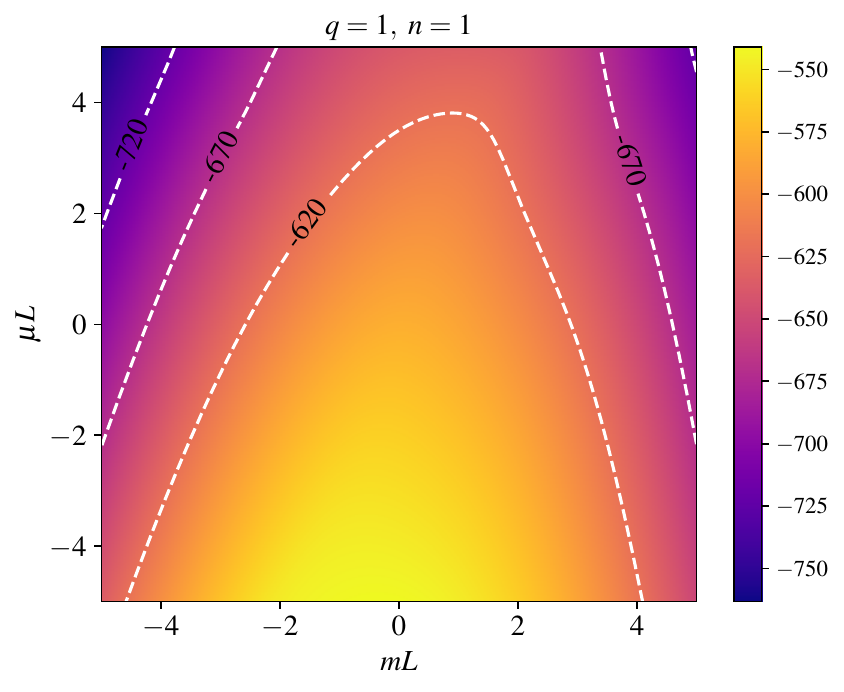}
    \caption{The energy spectrum into its three charge sectors
$q \in\{ -1,\;0,\;+1\}$.
We plot, for each sector, the ground level (top row, $n=0$) and the first excited level (bottom row, $n=1$) as a function of the dimensionless mass $mL$ (horizontal axis) and chemical potential $\mu L$ (vertical axis), at fixed horizon radius $r_h=10$ with $N=12$. For all plots, $L=1$ is used.}
    \label{fig:charged_energy}
\end{figure}

In Fig.~\ref{fig:charged_energy} we break out the energy spectrum into its three charge sectors
$q = -1,\;0,\;+1$.
We plot, for each sector, the ground level (top row, $n=0$) and the first excited level (bottom row, $n=1$) as a function of the dimensionless mass $mL$ (horizontal axis) and chemical potential $\mu L$ (vertical axis), at fixed horizon radius $r_h/L=10$.
In the top row we have the lowest-energy $E_{0,q}(mL,\mu L)$, and the three panels show $q=-1$ (left), $q=0$ (middle), $q=+1$ (right). Each heatmap is warped hill‐shaped rather than the diamond of the overall ground‐state energy 
(\ref{fig:E0_heatmap_BH_N12}). We see the combined symmetry under
$(q,m,\mu)\;\longrightarrow\;(-q,-m,-\mu)$,
by noting that the $q=+1$ plot is the point-reflection of the $q=-1$ plot, while the $q=0$ sector is symmetric under $(m,\mu)\to(-m,-\mu)$.
Physically, shifting $\mu$ and $m$ changes which charge sector minimizes the energy: for large positive $\mu$, $q=+1$ is favored, while for large negative $\mu$, $q=-1$ wins, with $q=0$ in between.

In the bottom row we plot the first excitation $E_{1,q}(mL,\mu L)$,
with the three charge sectors $q=-1,0,+1$.
Compared to the top row, these heatmaps are less smooth: the excitation energy in each sector varies more gently.
Where the ground‐state surfaces had their valley along $\mu=\pm m$, the excited‐level surfaces likewise show a trough near those lines. The same $(q,m,\mu)\to(-q,-m,-\mu)$ mapping relates the left and right panels, and the middle is self-invariant.

Thus, for both eigenstates, there is a spectral flow, where as we dial $\mu$ across $\pm m$, the energetically preferred charge sector switches. We also observe the redshift (\ref{redshift}) and curvature effects as  distortion of these heatmaps compared to the flat‐space results.

\begin{figure}[H]
    \centering
    \includegraphics[width=0.5\linewidth]{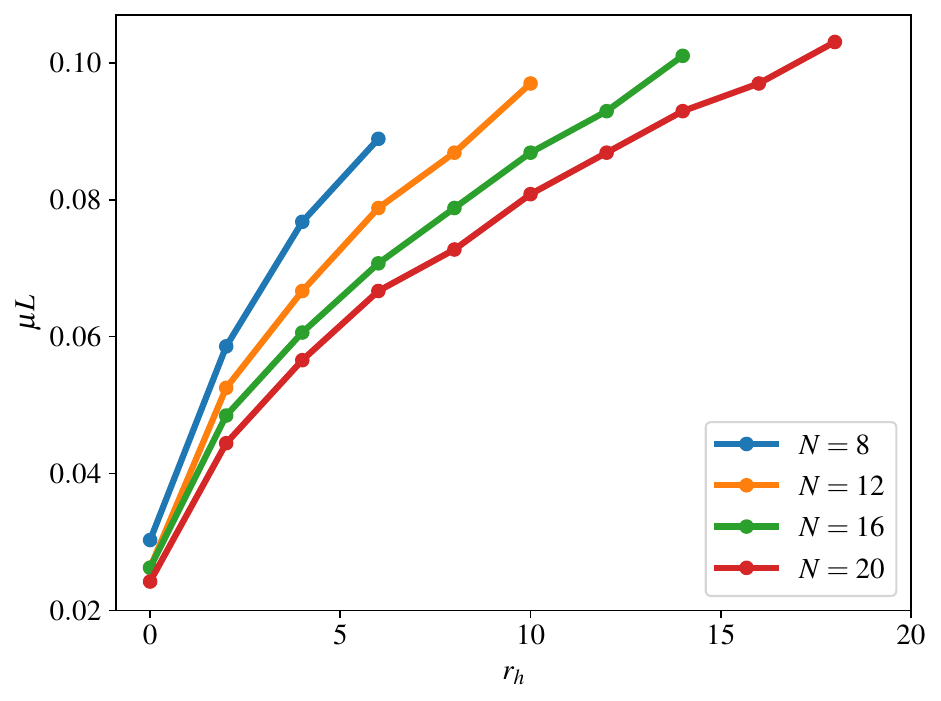}
    \caption{The transition point $\mu$ with $L=1$, when $m=0$,  where $E_{0,0}=E_{0,1}$, as a function of $r_h$ for different values of $N$.
We see that it  increases monotonically as a function of $r_h$.}
    \label{fig:charge gap}
\end{figure}

In order to further see the spectral flow, we plot in Fig.~\ref{fig:charge gap}
the transition point $\mu$ when $m=0$,  where $E_{0,0}=E_{0,1}$, as a function of $r_h$.
We see that it  increases monotonically as a function of $r_h$,  for different values of $N$. 

\subsection{The Continuum Limit}

\begin{figure}[H]
    \centering
    \includegraphics[width=0.32\linewidth]{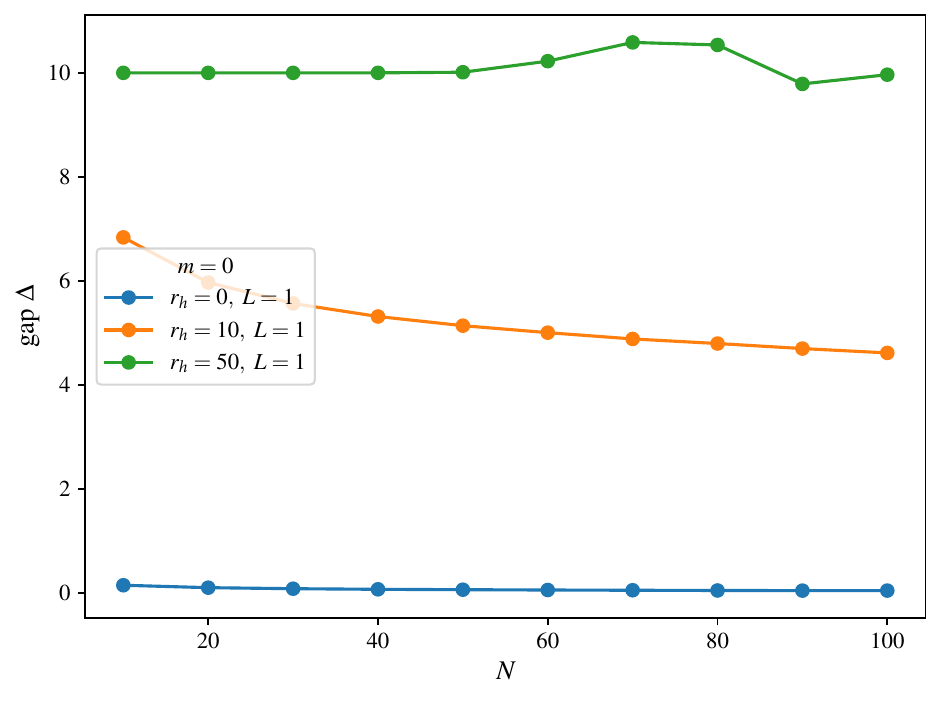}
    \includegraphics[width=0.32\linewidth]{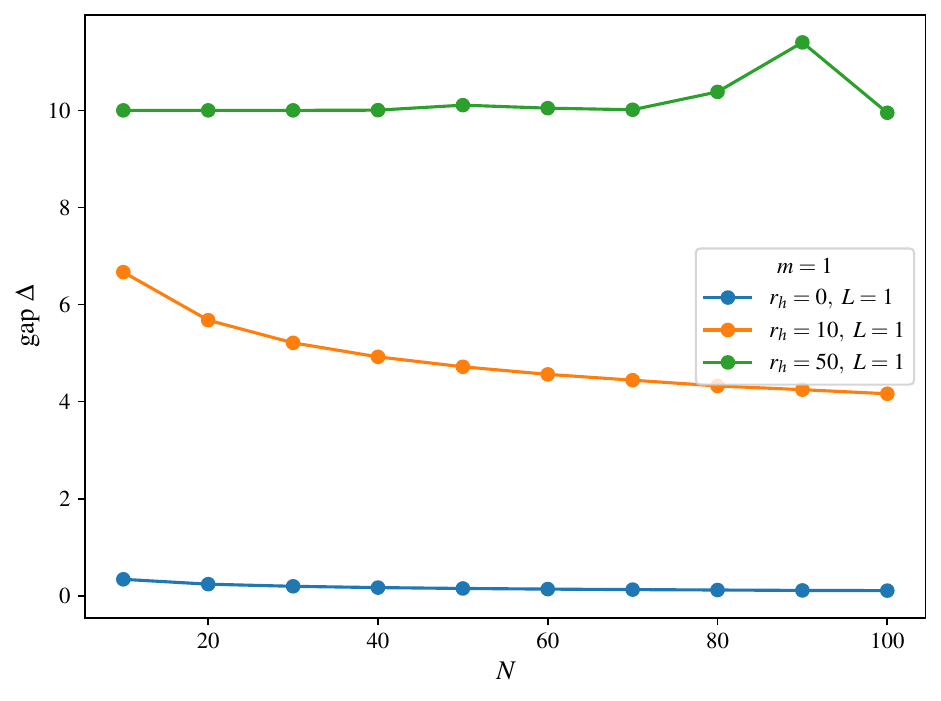}
    \includegraphics[width=0.32\linewidth]{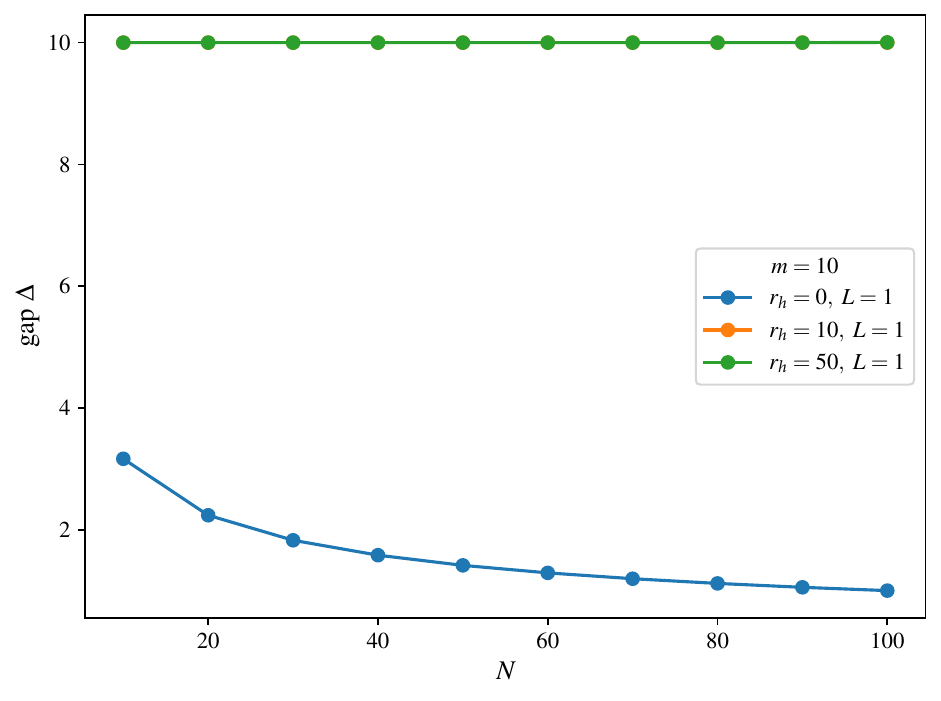}
    \includegraphics[width=0.32\linewidth]{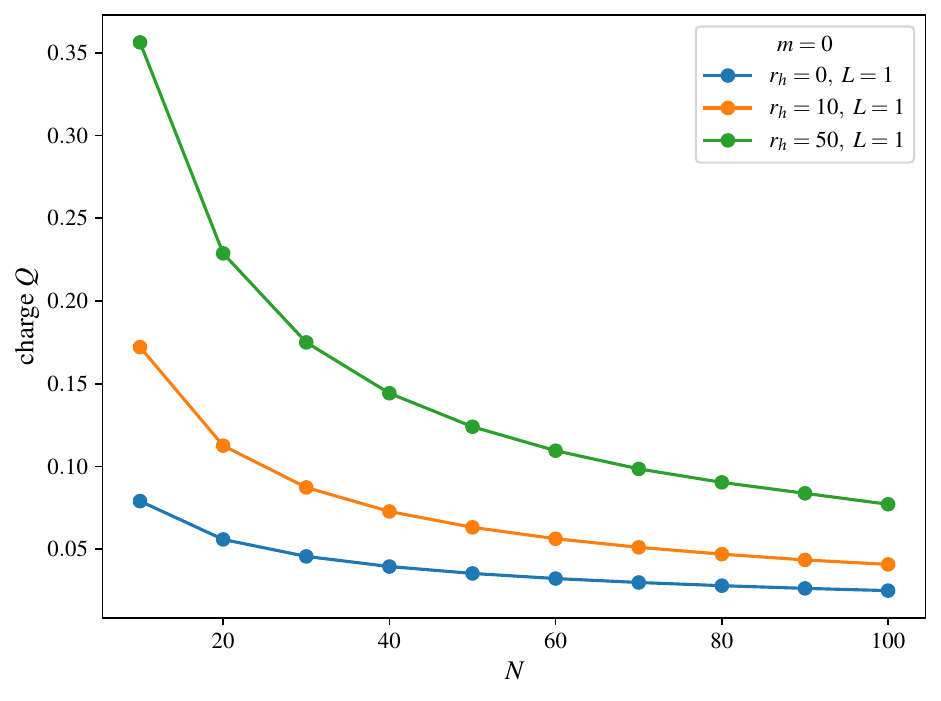}
    \includegraphics[width=0.32\linewidth]{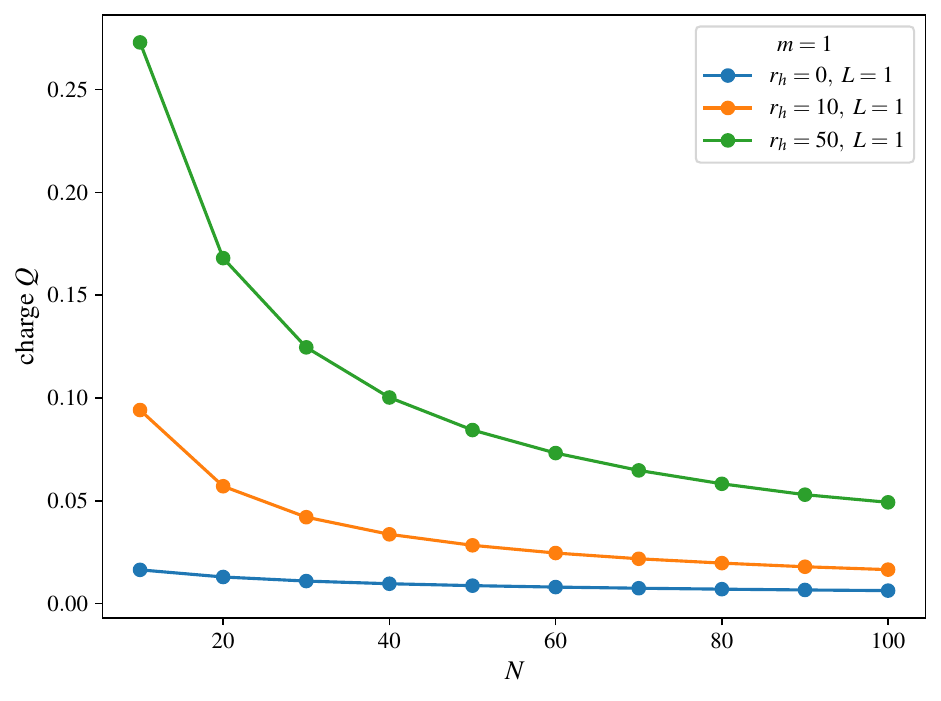}
    \includegraphics[width=0.32\linewidth]{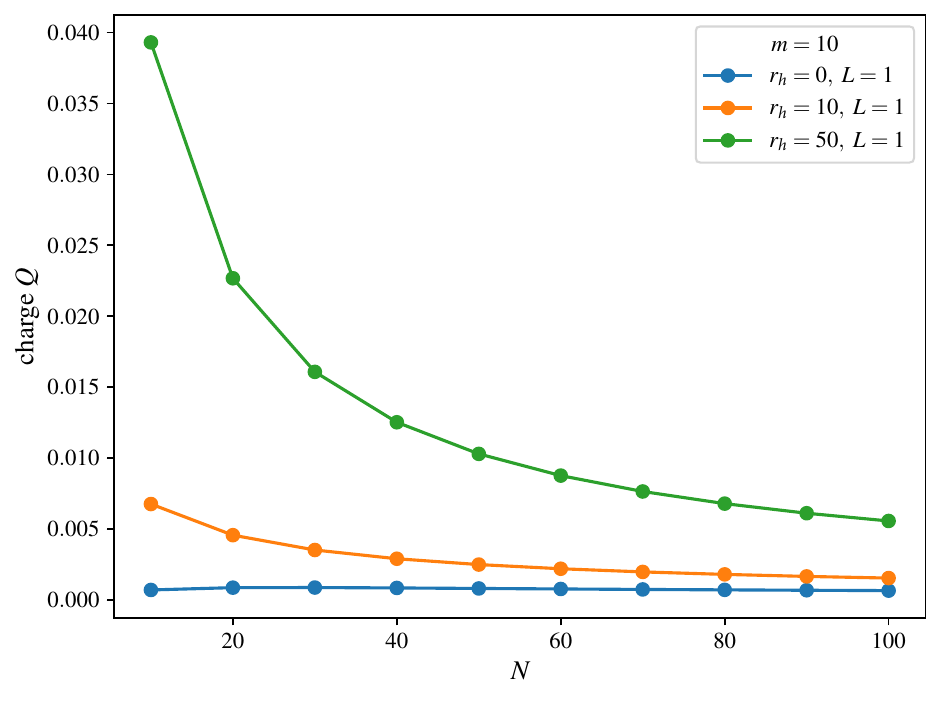}
    \caption{Continuum-limit check. With lattice spacing $a=1/\sqrt{N}$ and $\mu=0$, the energy gap $\Delta$ (top; $m\in \{0,1,10\}; r_h \in \{0,10,50\}; L=1$) rapidly converges as $N$ increases, while the net weighted charge $Q$ (bottom) decays to zero, confirming that vacuum polarization and finite-size imbalance vanish in the continuum. Holding the horizon radius $r_h$ fixed sets the discrete gap scale; in this limit the gap shows no residual $L$-dependence.}
    \label{fig:conti_Q}
\end{figure}

Fig.~\ref{fig:conti_Q} shows the continuum limit of energy gap (top) and the weighted charge (\ref{QC}) (bottom), for $L=1$.
We set the lattice spacing as $a=1/\sqrt{N}$, so that $r_N \;=\; r_h + Na \rightarrow \infty$ when $a\to 0,~N\to\infty$. The other parameters are chosen as $r_h=10, m=\mu=0$.
To make these quantities dimensionless, appropriate powers of $a=1/\sqrt{N}$ multiply both $\Delta$ and $Q$, causing them to rapidly diminish as $N$ increases.
In the top panels we see $\Delta$ as a function of $N$
for the $AdS_2$ scale $L=1$.  As $N$ grows (and hence $a\to0$), the curves rapidly settle toward their continuum values, demonstrating that the discretized gap converges to the analytic prediction in the limit $N\to\infty$.
Note that in our set-up the only length scale fixing a discrete gap is the horizon-to-boundary separation $r_h$. Since we held $r_h$ fixed, there is no dependence of the gap on $L$ in the continuum limit.
The bottom panels show the ground‐state net charge per length, $Q$, for $L=1$.  In all cases $Q$ decays toward zero as $N$ increases, confirming that vacuum polarization effects (and any finite‐size charge imbalance) vanish in the continuum.
Thus, Fig.~\ref{fig:conti_Q} provides a clear numerical demonstration that—with the scaling $a=1/\sqrt{N}$, both the energy gap and the net charge smoothly approach their expected continuum limits as $N\to\infty$.

\if{
\begin{equation}
    Q=\sum_{n=1}^Nw_n\frac{Z_n+(-1)^n}{2},
\end{equation}
where $w_n=\frac{L}{\sqrt{r^2_n-r^2_h}}, r_n=an+r_h$. 

}\fi

\section{\label{sec:CGEOTOC} Chiral Gravitational Effect and Information Scrambling}

\subsection{The effect of spin connection}

In the following we explore implications of the spin connection term 
$\left(X_nY_{n+1}-Y_nX_{n+1}\right)$ in \eqref{HBHF}. We introduce two operators,
$\kappa$ and $\chi$, that diagnose the emergence of spin‐current patterns and three‐spin chiral order in 
the system: 
\begin{align}
    \kappa&=\sum_{i=1}^{N-1}(S_i\times S_{i+1})_z=\frac{1}{4}\sum_{i=1}^{N-1}(X_iY_{i+1}-Y_iX_{i+1}),\nonumber\\
    \chi&=\sum_{i=1}^{N-2}S_i\cdot(S_{i+1}\times S_{i+2}) \ , 
\end{align}
where 
\begin{eqnarray}
S_i\times S_{i+1}&=&\left(\frac{Y_iZ_{i+1}-Z_iY_{i+1}}{4},\frac{Z_iX_{i+1}-X_iZ_{i+1}}{4},\frac{X_iY_{i+1}-Y_iX_{i+1}}{4}\right) \nonumber\\
S_i&=&(X_i/2,Y_i/2,Z_i/2) \ .
\end{eqnarray}
$\kappa$ corresponds to the local current without weight $w_n$ and vanishes unless spins are non-collinear in the plane perpendicular to the $z$-direction, whereas $\chi$ tracks the three-spin solid angle and vanishes unless the triad is non-coplanar. 

If $\langle\kappa\rangle>0$, then on every bond $i\to i+1$ the spin at $i+1$ is canted a little counter-clockwise (in the $XY$ plane) relative to the spin at $i$, while if $\langle\kappa\rangle<0$, it is canted clockwise. When $\langle \chi\rangle \;\neq\;0$,
it means that the spin chain has developed a non-coplanar, chiral ordering of triples of spins, rather than all lying flat in a single plane. 
If $\langle\chi\rangle>0$, then on average each spin triple $(i,i+1,i+2)$ twists in a right-handed sense (e.g., from $i$ to $i+1$ to $i+2$). If $\langle\chi\rangle<0$, the twist is left-handed.

Define the time-reversal operator $\mathcal{T}$ via its action on the Hilbert space:
\begin{equation}
    \mathcal{T}S^\alpha_i\mathcal{T}^{-1}=-S^\alpha_{i} \ . 
\end{equation}
$\kappa$ is even under $\mathcal{T}$, while $\chi$ is odd: $\mathcal{T}\kappa\mathcal{T}^{-1}=\kappa$, $\mathcal{T}\chi\mathcal{T}^{-1}=-\chi$. Define also the parity operator $\mathcal{P}$:
\begin{equation}
    \mathcal{P}S^\alpha_i\mathcal{P}^{-1}=S^\alpha_{N+1-i} \ . 
\end{equation}
$\kappa$ is even under $\mathcal{P}$, while $\chi$ is odd:
\begin{equation}
\mathcal{P}\kappa\mathcal{P}^{-1}=\kappa,~~ \mathcal{P}\chi\mathcal{P}^{-1}=-\chi \ .     
\end{equation}

Under a parity or mirror reflection in the chain, $\kappa$ is even, but under a global spin‐reflection 
$Y\to -Y$ (or time‐reversal acting on spins) it flips sign. Thus, a nonzero $\langle\kappa\rangle$ means that one of the two “handed” patterns (clockwise vs.\ counterclockwise) has been spontaneously chosen, breaking that discrete reflection symmetry. As a consequence, the system supports a persistent spin current $j^z_i\propto\bigl(\mathbf S_i\times\mathbf S_{i+1}\bigr)_z$ flowing around the chain.
Note that there is no conventional magnetic order,$\langle X\rangle=\langle Y\rangle=0$, yet the ground state is chiral.
Under parity or time-reversal, $\chi$ changes sign.  Hence, a nonzero $\langle\chi\rangle$ means one of the two mirror-related, time-reversed patterns has been chosen spontaneously—the system breaks those discrete symmetries in favor of a particular chirality. As a consequence, 
we have a chiral spin liquid–like order: no conventional magnetic order $(\langle\mathbf S\rangle=0)$, but a uniform twist in every triple of sites.

\begin{figure}[H]
    \centering
    \includegraphics[width=0.32\linewidth]{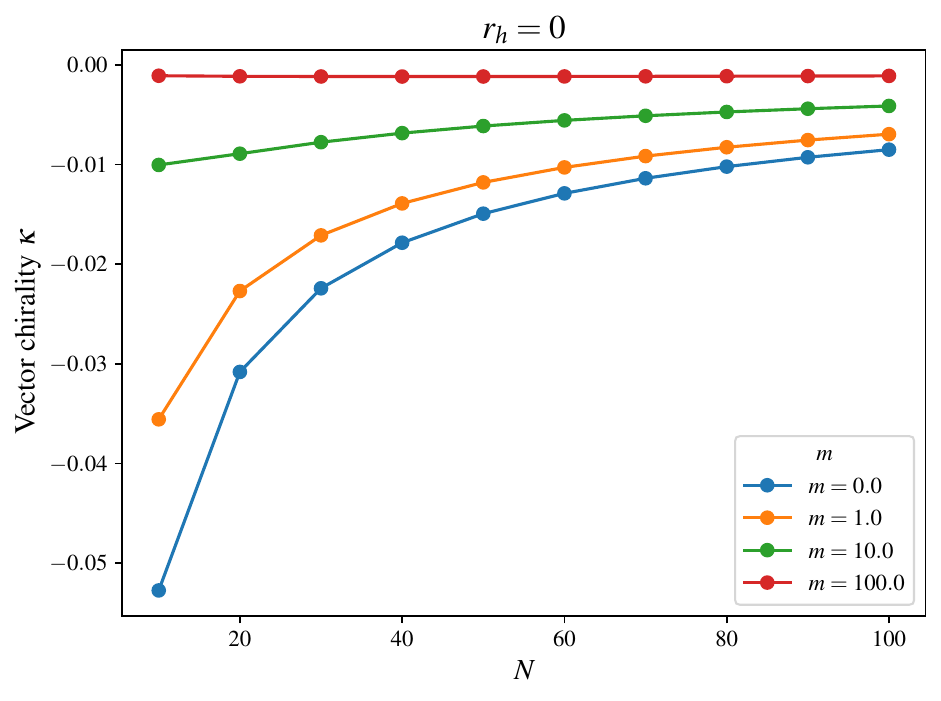}
    \includegraphics[width=0.32\linewidth]{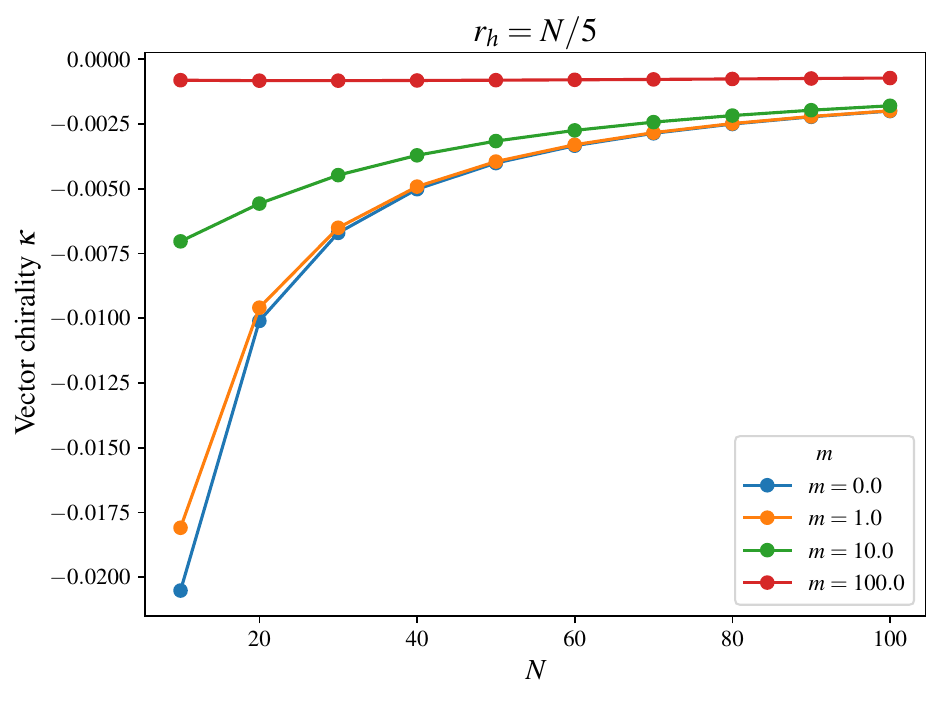}
    \includegraphics[width=0.32\linewidth]{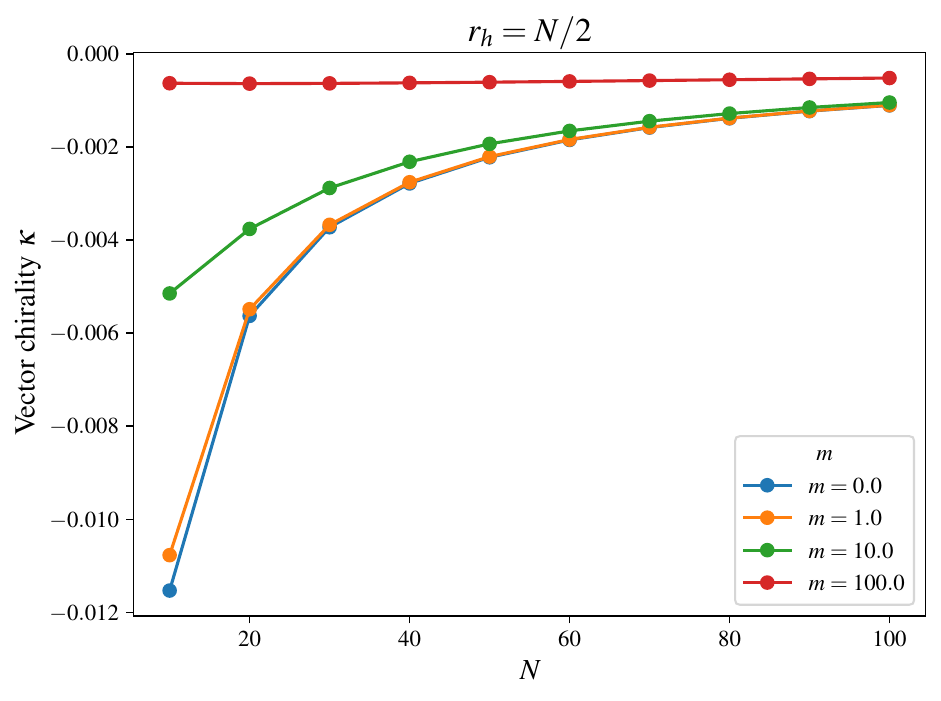}
    \caption{The expectation value of the vector chirality, which measures the handed twist on each nearest-neighbor bond 
$\langle \kappa\rangle$ in the ground state vs. the system size $N$.
The three panels correspond to horizon radii $r_h \in \{0, N/5, N/2\}$.
The colored curves track four values of the mass $m \in\{0,\;1,\;10,\;100\}$ with $L=1$.}
    \label{fig:Vector_chirality}
\end{figure}

In Fig. \ref{fig:Vector_chirality},  we plot the expectation value of the vector chirality, which measures the handed twist on each nearest-neighbor bond 
$\langle \kappa\rangle$ in the ground state vs. the system size $N$.
The three panels correspond to horizon radii $r_h=0, r_h=N/5$, and $r_h=N/2$.
The colored curves track four values of the mass $mL \in \{0,\;1,\;10,\;100\}$.
For $r_h=0$, $\langle\kappa\rangle$ is slightly negative and grows (in absolute value) as $1/N$, indicating a small uniform twist even in pure $AdS_2$ (no black hole).
As the horizon appears $(r_h=N/5, r_h=N/2)$, the magnitude of $\langle\kappa\rangle$ decreases—strong redshift tends to oppose the two-site canting direction seen at small or zero $r_h$.
Heavier masses $(mL\gtrsim10)$ suppress the chirality less, so the curves fan out slightly at large $mL$.

\begin{figure}[H]
    \centering
    \includegraphics[width=0.32\linewidth]{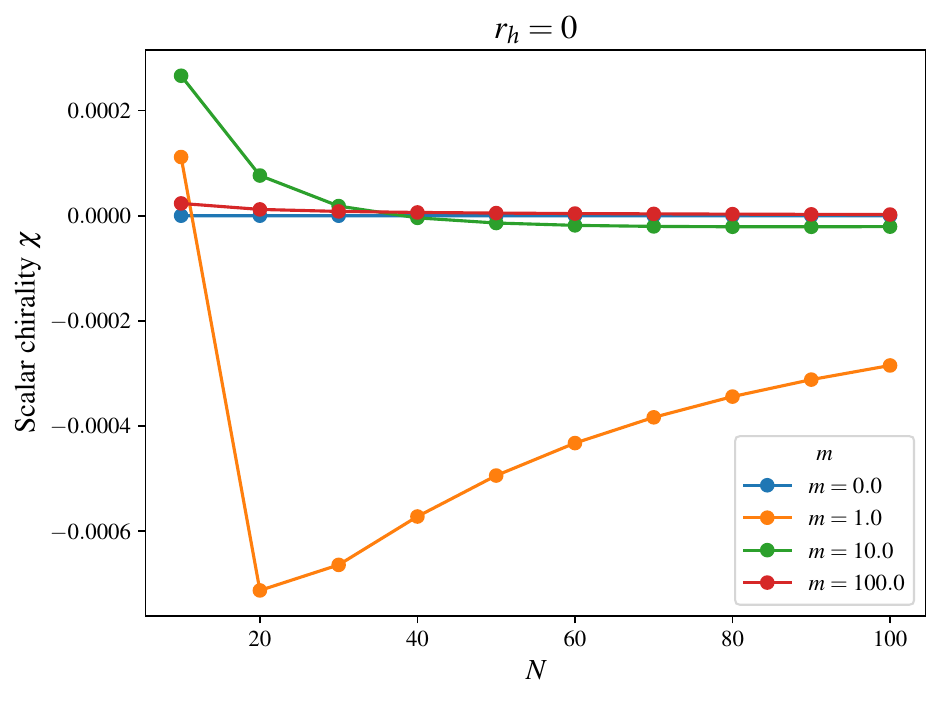}
    \includegraphics[width=0.32\linewidth]{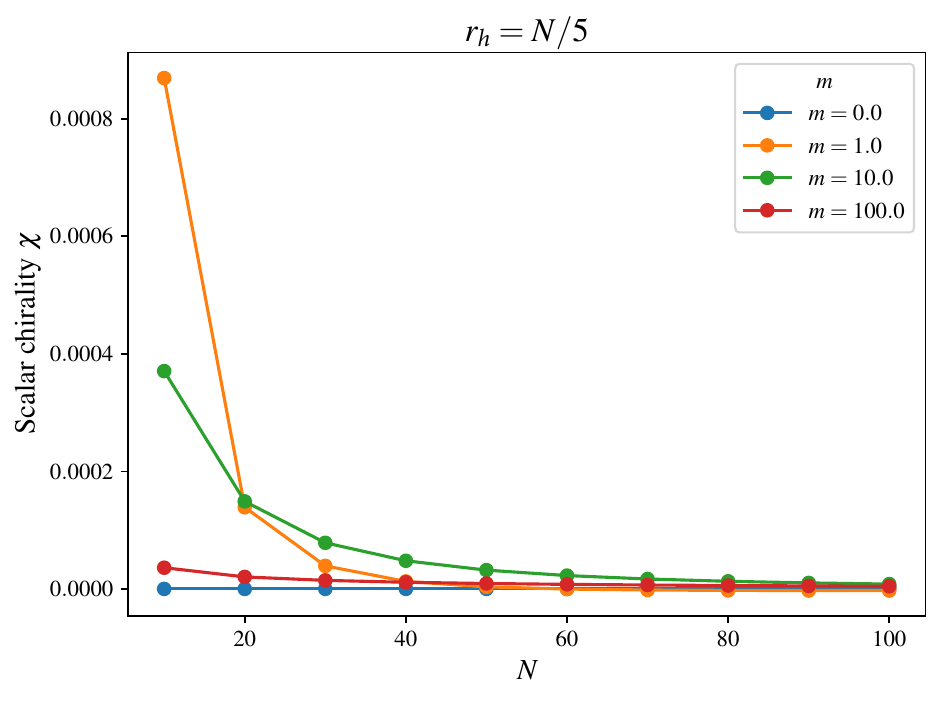}
    \includegraphics[width=0.32\linewidth]{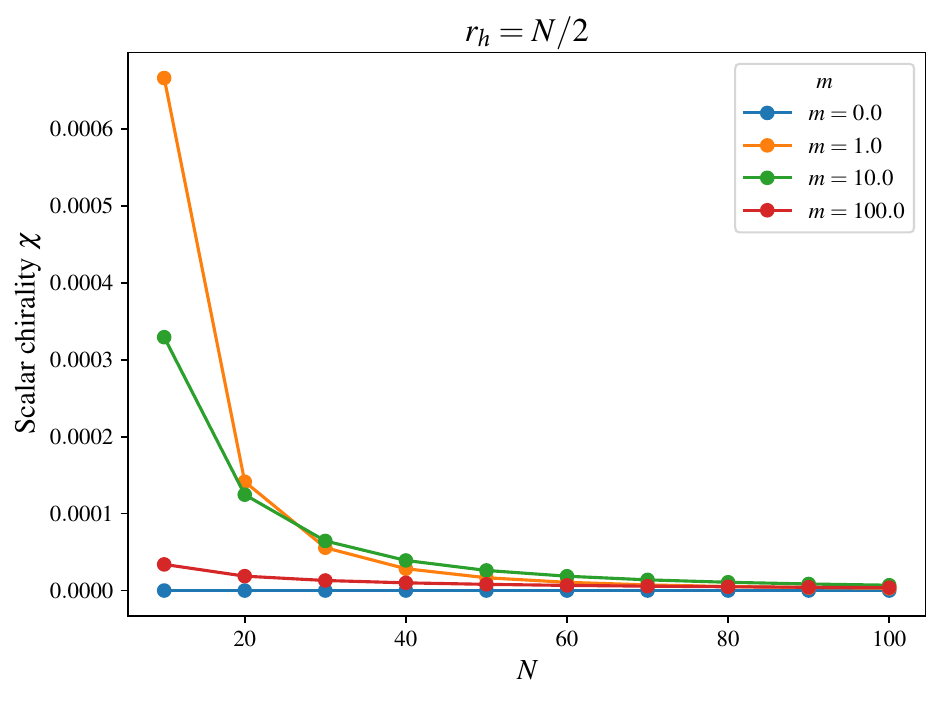}
    \caption{The expectation value of the scalar chirality $\langle\chi\rangle$ in the ground state vs. $N$.
 The three panels correspond to horizon radii $r_h=0, r_h=N/2$, and $r_h=N/5$.
The colored curves track four values of the mass $m \in\{0,\;1,\;10,\;100\}$ with $L=1$.}
    \label{fig:Scalar_chirality1}
\end{figure}

In Fig. \ref{fig:Scalar_chirality1} we plot 
the expectation value of the scalar chirality $\langle\chi\rangle$ in the ground state vs. $N$.
The Layout mirrors that of Fig. \ref{fig:Vector_chirality}.
The overall magnitude of $\langle\chi\rangle$ is about $10\times{}$ smaller than $\langle\kappa\rangle$, reflecting that the chain remains nearly coplanar.
The sign flips when we go from $r_h=0$ to finite $r_h$ means that the three-site volume’s handedness is opposite to the bond twist.
$\langle\chi\rangle$ decays toward zero as $N\to\infty$, showing that these chiral effects are finite-size edge phenomena that vanish in the strict continuum limit.
It should be emphasized that $\chi\neq 0$ even at $m=0$, indicating that there is a (static) current. This is a significant difference from the flat case ($L\to\infty$), where the chiral symmetry is protected. When $m\neq0$, the current is dynamical (see Sec.~\ref{sec:CGE}). 

\if{

$\langle \kappa \rangle$ and $\langle chi\rangle$ are shown in Figs. \ref{fig:Vector_chirality} and \ref{fig:Scalar_chirality}, respectively. They were computed up to $N=100$ for different values of $r_h$ and $m$. It should be emphasized that $\chi\neq0$ even at $m=0$, indicating that there is a (static) current. This is a significant difference from the flat case ($L\to\infty$), where the chiral symmetry is protected. When $m\neq0$, the current is dynamical (see Sec.~\ref{sec:CGE}). 

When $r_h=0$ (see left panels of Figs \ref{fig:Vector_chirality} and \ref{fig:Scalar_chirality}), both signs are the same. This means that the way each nearest-neighbour bond twists and the way every consecutive three-spin cluster subtends a solid angle share the same handedness.

It is interesting that their signs differ when  $r_h=N/5,N/2$, as shown in the middle and right panels of Figs \ref{fig:Vector_chirality} and \ref{fig:Scalar_chirality}. This indicates that the spin texture twists one way on the scale of two sites but the three-site volume it carves out is oriented the opposite way. 

}\fi

\begin{figure}[H]
    \centering
    \includegraphics[width=0.32\linewidth]{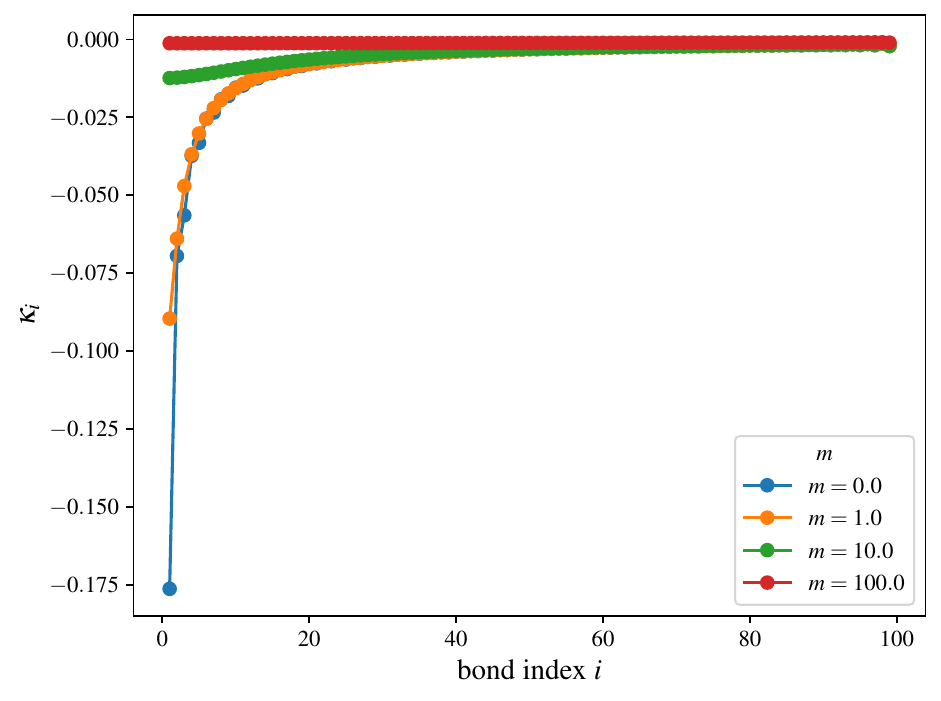}
    \includegraphics[width=0.32\linewidth]{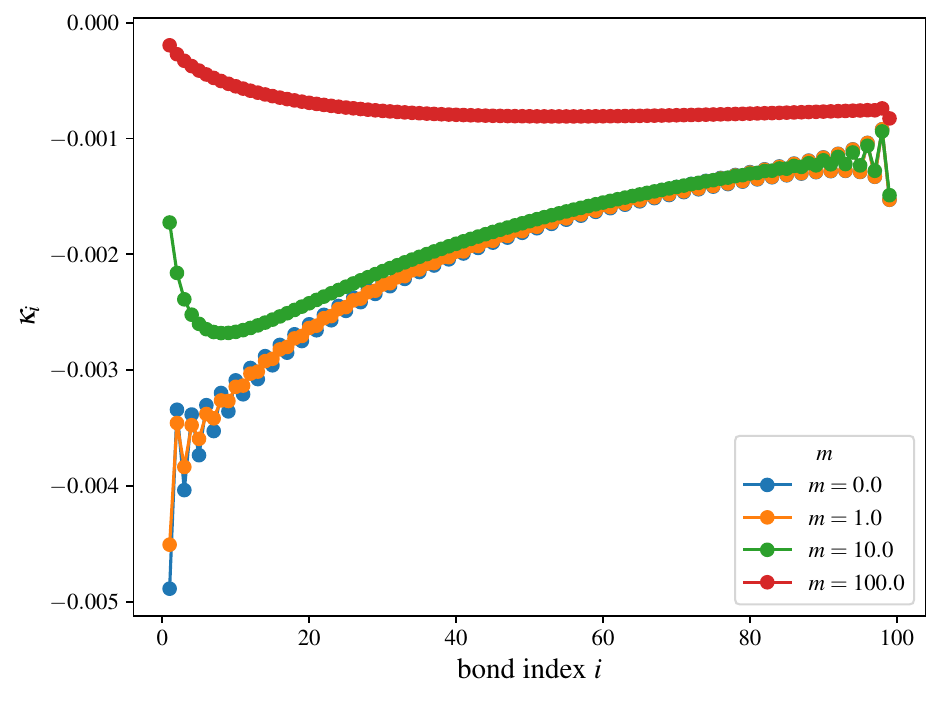}
    \includegraphics[width=0.32\linewidth]{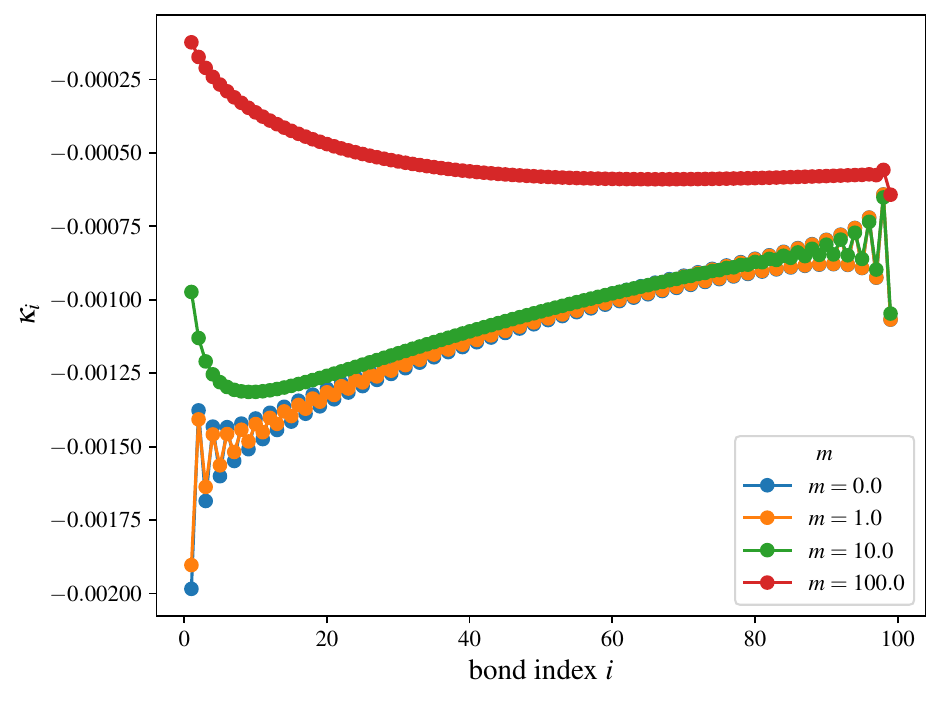}
    \includegraphics[width=0.32\linewidth]{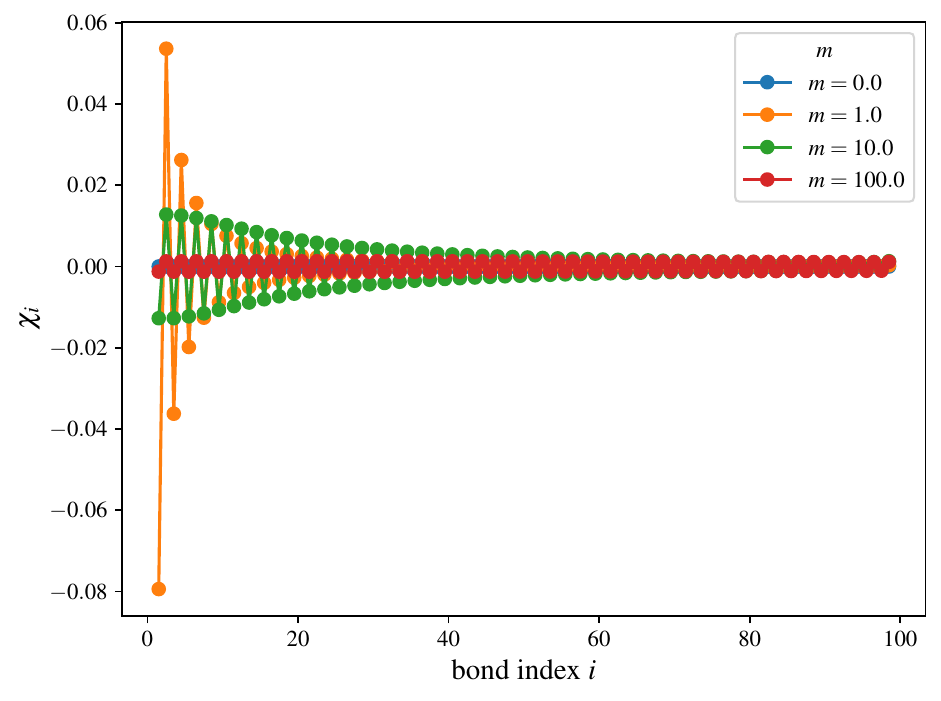}
    \includegraphics[width=0.32\linewidth]{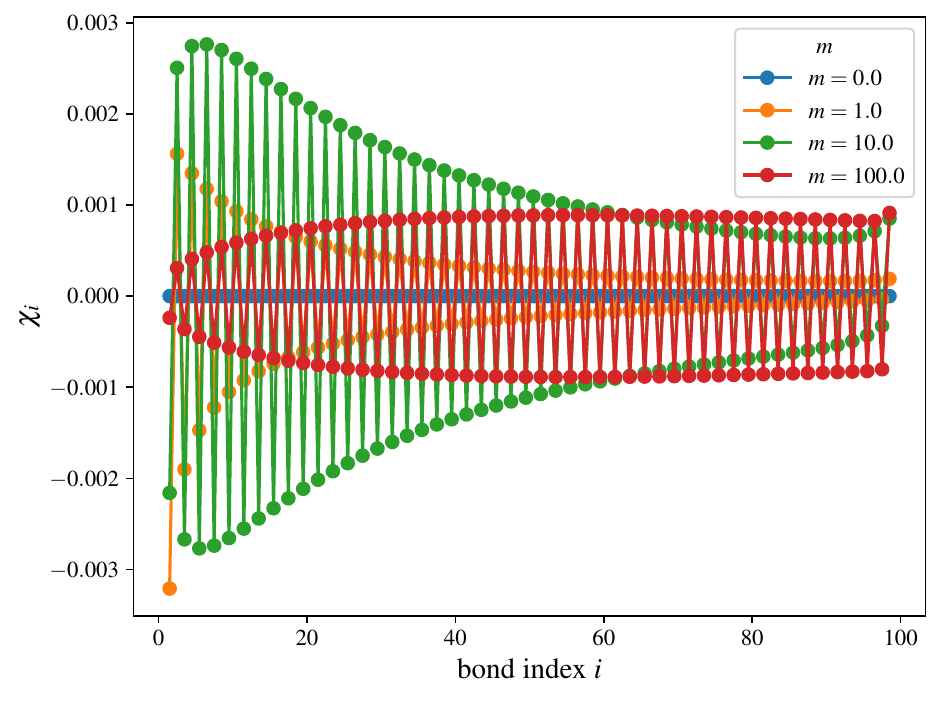}
    \includegraphics[width=0.32\linewidth]{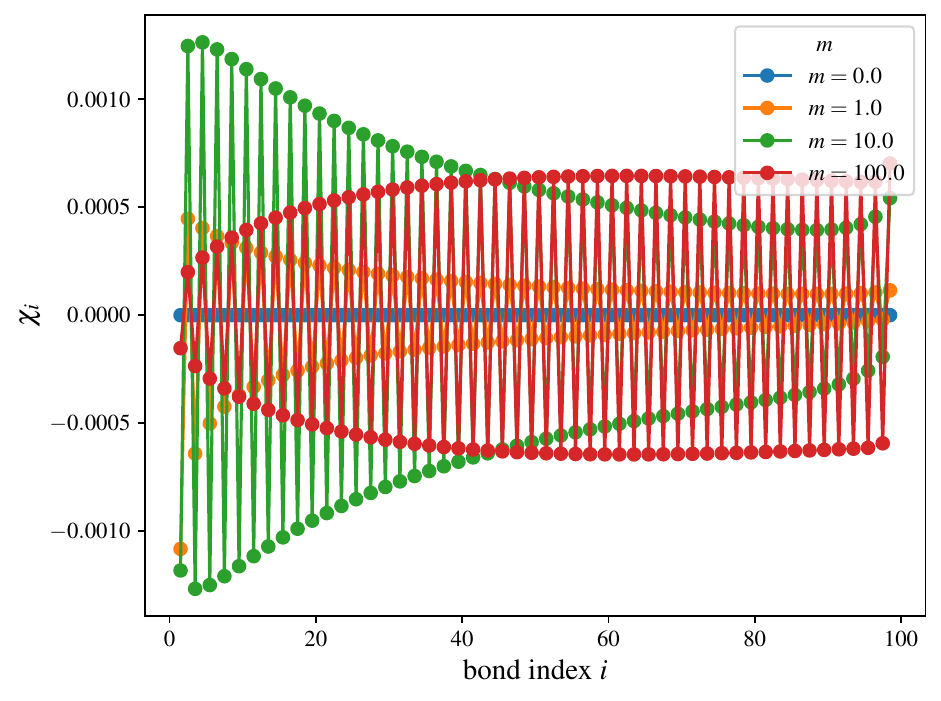}
    \caption{Local profiles $\kappa_i$ and $\chi_i$ for $N=100$, where
the horizontal axis is the bond index $i\in[1,\,N]$.
From left to right $r_h/N \in \{0,1/5, 1/2\}$. For all plots $L=1$ is used.}
    \label{fig:BH_local_chirality}
\end{figure}

In Fig. \ref{fig:BH_local_chirality} we plot the local profiles $\kappa_i$ and $\chi_i$ for $N=100$, where
the horizontal axis is the bond index $i\in[1,\,N]$.
The curves are:
\begin{equation}
\kappa_i = (\mathbf S_i\times\mathbf S_{i+1})_z,~~~ \chi_i = \mathbf S_i\!\cdot\!(\mathbf S_{i+1}\times\mathbf S_{i+2}) \ .  
\label{ki}
\end{equation}
Both $\kappa_i$ and $\chi_i$ peak near the center of the chain and fall off toward the ends—edge effects dominate the chiral ordering. $\kappa_i$ oscillates smoothly (bond by bond), while $\chi_i$ is smaller and more sharply localized (only a few triangles carry appreciable volume).
Larger $mL$ slightly reduces the oscillation amplitude but doesn’t qualitatively change the spatial pattern.
Taken together, these three figures show that: (i) chirality in the ground state is a finite-size, edge-dominated phenomenon that flips sign under strong $AdS_2$ redshift, (ii)
Bond twists ($\kappa$) are an order of magnitude larger than triangular volumes ($\chi$), but both vanish as $N\to\infty$, (iii) Local profiles confirm that the chiral order lives mainly in the chain’s bulk region (peaking at mid-chain) and decays toward the boundaries.

\subsection{\label{sec:CGE}Chiral Gravitational Effect}

{\if
The spin connection induces a
chiral imbalance between the left- and right-handed fermions, therefore it plays a role of chiral chemical potential $\mu_5$. In the curved space, $\mu_5$ is position-dependent: $\mu_5(r)=\frac{r}{2L^2f(r)}$, as it can be read from 
the term $\frac{r}{2L^2f(r)}\bar{\psi}\gamma^1\psi$ in the Hamiltonian. Importantly, such a chiral imbalance originates from the spacetime geometry and presents regardless of the fermion mass.\footnote{This chiral imbalance is intrinsic to the AdS background (with any choice of coordinates, eg. Poincare, Global and Schwarzschild) and does not occur for a massive or massless free fermion in dS or flat backgrounds.}

In the absence of a background classical electric field, the chiral anomaly in (1+1) dimension is explained by
\begin{equation}
    \partial_\mu J^\mu_5\;=\;2im\bar{\psi}\gamma_5\psi. 
\end{equation}
Notably, the vector current $J\equiv J^0_5=\bar{\psi}\gamma^1\psi$ of a massive fermion is not conserved, and this non-conservation is key to understanding chiral dynamics in (1+1) dimensions \cite{PhysRevResearch.2.023342,PhysRevD.108.074001,Ikeda:2024rzv}. In the AdS background, the commutation relation is:
\begin{equation}
    [H,J]\;=\;2m\int dz \frac{L}{z}\bar{\psi}\gamma_5\psi. 
\end{equation}

In four dimensions, there is a chiral gravitational anomaly,
classically, the axial current \( J_5^\mu \) for a massless fermion field is defined as:
\begin{equation}
J_5^\mu = \bar{\psi} \gamma^\mu \gamma^5 \psi, 
\end{equation}
where $\gamma^5$ is the chirality matrix.
At the quantum level, the conservation of this current is broken in the presence of a gravitational field, leading to the chiral gravitational anomaly:
\begin{equation}
\nabla_\mu J_5^\mu = \frac{1}{384 \pi^2} R_{\alpha \beta \gamma \delta} \tilde{R}^{\alpha \beta \gamma \delta}
\ ,
\end{equation}
where  $R_{\alpha \beta \gamma \delta}$ is the Riemann curvature tensor,
$\tilde{R}^{\alpha \beta \gamma \delta}$
is the dual Riemann tensor, defined as $\tilde{R}^{\alpha \beta \gamma \delta} = \frac{1}{2} \epsilon^{\gamma \delta \rho \sigma} R^{\alpha \beta}_{\ \ \rho \sigma}$.
The product $R_{\alpha \beta \gamma \delta} \tilde{R}^{\alpha \beta \gamma \delta}$ is known as the Pontryagin density in four dimensions, which can be expressed as:
\begin{equation}
R_{\alpha \beta \gamma \delta} \tilde{R}^{\alpha \beta \gamma \delta} = \frac{1}{2} \epsilon^{\alpha \beta \gamma \delta} R^{\rho \sigma}_{\ \ \alpha \beta} R_{\rho \sigma \gamma \delta} \ .
\end{equation}
Thus, the expression for the chiral gravitational anomaly in four dimensions is:
\begin{equation}
\nabla_\mu J_5^\mu = \frac{1}{384 \pi^2} R_{\alpha \beta \gamma \delta} \tilde{R}^{\alpha \beta \gamma \delta}.
\end{equation}

In two dimensions there is no equivalent of the four-dimensional Pontryagin density. However, a massive fermion, even in the absence of interactions, exhibits chiral imbalance in an AdS background.

}\fi

In the following we will discuss a lattice‐chirality phenomenon, which we call
a chiral gravitational effect, since it vanishes in flat space and only appears once we turn on the 
$AdS_2$ black hole background. 
On a curved spatial slice the Dirac fermion picks up a coupling to the background spin connection,
which when discretized becomes the bond‐chirality operator $\kappa_i$. 
Consider the total vector‐chirality current operator:
\begin{equation}
J \;=\;\sum_{i=1}^{N-1}\kappa_i  \ .
\end{equation}
$\kappa_i$ lives on the bond between site $i$ and site $i+1$, and measures
the twist between sites $i$ and $i+1$. Note that in the
open $N$-site chain we have $N\!-\!1$  nearest-neighbor bonds, hence the summation is up to $N-1$.
When the fermion is massless, $[H,J]=0$ (under the periodic boundary condition) and $\langle J(t)\rangle$ is constant.  A nonzero mass $m$
implies that $[H,J]\sim m\neq0$, hence $\langle J(t)\rangle$ is time dependent.

Consider the real-time evolution of the current: 
\begin{equation}
\label{eq:current_time}
    J(t)=e^{+i\int_0^tHd\tau}Je^{-i\int_0^tHd\tau} \ .
\end{equation}
Here the time-ordering should be imposed on the integrals. Time-evolution of $J(t)$ is proportional to $\kappa(t)$.

In the continuum limit of the open chain, $a\to0, N\to\infty$ at fixed physical length
the lattice spacing goes to zero and the number of bonds $N-1$ goes to infinity, as seen
in Fig.~\ref{fig:BH_local_chirality}:
\begin{equation}
\langle\kappa\rangle
= \frac1{N-1}\sum_{i=1}^{N-1}\langle\kappa_i\rangle
\;\sim\;O(1/N)
\;\longrightarrow\;0    \ .
\end{equation}
Thus, the average bond‐chirality vanishes in the continuum.
More precisely, $\kappa_1$ is large (nonzero) because bond $1–2$ sits where the spin connection 
$\omega$ effect is large, while $\kappa_{N-1}$ (bond 99–100) is almost zero because right at the horizon the redshift factor vanishes and there is no further change of geometry to induce chirality.
The reason being that all of the nonzero chirality is sourced by the boundaries, and in an infinite, translation‐invariant continuum there are no edges, so there’s nowhere for a net chirality‐current to reside.
Locally we still have a nonzero spin‐connection term in the Hamiltonian, so at any finite lattice spacing we
see a small $\kappa_i$, but when we smear that over a continuum interval, those local tilts average out to zero unless we explicitly keep a boundary.
Indeed, if instead we would have taken periodic boundary conditions, there would have been no
net $\sum_i\kappa_i$ even at finite $a$: every bond’s spin‐connection phase cancels once around the loop. Thus,
the open‐chain result is purely a finite‐size, boundary‐induced (edge phenomenon) chiral gravitational effect.

Note that in a translationally symmetric flat‐space open chain, both ends are identical and we should have seen equal effects on bonds $1$ and $N\!-\!1$.  Here, because the geometry itself is inhomogeneous (it interpolates from flat boundary to horizon), the only edge that matters for the chiral current is the boundary side.  The horizon side is a smooth cap, where the connection dies off.
The edge effect is localized where the background geometry changes abruptly from flat to curved space
near the $AdS_2$ boundary. 

\begin{figure}[H]
    \centering
    \includegraphics[width=0.24\linewidth]{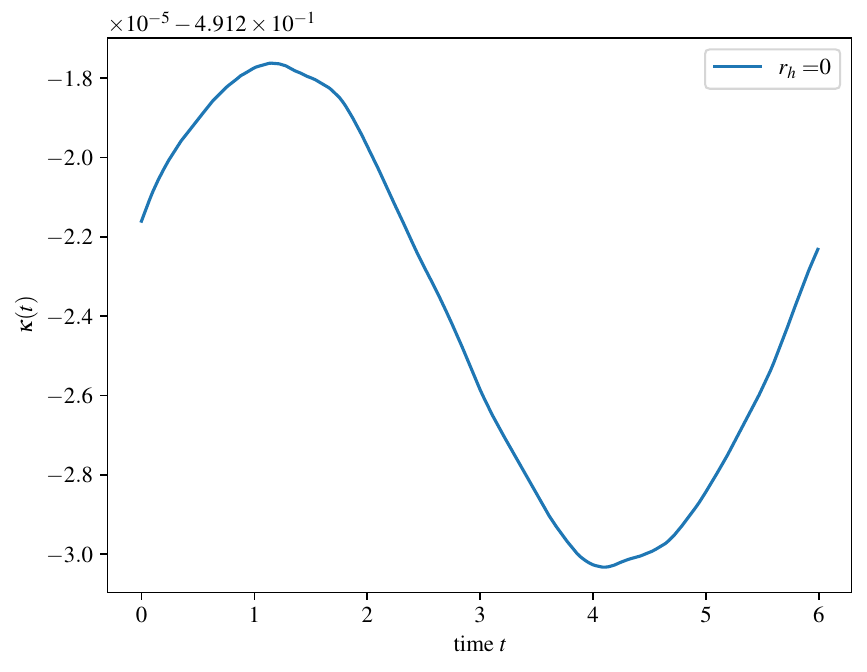}
    \includegraphics[width=0.24\linewidth]{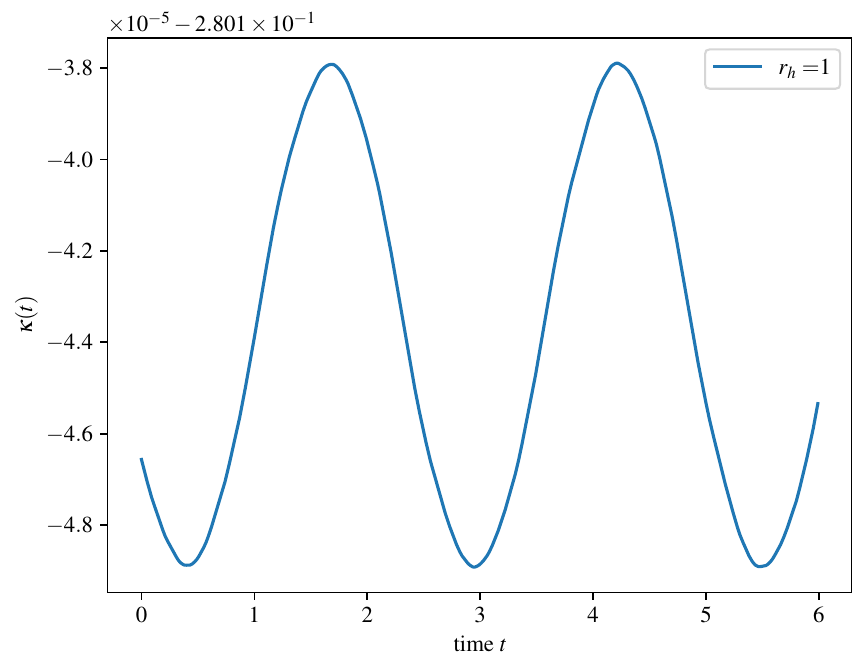}
    \includegraphics[width=0.24\linewidth]{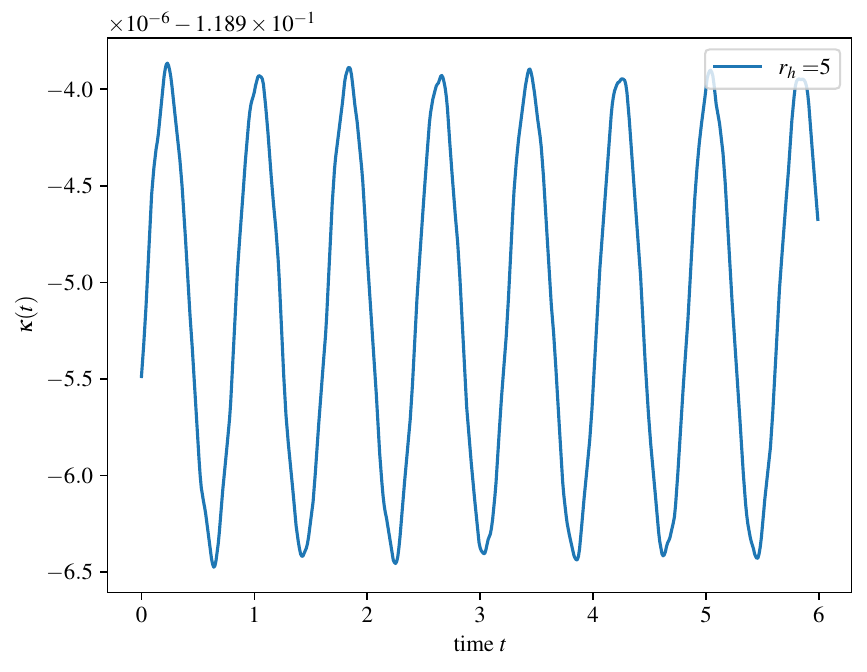}
    \includegraphics[width=0.24\linewidth]{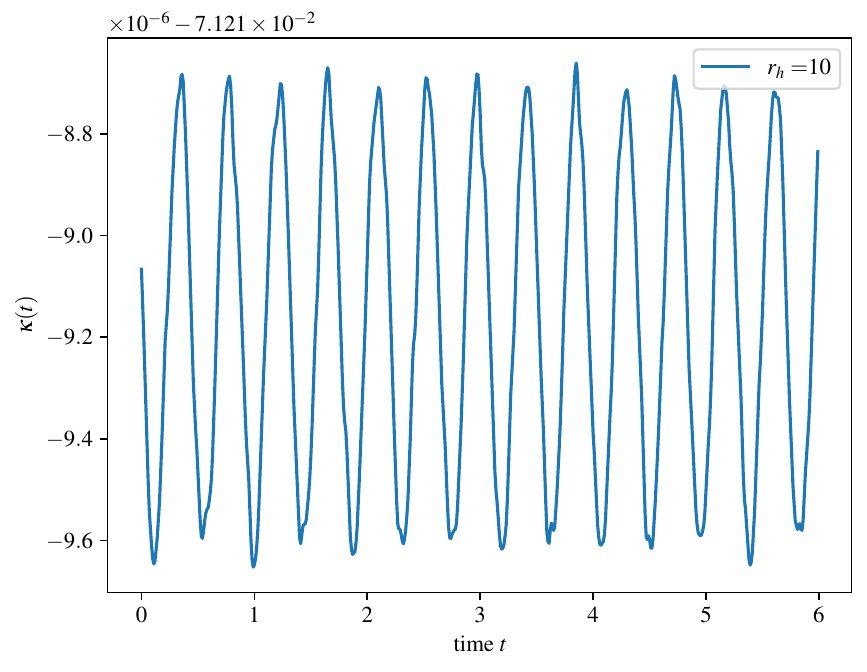}
    \includegraphics[width=0.24\linewidth]{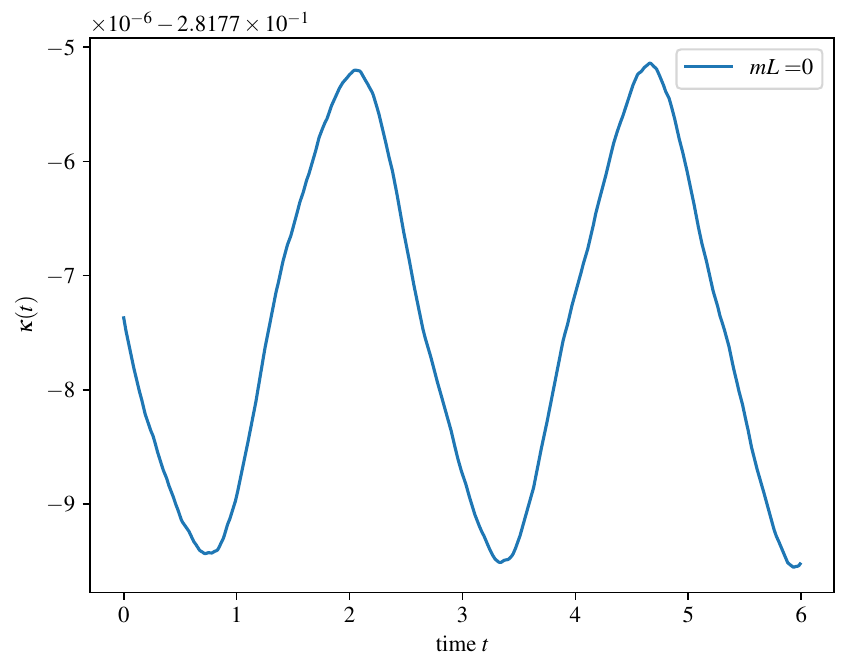}
    \includegraphics[width=0.24\linewidth]{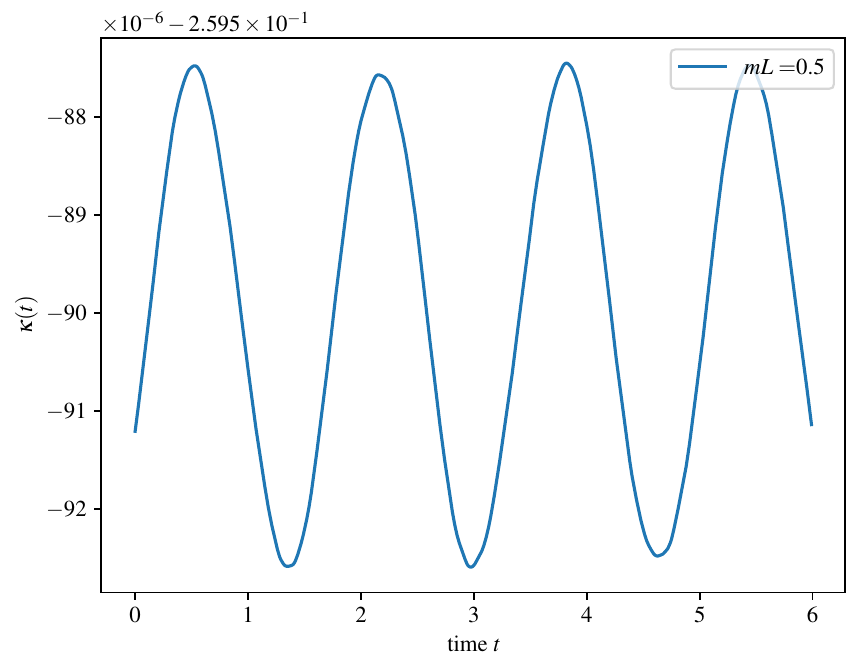}
    \includegraphics[width=0.24\linewidth]{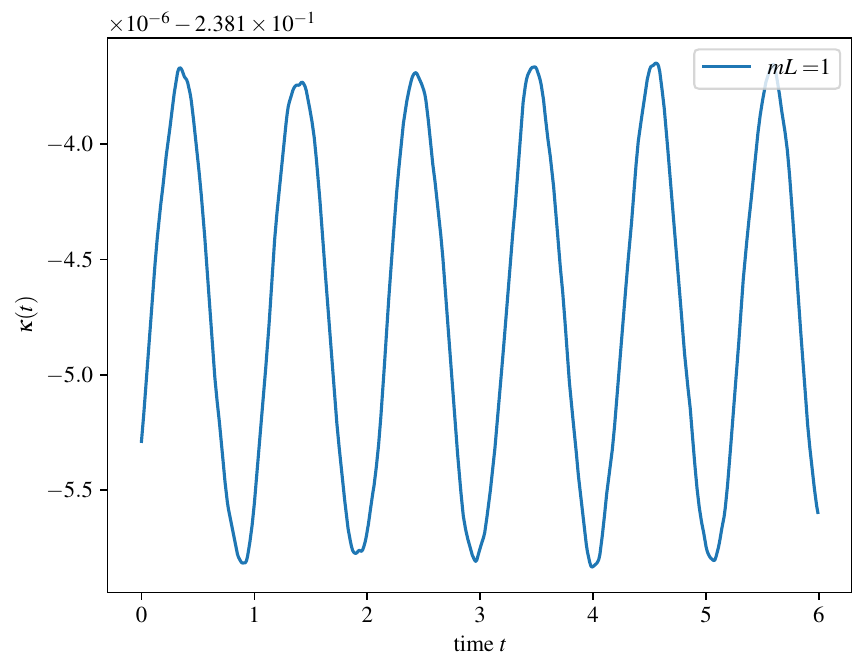}
    \includegraphics[width=0.24\linewidth]{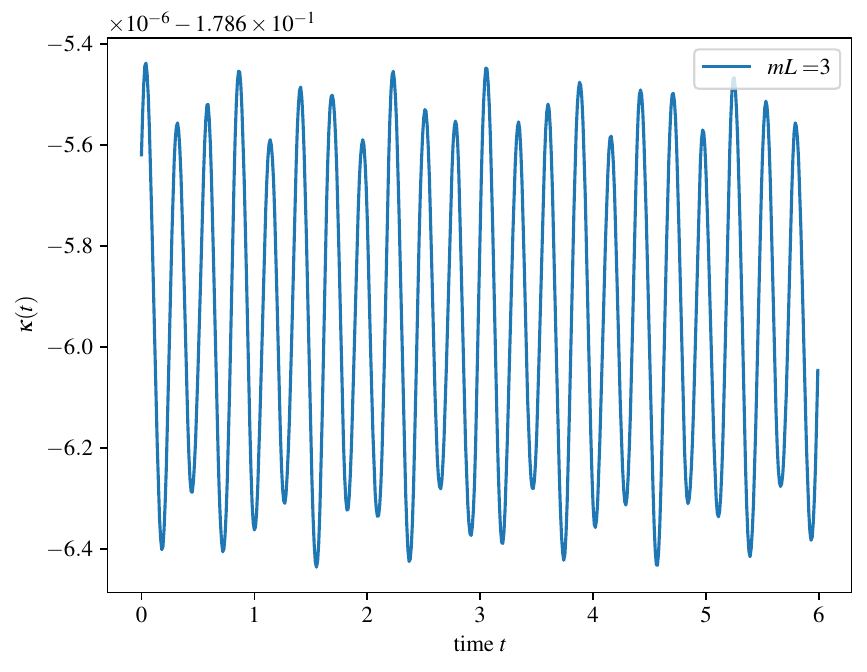}
    \caption{Chiral gravitational effect in time. Time evolution of the vector-chirality current
    $\kappa(t)=\langle J(t)\rangle$ for $N=12$ at $\mu L=0.01, L=1$. Top: increasing horizon size $(r_h \in \{0,1,5,10\};\ m=0.1)$ suppresses the baseline and oscillation envelope via gravitational redshift. Bottom: increasing mass $(m \in \{0,0.5,1,3\};\ r_h=1)$ raises the oscillation frequency (since $[H,J]\!\sim\! m$), highlighting curvature-induced, parity-odd spin currents absent in flat space.}
\label{fig:evolv_current_BH}
\end{figure}

In Fig. \ref{fig:evolv_current_BH}  we plot $\kappa(t)=\langle J(t)\rangle$ as a function of time,
for a chain of $N=12$ qubits at a small chemical potential $\mu L=0.01$, with four choices of horizon radii $r_h \in \{0,\;1,\;5,\;10\}$ (top panel) and four masses $mL \in \{0,\;0.5,\;1,\;3\}$ (bottom panel). The exact conservation is broken for every mass, so $\kappa(t)$ oscillates.
The oscillation frequency grows with $m$, reflecting the increasing commutator $[H,J]\sim m$. (Equivalently, the frequency is proportional to the energy gap, $\omega \sim \Delta$ \cite{Ikeda:2023vfk,Ikeda:2024rzv}, and $\Delta$ increases monotonically as $m$ increases.) For an open chain, the current is not strictly conserved even in flat space. However, when a curved space background is introduced, the violation of current conservation becomes even more pronounced due to the effects of $\alpha_n$. The amplitude of these oscillations also ramps up over time, in accordance with the short-time expansion
\begin{equation}
J(t)=J(0)+it[H,J]-\tfrac{t^2}{2}[H,[H,J]]+\cdots \ .   
\end{equation}
The horizon-radius dependence is seen by comparing the two panels.
The overall scale of $\kappa(t)$, both its constant baseline and oscillation envelope,
shrinks as $r_h$ increases, because the redshift factor (\ref{redshift}) dilutes the strength of the spin-connection–induced chiral current.

In two dimensions there is no chiral-gravitational anomaly.
Here we see a static, zero‐temperature ground‐state current induced by the spatial curvature (the redshift factor).
This effect is in the same family as the chiral vortical effect \cite{PhysRevLett.106.062301,PhysRevD.98.096011} and the gravitational spin Hall effect, where background geometry sources an equilibrium spin current.  
We observe the one–dimensional $AdS_2$ analog: the horizon’s presence, and the associated spin connection, pumps a steady, parity‐odd current around the chain.
It is a chiral gravitational effect in a one‐dimensional lattice setting: a geometric/gravitational chirality, where a ground state spin current is sourced purely by the curvature/red‐shift of the $AdS_2$ black hole, and is absent in flat space.

\begin{figure}[H]
    \centering
    \includegraphics[width=0.24\linewidth]{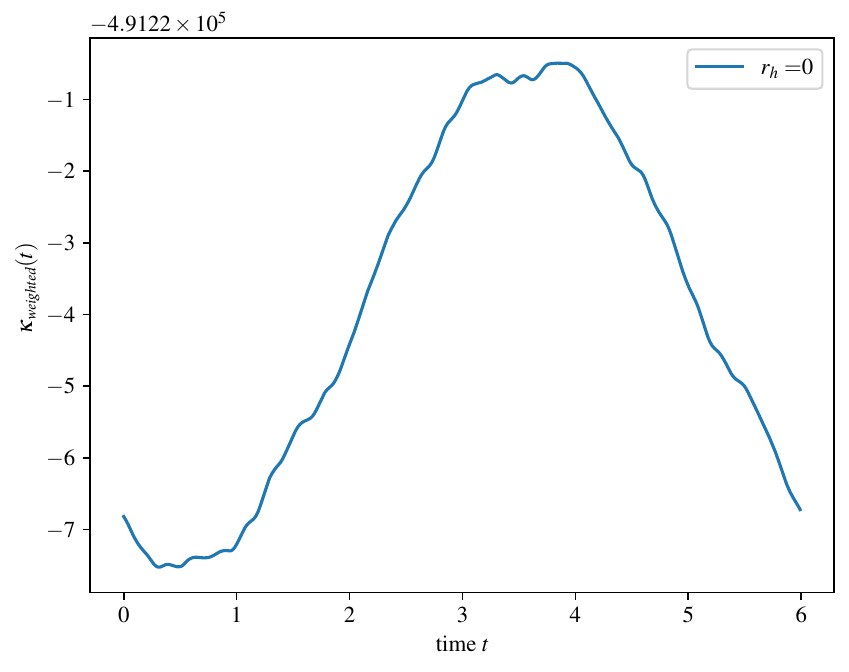}
    \includegraphics[width=0.24\linewidth]{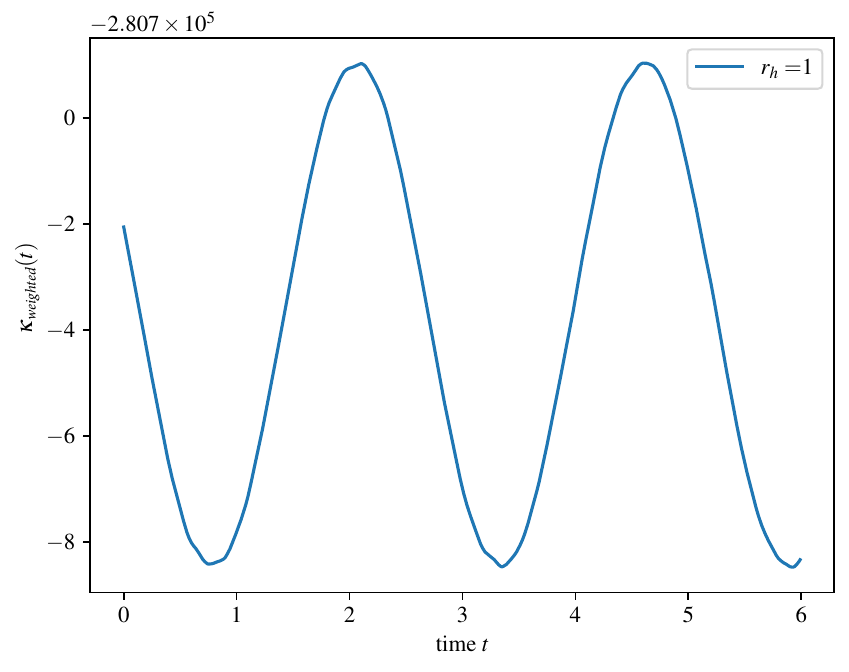}
    \includegraphics[width=0.24\linewidth]{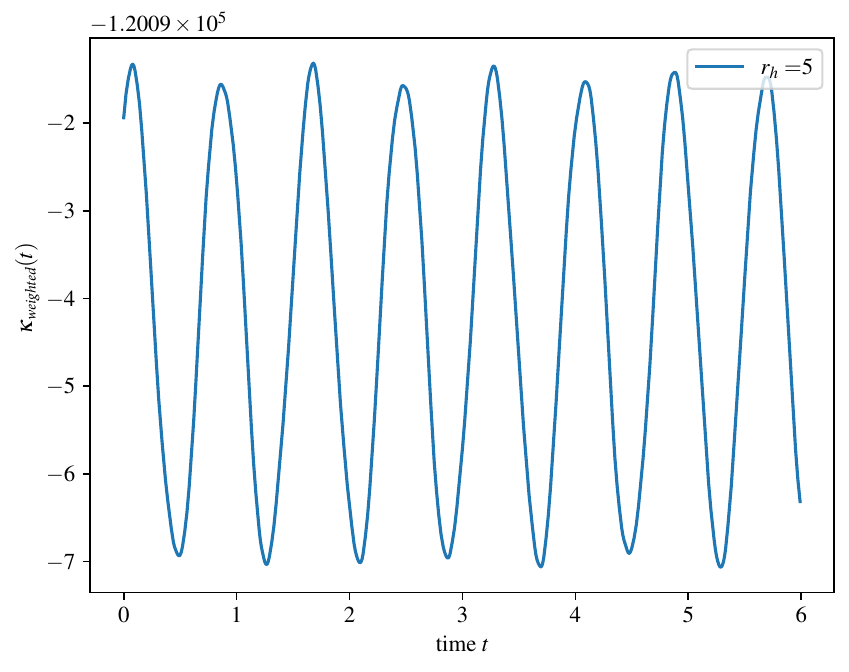}
    \includegraphics[width=0.24\linewidth]{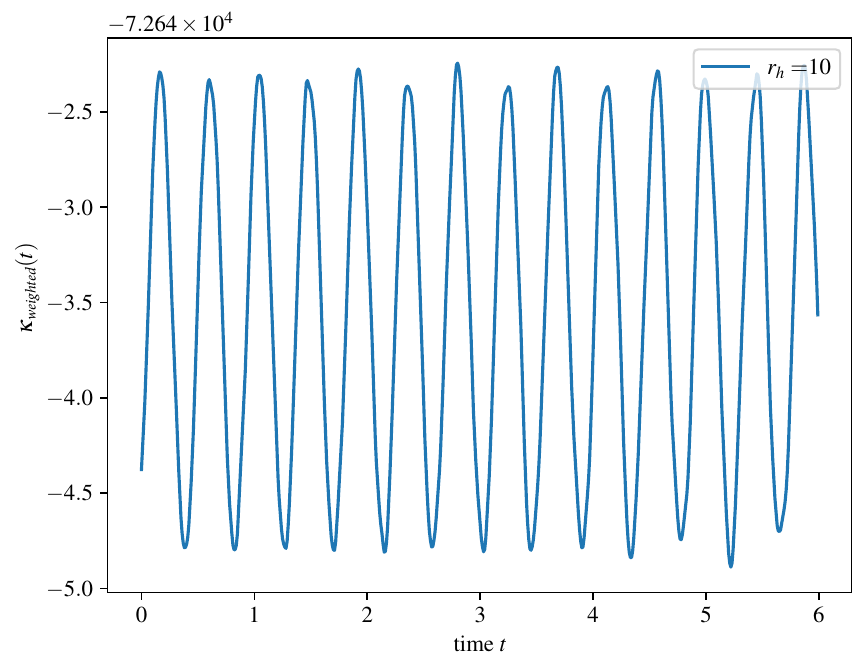}
    \includegraphics[width=0.24\linewidth]{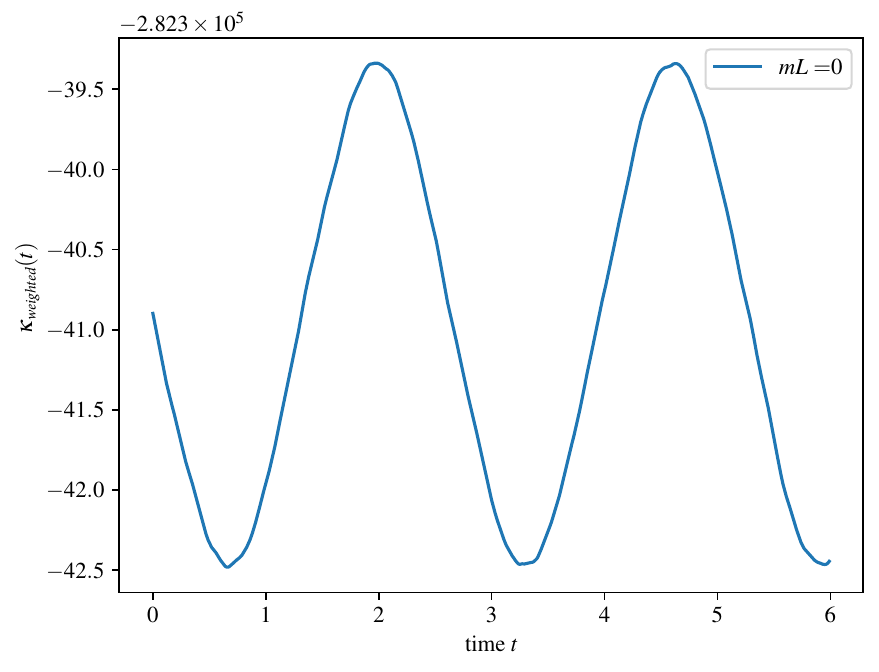}
    \includegraphics[width=0.24\linewidth]{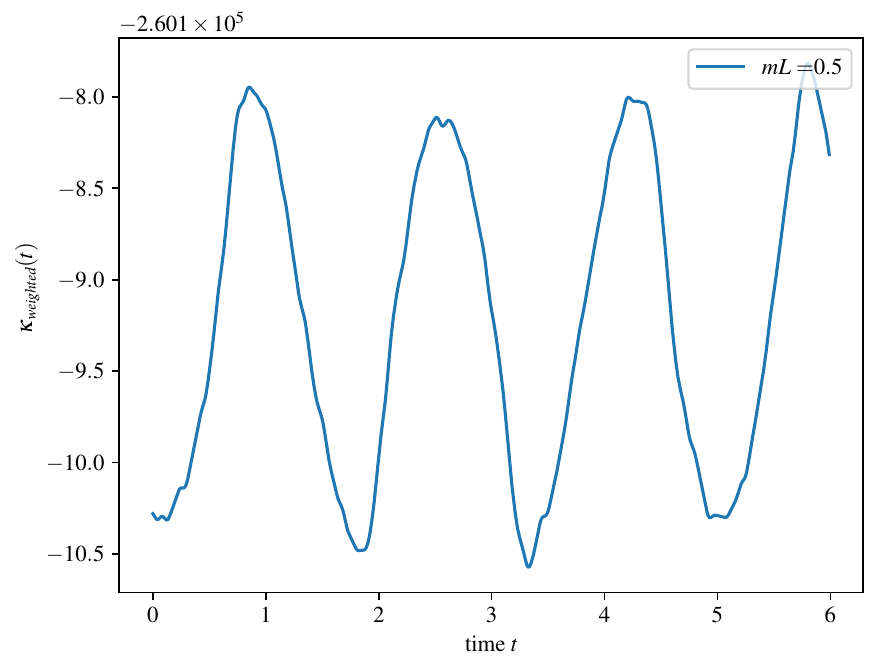}
    \includegraphics[width=0.24\linewidth]{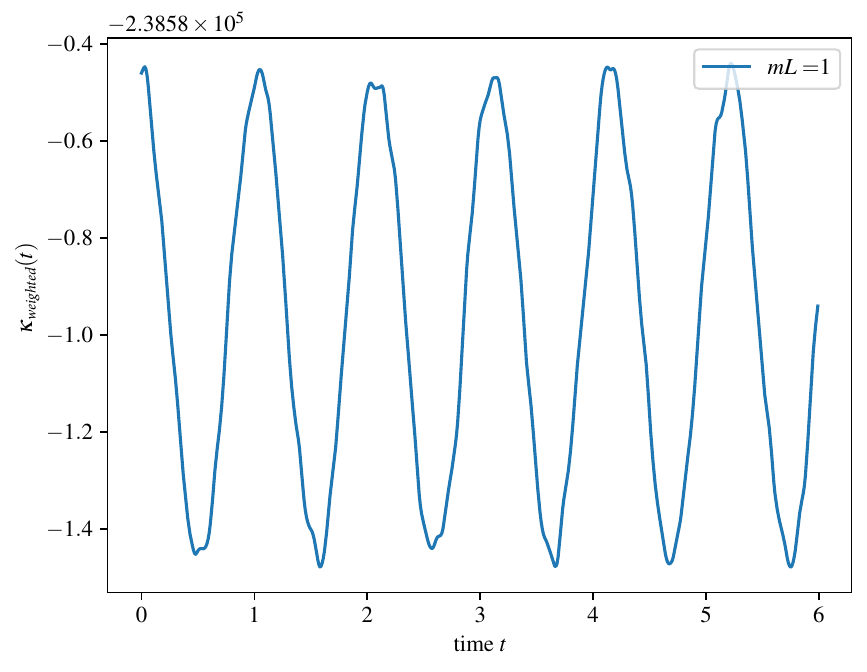}
    \includegraphics[width=0.24\linewidth]{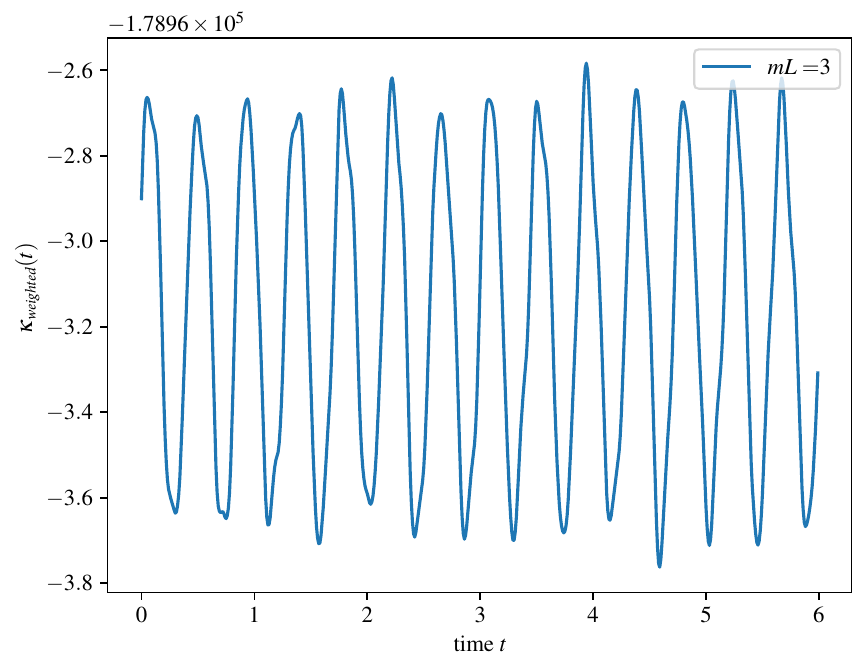}
    \caption{Chiral gravitational effect in time. Real-time evolution of the vector-chirality current
    $\kappa_\text{weighted}(t)=\langle J_\text{weighted}(t)\rangle$ for $N=12$ at $\mu=0.01, L=1$. Top: increasing horizon size $(r_h \in \{0,1,5,10\};\ m=0.1)$ suppresses the baseline and oscillation envelope via gravitational redshift. Bottom: increasing mass $(m \in \{0,0.5,1,3\};\ r_h=1)$ raises the oscillation frequency (since $[H,J]\!\sim\! m$), highlighting curvature-induced, parity-odd spin currents absent in flat space.}
\label{fig:evolv_weighted_current_BH}
\end{figure}

In Fig.~\ref{fig:evolv_weighted_current_BH}, we show the current reflecting the background AdS geometry:
\begin{equation}
J_\text{weighted} \;=\;\sum_{i=1}^{N-1}\alpha^2_i\kappa_i  \ .
\end{equation}
Its time evolution is defined in the same manner as in eq.~\eqref{eq:current_time}. We observe a trend similar to that in Fig.~\ref{fig:evolv_current_BH}. The value becomes large due to the influence of $\alpha_n$.

\subsection{OTOC and Information Scrambling}
\begin{figure}[H]
    \centering
    \includegraphics[width=0.49\linewidth]
    {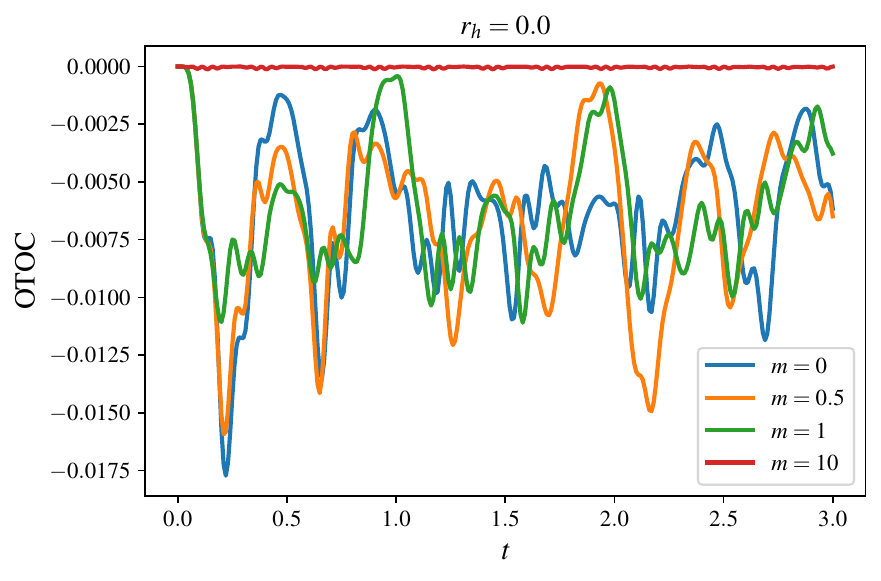}
    \includegraphics[width=0.49\linewidth]
    {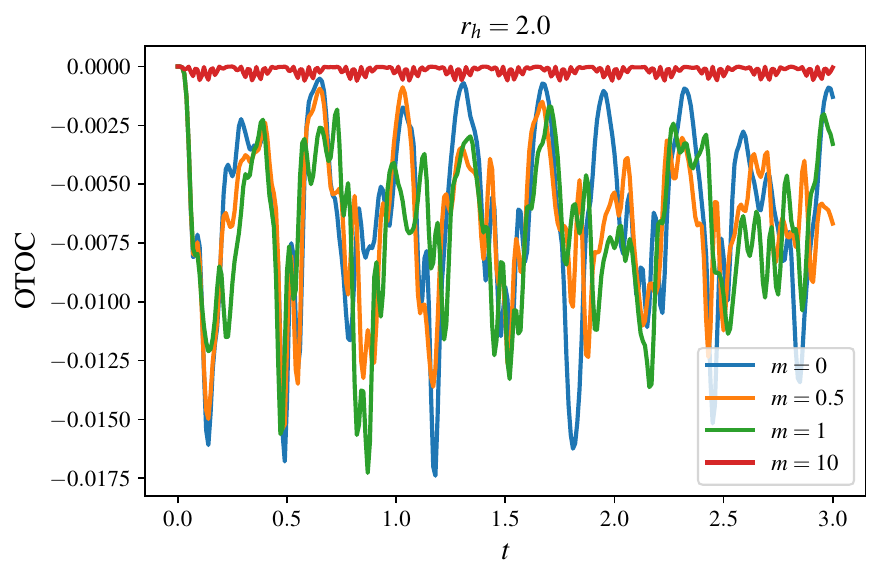}
    \caption{The OTOC (\ref{OTOC})
for a chain of $N=12$ qubits at a small chemical potential ($\mu=0.1,L=1$), using sites 
$(i,j)=(4,8)$ for $C_{ij}(t)$.
In the left panel ($r_h/N = 0$), there is
no horizon (pure $AdS_2$ limit), so scrambling arises solely from the lattice dynamics without gravitational redshift.
In the right panel ($r_h/N = 1/6$), 
the black hole horizon induces a nontrivial spin-connection and enhances the operator growth.}
    \label{fig:OTOC}
\end{figure}

In Fig. \ref{fig:OTOC} we plot the OTOC: 
\begin{eqnarray}
    C(t)_{ij}=\langle \kappa_i(0)\kappa_j(t)\kappa_i(0)\kappa_j(t)\rangle \ , 
    \label{OTOC}
\end{eqnarray}
for a chain of $N=12$ qubits at a small chemical potential ($\mu L=0.1$), using sites 
$(i,j)=(4,8)$ for $C_{ij}(t)$.
In the left panel ($r_h/N = 0$), there is
no horizon (pure $AdS_2$ limit), so scrambling arises solely from the lattice dynamics without gravitational redshift.
In the right panel ($r_h/N = 1/6$), 
a black hole horizon is present at half the chain length, inducing a nontrivial spin-connection and enhancing operator growth.
Each curve (in both panels) corresponds to a different fermion mass $m$ (here $m=0.5,\,1,\,10$), and time $t$ runs along the horizontal axis. The vertical axis shows $C_{ij}(t)$, which starts at zero and can become negative as operators fail to commute at later times—consistent with information scrambling.

In the horizon case ($r_h/N=1/6$), the OTOC decays more rapidly (more negative) than in the pure-$AdS_2$ case, indicating stronger scrambling. Increasing the mass $m$ tends to slow down the decay (less negative OTOC), reflecting that heavier fermions scramble more slowly.
Overall, the figure demonstrates how the presence of a black hole horizon accelerates the spread of quantum operators (information scrambling), with a systematic dependence on the fermion mass.
Note that while we see an enhanced scrambling when a horizon is present on the finite, free-fermion lattice,  the model remains quadratic (integrable), and there is no genuine exponential Lyapunov regime. Instead the OTOC decay is ultimately power-law or oscillatory—scrambling without chaos.
This is the behavior we expect to persist as we send the lattice spacing to zero and $N\to\infty$.  
In the continuum fixed $AdS_2$ black hole background, 
we have free (quadratic) matter on a curved geometry, and while we see operator scrambling, there is no true exponential Lyapunov growth. True chaos is a property of the dynamical $AdS_2$ black hole. 

\if{
that is 
in the presence of a horizon we have a nonzero Lyapunov exponent
\begin{equation}
C(t)\sim 1 - \#\,e^{\lambda_L\,t} + \cdots,
\quad \lambda_L\le \frac{2\pi}{\beta} \ ,    
\end{equation}
which in a true black hole at temperature $T=1/\beta = \frac{r_h}{2\pi\,L^2}\,$ saturates the $\lambda_L=2\pi T$.  On the lattice we already see the enhanced, faster decay of the OTOC when $r_h/N=1/2$.

In the continuum limit that turn-over becomes a clear exponential regime with the expected horizon-set rate.
In the case of pure $AdS_2$ without a horizon, we expect no chaos
and the system is integrable or near-integrable.
Thus, scrambling is much weaker—precisely what you see in the flatter, delayed decay of the OTOC.  In the continuum pure $AdS_2$ there’s no exponential growth at all, just power-law or oscillatory behavior.
Heavier fermions carry less thermal energy into the scrambling dynamics, so increasing $m$ slows down the onset of operator growth.  That too matches continuum intuition: large conformal dimensions suppress the size-expansion rate of operators in the black hole background.
Thus, the lattice calculation is pointing in the right direction: as we will refine the lattice and 
send $N\to\infty$, we expect that the horizon curves will settle into a clean exponential decay with $\lambda_L\approx 2\pi T$, while the pure $AdS_2$ curves will remain non-chaotic.

}\fi

\if{

As a consequence of chiral symmetry breaking induced by the spin connection, we can discuss operator growth in the AdS background. In the previous subsection, we examined the real-time evolution of the current. Here, we focus on the growth of local terms. To this end, we consider the out-of-time-order correlation (OTOC) function:
\begin{eqnarray}
    C(t)_{ij}=\langle \kappa_i(0)\kappa_j(t)\kappa_i(0)\kappa_j(t)\rangle. 
    \label{OTOC}
\end{eqnarray}

}\fi

\subsection{$r$-Statistics}
To further investigate the ``chaotic'' properties of the model we analyze the level statistics \cite{OganesyanHuse2007}, and 
examine whether they behave more like an integrable (Poisson) or a chaotic (Wigner–Dyson) ensemble.
Specifically, we consider the following statistic for each charge sector $q$:
\begin{equation}
    \langle r\rangle_q=\frac{1}{M_q}\sum_{i=1}^{M_q}r^{(q)}_i,\quad r^{(q)}_i=\frac{\min (s_i,s_{i-1})}{\max (s_i,s_{i-1})} \ ,
\end{equation}
where $M_q$ denotes the number of samples in the charge sector $q$ designated by the flat charge operator \eqref{flat}, and $s_i = \epsilon_{i+1} - \epsilon_i$ represents the level spacing. The unfolded energy levels are given by $\epsilon_i = \bar{N}(E_i)$, where $\bar{N}(E_i)$ is obtained via polynomial fitting to the energy distribution $\{E_i\}$ following exact diagonalization.
We will study the average $r$-statistics, which is defined as:
\begin{equation}
    \langle r\rangle=\frac{\sum_{q}M_q\langle r\rangle_q}{\sum_qM_q} \ .
\end{equation}

In our $AdS_2$ black hole discretization, the Hamiltonian is multiplied by position-dependent weight $\alpha_n$. Near the horizon $\alpha_n$ is small, and it increases toward the boundary. This spatial variation is a redshift gradient. When we increase $r_h$, while
keeping the system size/scale fixed, the contrast between the horizon region and the far region grows and we have a steeper $\alpha_n$ profile.
While a uniform quadratic chain is diagonal in plane waves $|k\rangle$, the spatially varying factor $\alpha_n$  acts like an inhomogeneous potential/coupling, the momentum is no longer a good quantum number and eigenstates become hybrids of many $k$’s, which leads to stronger level repulsion in the spectrum.
More mixing implies that spacings repel more, and the average adjacent-gap ratio rises above $\langle r \rangle \sim 0.386$ (Poisson distribution). However, because the model is still quadratic, it does
not reach fully chaotic values $\langle r \rangle \sim 0.5307$ (Wigner distribution).

Consider the dependence of $\langle r\rangle$ on the model parameters.
Increasing $\frac{r_h}{L}$ strengthens the redshift gradient and the modes mixing and increases $\langle r\rangle$, still below GOE value.
As the mass $|m|\!\to\!0$, the eigenspectrum has the structure $\{\pm E_i\}$  (particle-hole symmetry), which produces pairs of levels clustered symmetrically around zero, and can generate quasi zero modes of very small $|E|$. These pairings and near-zeros create a large number of very small spacings in the ordered list $\{\pm E_i\}$, which reduces $\langle r\rangle$ near, or even below, Poisson.
For intermediate $|m|$ with $|m|L\sim\mathcal O(1)$ and $|m|r_h\sim\mathcal O(1)$, the
degeneracies are lifted and states hybridize most, hence we expect a peak in $\langle r\rangle$.
When $|m|L\gg 1$ and $|m|r_h \gg 1$, we have two weakly-mixed bands at $\pm m$ leading to a Poisson-like behaviour.
At $\mu=0$, particle–hole pairings reduce $\langle r\rangle$. A small nonzero $|\mu|$ breaks these pairings and raises $\langle r\rangle$. Very large $|\mu|$ tends to reduce it again.

To further investigate this, we also examine the Brody distribution, defined as
\begin{equation}
    P_\beta(s)=(\beta+1)bs^\beta\exp\left(-bs^{\beta+1}\right),\quad b=\left[\Gamma\left(\frac{\beta+2}{\beta+1}\right)\right]^{\beta+1}\;.
\end{equation}
Here, $\Gamma$ is the gamma function and $\beta$ ranges from 0 to 1: $\beta = 0$ corresponds to the Poisson distribution, while $\beta = 1$ approaches the Wigner distribution. Theoretically, the relation between $\beta$ and $\langle r\rangle$ can be approximated as $\langle r\rangle\approx 0.39+0.26\beta$. The fitted parameter $\beta\in[0,1]$ increases with $r_h$, showing a continuous crossover from Poisson 
toward Wigner–Dyson.
\if{
\begin{figure}[H]
    \centering
    \includegraphics[width=0.32\linewidth]{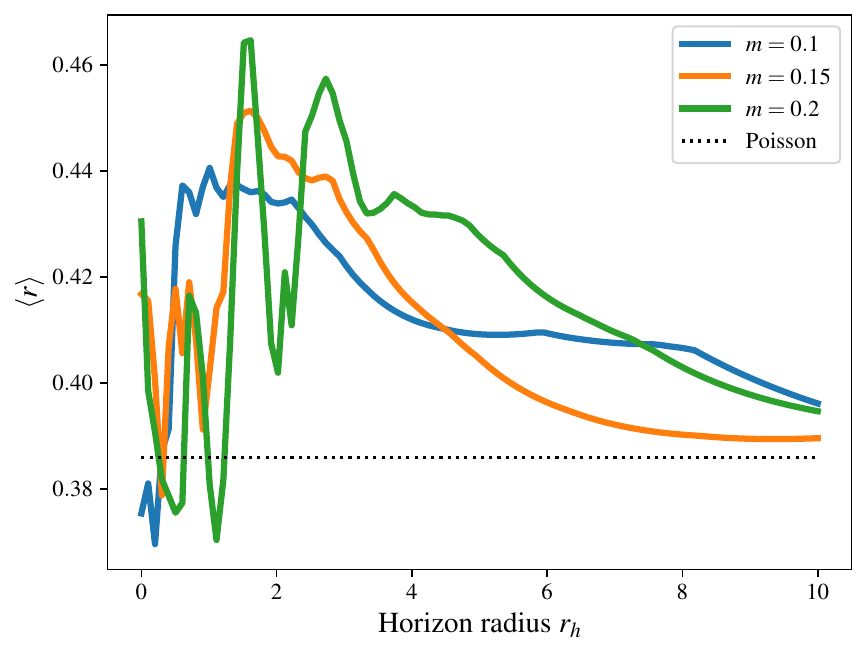}
    \includegraphics[width=0.32\linewidth]{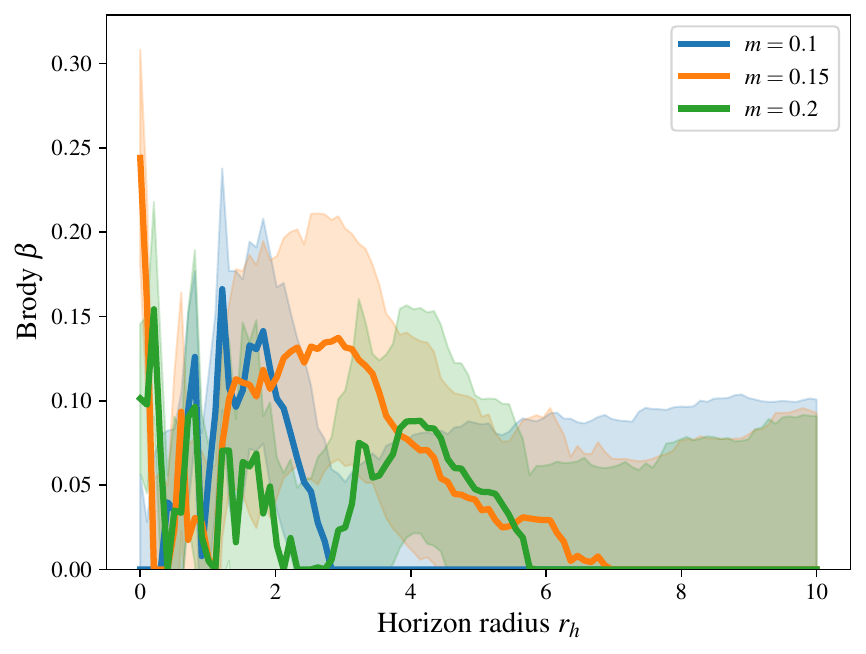}
    \includegraphics[width=0.32\linewidth]{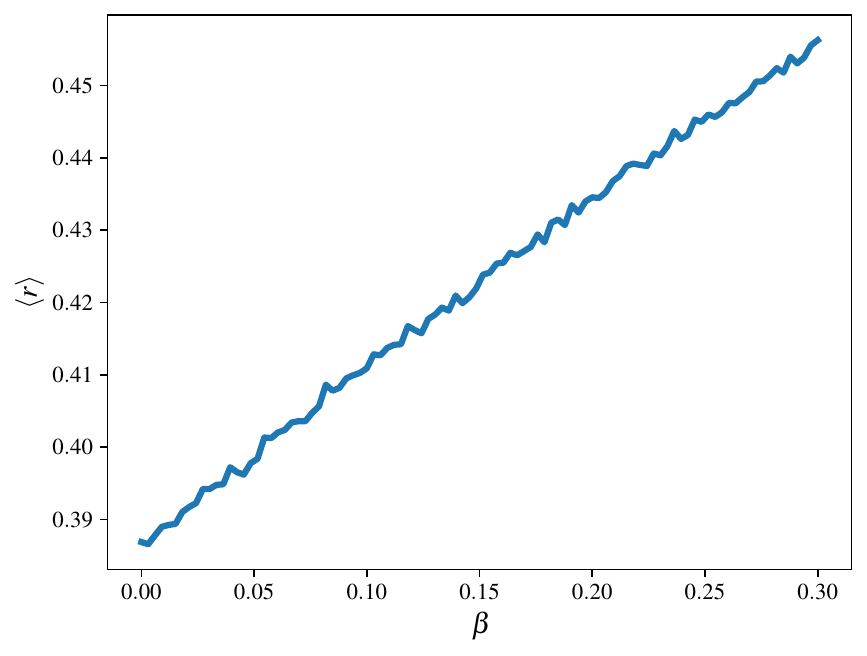}
    \caption{The average $r$-statistics (left), the Brody $\beta$ (middle), $\langle r\rangle$ vs $\beta$ (right). For the left and middle panels, $N=10,\mu L=0.1$ are used. In the middle, the shadows illustrate fitting errors.\ki{to be updated}}
    \label{fig:Brody_distribution}
\end{figure}
}\fi

\begin{figure}
    \centering
    \includegraphics[width=0.32\linewidth]{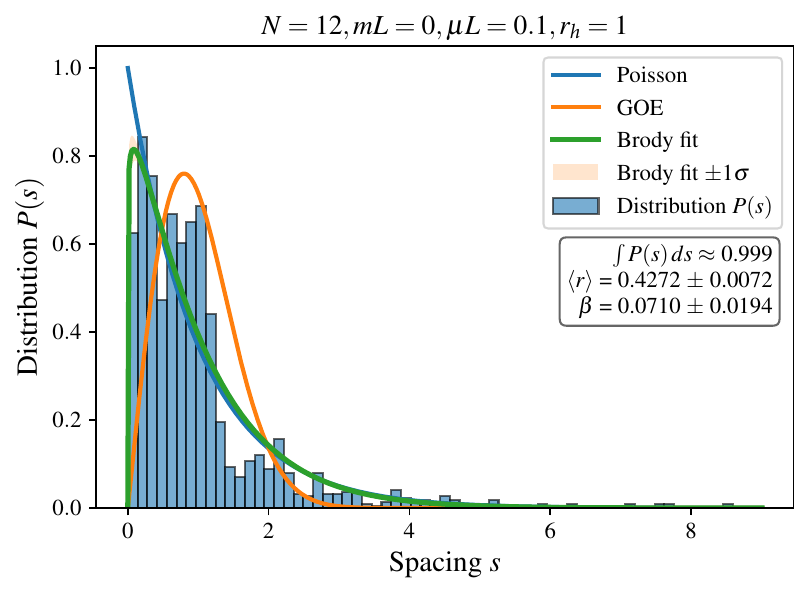}
    \includegraphics[width=0.32\linewidth]{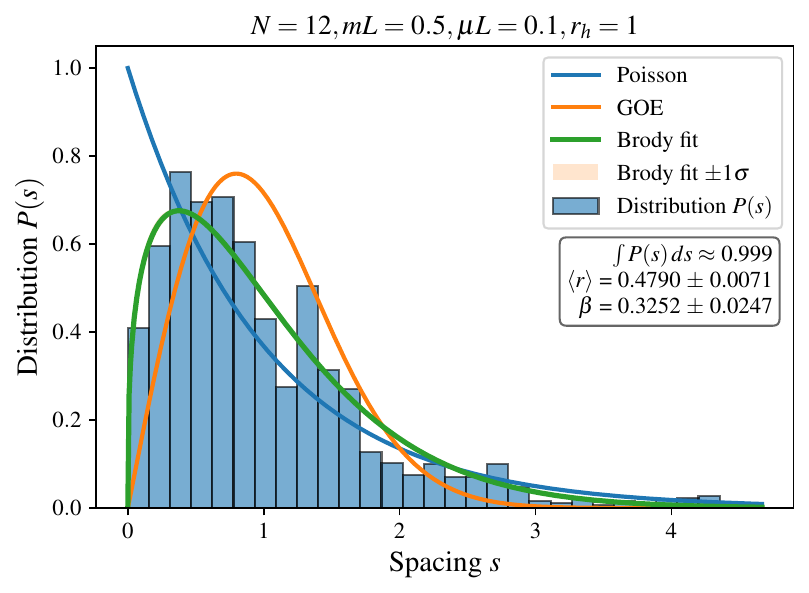}
    \includegraphics[width=0.32\linewidth]{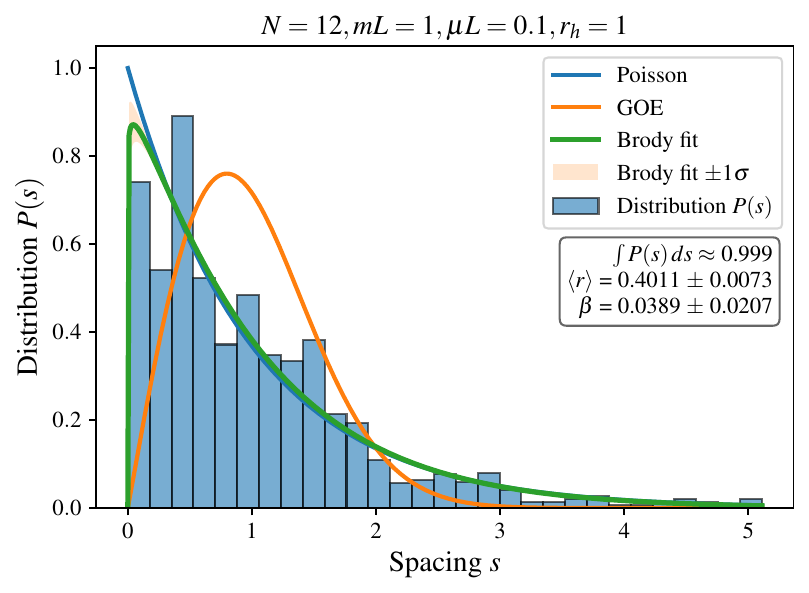}
    \includegraphics[width=0.32\linewidth]{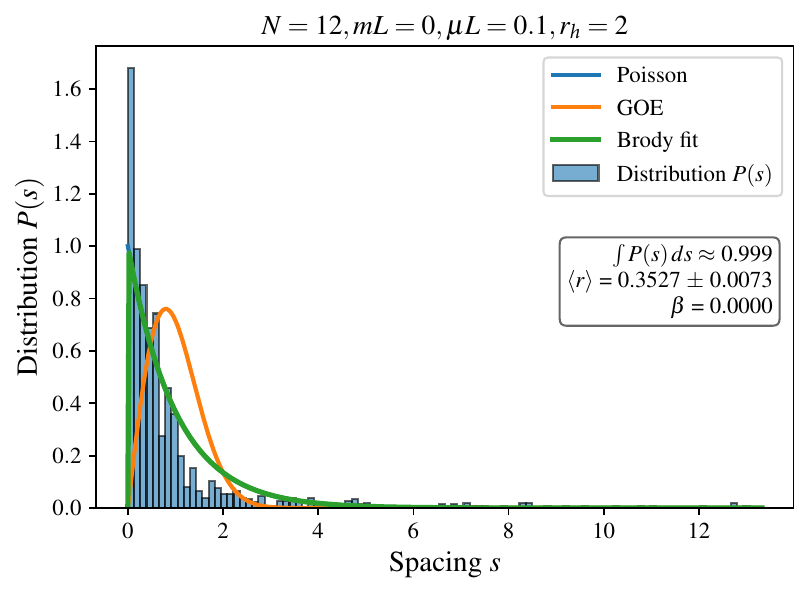}
    \includegraphics[width=0.32\linewidth]{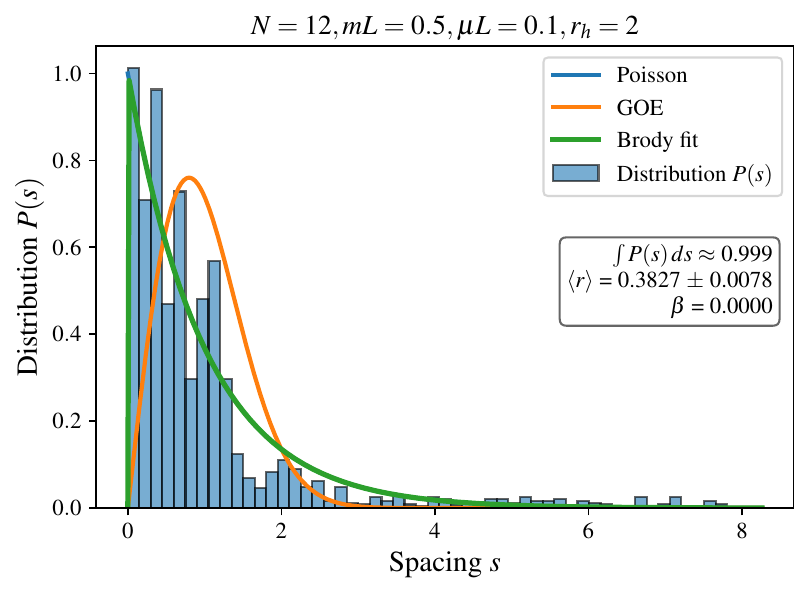}\includegraphics[width=0.32\linewidth]{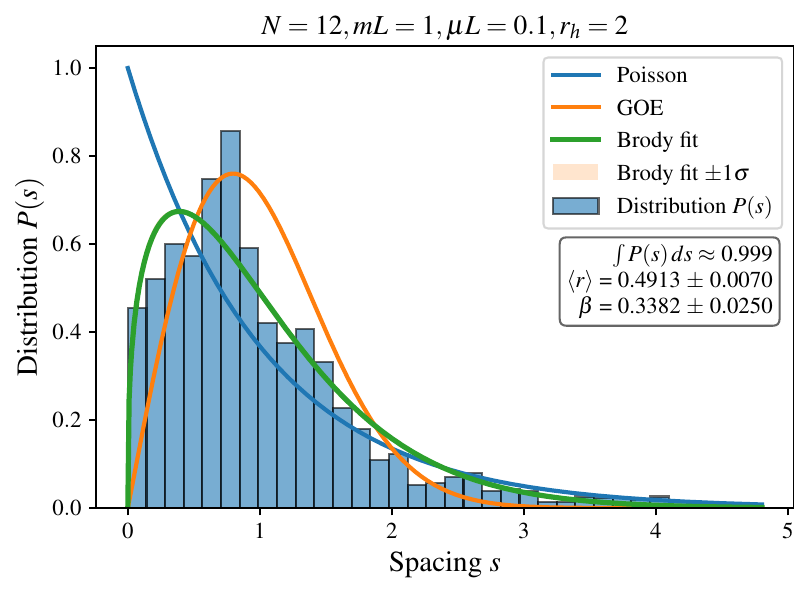}
    \caption{The level spacing distributions. $\mu=0.1, L=1, N=12$ are common parameters. From left to right: $m \in \{0,0.5,1\}$.}
    \label{fig:r_beta}
\end{figure}
Fig.~\ref{fig:r_beta} presents the distributions of level spacing $s$ for $mL=0,0.5,1$ and $r_h=1,2$, alongside benchmark comparisons with the Poisson and Gaussian Orthogonal Ensemble (GOE) cases. All panels show properly normalized histograms with $\int_0^\infty P(s)\,ds \approx 0.999$ for $N=12$ and $\mu L=0.1$. The top row corresponds to $r_h=1$ with $mL\in\{0,\,0.5,\,1\}$; the bottom row to $r_h=2$ with the same $mL$ values. None of the cases reaches the GOE benchmark $\langle r\rangle_{\mathrm{GOE}}\approx 0.5307$; the spectra remain in the Poisson $\leftrightarrow$ chaotic crossover.

For $r_h=1$, the spectra are near–Poisson at $mL=0$ and $mL=1$ (tall first bin, Brody $\beta\simeq 0$ and $\langle r\rangle\approx 0.41$ and $0.40$ respectively), while $mL=0.5$ shows the clearest level repulsion in this row with $\langle r\rangle\approx 0.479$ and $\beta\approx 0.33$. For $r_h=2$, the $mL=0$ case exhibits a very strong spike near $s=0$ (effective degeneracies) with $\langle r\rangle\approx 0.353$ and $\beta=0$, the $mL=0.5$ case is again close to Poisson with $\langle r\rangle\approx 0.383$, and the most chaotic spectrum among the six appears at $mL=1$ with $\langle r\rangle\approx 0.491$ and $\beta\approx 0.34$—still below GOE.

The prominently tall first bin in several panels is expected near integrable limits and reflects a discrete component at zero spacing. Mathematically,
\[
P(s)=p_0\,\delta(s)+(1-p_0)\,P_{\mathrm{cont}}(s),
\]
so with a finite first bin of width $\Delta s$ the bar height is $\simeq p_0/\Delta s+(1-p_0)\,\overline{P}_{\mathrm{cont}}(0)$, which can exceed $1$ even though the total area remains unity. This discrete spike drives the Brody fit toward $\beta\simeq 0$ while the ratio statistic $\langle r\rangle$ often stays above the Poisson value $2\ln2-1\approx 0.386$, explaining the mild mismatch between $\beta$ and $\langle r\rangle$ in near–integrable cases.

\subsection{Ergodic to Many-Body Localization Crossover}
We consider a transition from an ergodic phase to a many‑body localized (MBL) phase, by introducing a local disorder term to the Hamiltonian:
\begin{equation}
\label{eq:disorder}
    H_\text{disorder}=\sum_{n=1}\frac{h_nZ_n}{2}\;,
\end{equation}
where $h_n$ obeys a uniform random distribution in the range $[-W, W]$ with $W>0$. A large $W$ corresponds to strong disorder. This term affects only the diagonal elements of the Hamiltonian matrix. We consider the time-evolution of the imbalance
\begin{equation}
\label{eq:CC_flat}
    \mathcal{I}=\frac{1}{2N}\sum_{n=1}^N(-1)^nZ_n\;,
\end{equation}
which corresponds to the chiral condensate density $\overline{\psi}\psi_\text{flat}/N$ in the flat ground (see Table~\ref{tab:dic}). 

\begin{figure}[H]
    \centering
    \includegraphics[width=0.32\linewidth]{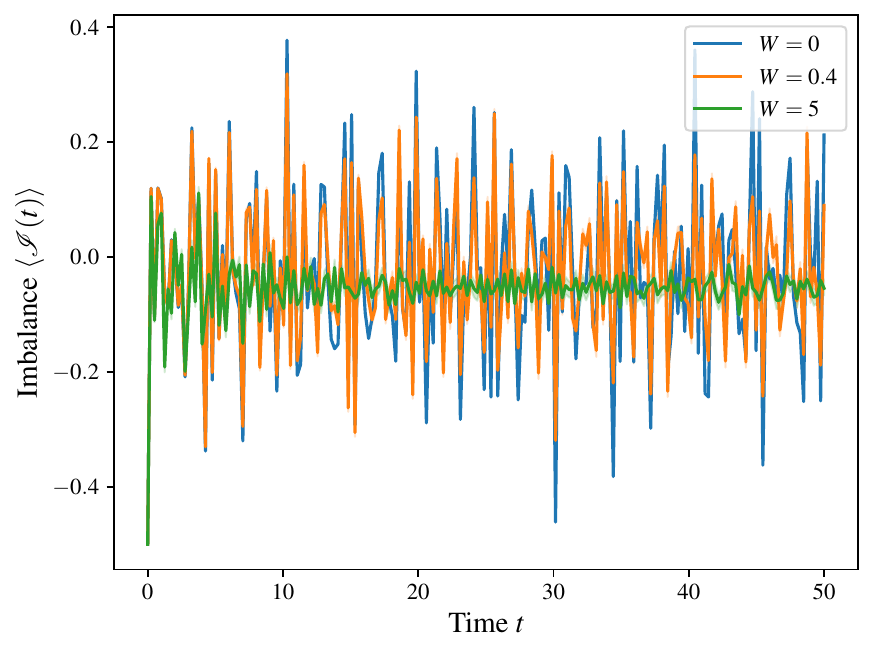}
    \includegraphics[width=0.32\linewidth]{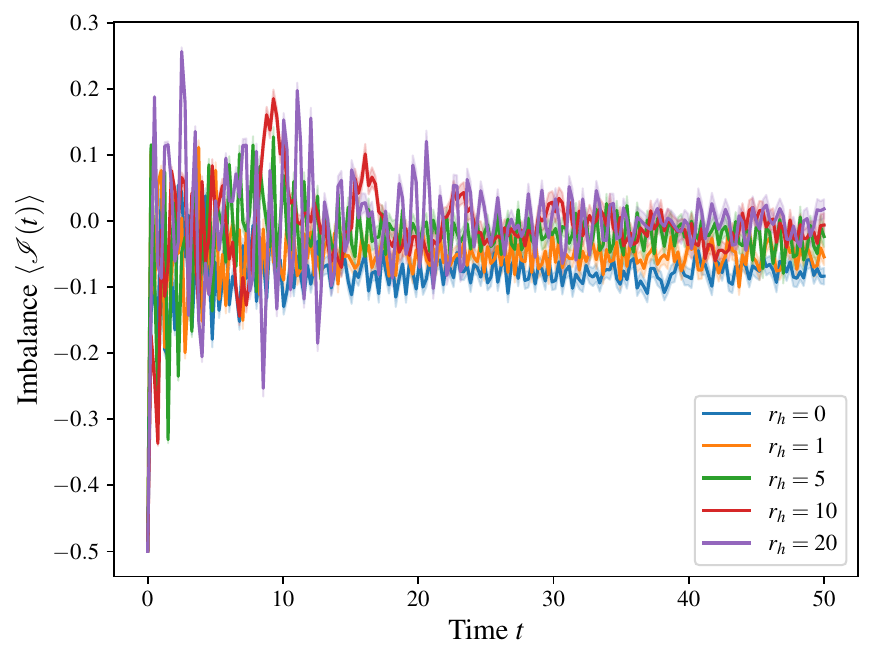}
    \includegraphics[width=0.32\linewidth]{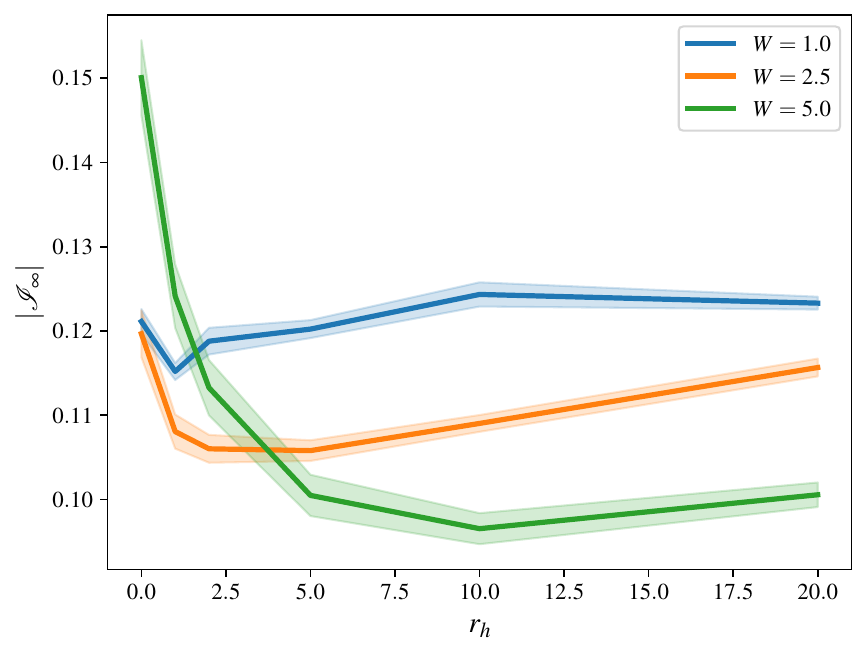}
    \caption{Quench dynamics of the sampling-average $\langle \mathcal{I}(t)\rangle=\frac{1}{N_\text{samp}}\sum_{k}^{N_\text{samp}}\mathcal{I}_k(t)$ of imbalance, with eq.~\eqref{eq:CC_flat} and disorder term is given by eq. \eqref{eq:disorder}. For left: $m =0.25, 
\mu = 0.1, r_h = 1$ and for right: $m=0.25,\mu=0.1, W=5$. The other parameters are commonly set to $N = 10, L= 1, N_\text{samp}=100$. The mean values are presented as solid lines, and the standard deviations are shown as shadows.}
    \label{fig:MBL}
\end{figure}

We analyze the quench dynamics of $\mathcal{I}$, starting from the Néel state $\ket{0101\cdots01}$ as the initial condition. With this choice, the initial value of $\mathcal{I}$ is $-\tfrac{1}{2}$, regardless of $r_h$. In Fig.~\ref{fig:MBL} (left), we present the sampling average of $\mathcal{I}(t)$ with various disorders $W$. Without disorder ($W=0$), the model exhibit a rapid oscillation around 0 with a large magnitude of amplitude, indicating the integrability of the system. For weak disorder ($W=0.4$), the imbalance decays quickly to zero and thereafter only small fluctuations around zero remain.
This behavior is characteristic of the ergodic (thermalizing) phase, where the initial Néel pattern is completely washed out. For strong disorder ($W=5$), the imbalance never reaches zero but instead settles into a non‑zero “frozen” plateau at late times. This is the hallmark of the MBL  phase, where local memory of the Néel order is preserved indefinitely. 

In Fig.~\ref{fig:MBL} (right), we show the $r_h$-dependence in the MBL phase with $mL=0.25,\mu L=0.1, W=5$. We see that regardless of $r_h$, it converges into almost the same value $\sim-0.3$. One can also confirm a similar behavior with different $mL$. 

\begin{figure}[H]
    \centering
    \includegraphics[width=0.32\linewidth]{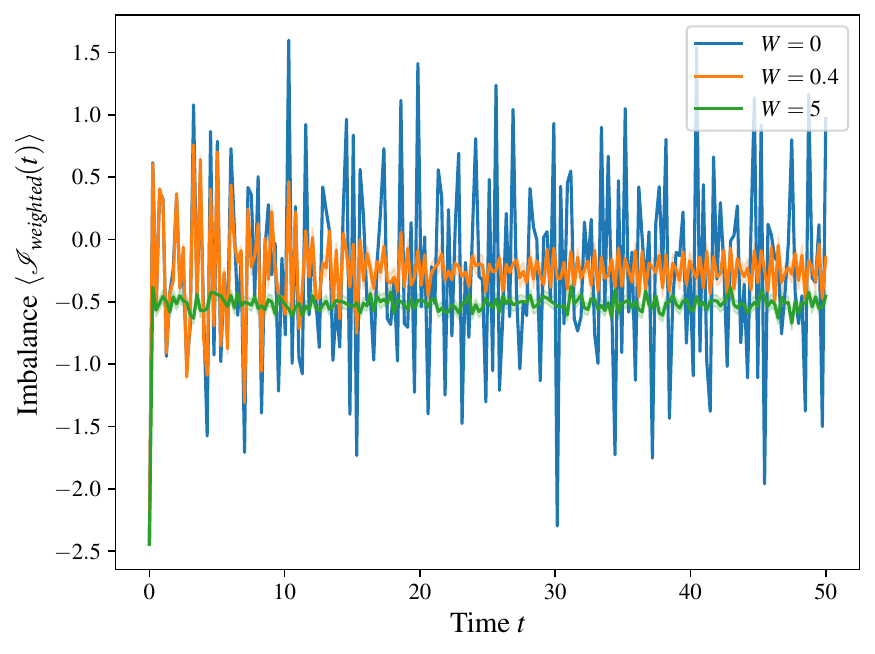}
    \includegraphics[width=0.32\linewidth]{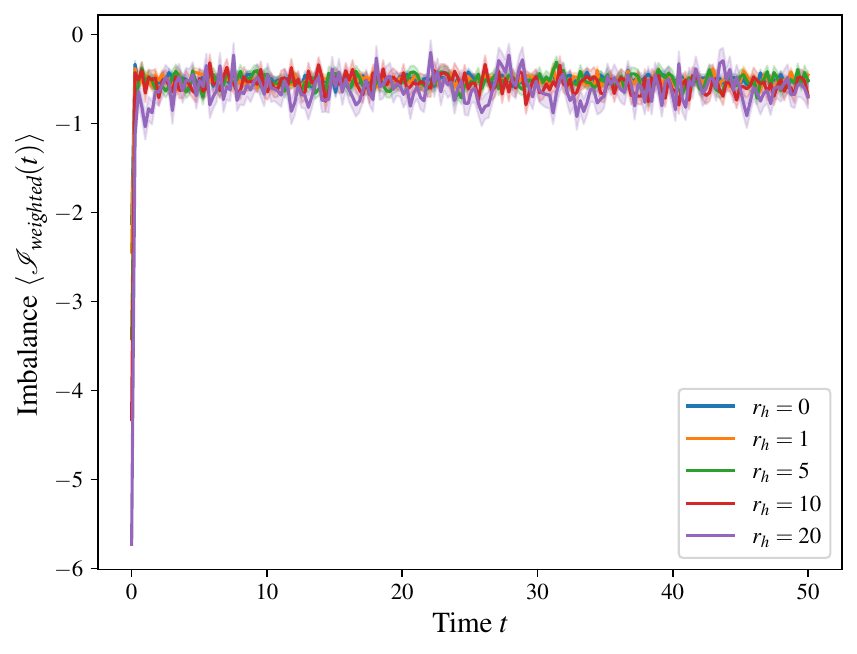}
    \includegraphics[width=0.32\linewidth]{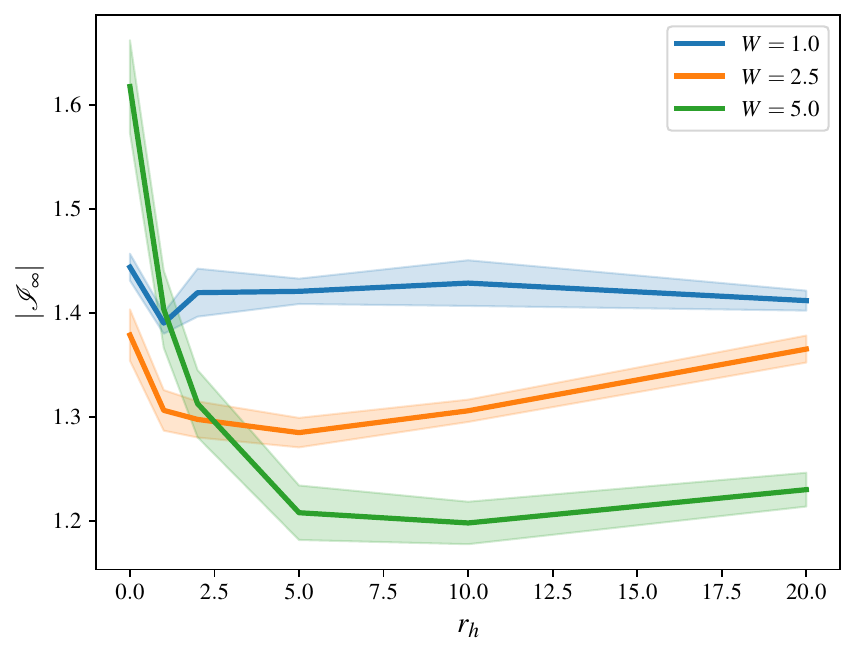}
    \caption{Quench dynamics of the sampling-average $\langle \mathcal{I}_\text{weighted}(t)\rangle=\frac{1}{N_\text{samp}}\sum_{k}^{N_\text{samp}}\mathcal{I}_k(t)$ of imbalance, with eq.~\eqref{eq:CC_AdS} and the disorder term \eqref{eq:disorder_phys}. For left: $r_h = 1$, for middle: $W=5$. $m = 0.25, 
\mu  = 0.1$ are used for all plots. The other parameters are commonly set to $N = 10, L=1, N_\text{samp}=100$. The mean values are presented as solid lines, and the standard deviations are shown as shadows.}
    \label{fig:MBL_phys}
\end{figure}

In Fig.~\ref{fig:MBL_phys} (left), we present the sampling-average  dynamics of the physical chiral condensate density $\mathcal{I}_\text{weighted}=\overline{\psi}\psi_\text{weighted}/N$ with weight of spin-connection $w_n$, reflecting the curved space background:
\begin{equation}
\label{eq:CC_AdS}
    \mathcal{I}_\text{weighted}=\frac{1}{2N}\sum_{n=1}^N(-1)^n\alpha_nZ_n\;. 
\end{equation}
We also modify the disorder term \eqref{eq:disorder} by reflecting the curved space geometry as
\begin{equation}
\label{eq:disorder_phys}
    H_\text{disorder}=\sum_{n=1}\frac{\alpha_nh_nZ_n}{2}\;.
\end{equation}
Without disorder, the system oscillates around zero and does not form a plateau, as previously confirmed -- this is a sign of an integrable regime. With weak disorder, $W=0.4$, the system effectively thermalizes to its microcanonical expectation, marking the ergodic regime where disorder is sufficient to break integrability and induce thermalization. For strong disorder, $W=5$, a new non-zero plateau emerges below -0.1, indicating the MBL phase: sufficiently large diagonal randomness localizes the system and preserves a significant memory of the initial Néel pattern.

Unlike Fig.~\ref{fig:MBL} (right), Fig.~\ref{fig:MBL_phys} (right) shows a clear $r_h$ dependence because $I_\text{weighted}$ weights each site by $\alpha_n$, where geometry changes those weights and hence both the initial value and the late-time plateau. $I_\text{flat}$ has no such weights, so its MBL plateau is nearly $r_h$-independent.
Fig.~\ref{fig:MBL_phys} (right) shows the disorder–averaged frozen memory $|\mathcal{I}_\infty|$:
\begin{equation}
    \mathcal{I}_\infty=\frac{1}{T_2-T_1}\int_{T_1}^{T_2}\mathcal{I}_\text{weighted}(t)dt
\end{equation}
as a function of the geometric parameter $r_h$ for three disorder strengths $W\in\{1.0,\,2.5,\,5.0\}$. For each data point, $\mathcal{I}_\infty$ is defined as the time–average of the imbalance over the last portion of the simulation window (tail fraction), then averaged over disorder realizations; shaded bands denote the standard error of the mean. For the plot, $T_2$ is last time point, and $T_1$ is defined in a way that we average over the last 40\% of the simulation. Across all $W$, $|\mathcal{I}_\infty|$ drops rapidly as $r_h$ increases from small values and then saturates to a small but nonzero baseline at large $r_h$, indicating that geometry weakens the memory of the initial Néel pattern. At small $r_h$ the curves are ordered by disorder, with stronger $W$ producing a larger frozen value (stronger localization); at large $r_h$ the three curves nearly coalesce. A slight upturn of the weaker-disorder curves at the largest $r_h$ is within the uncertainty band and is consistent with finite–size/time effects together with plotting the magnitude $|\mathcal{I}|$ (which leaves a small positive offset when the signed plateau fluctuates around zero).

\if{
\subsection{Tensor network and Large $N$}

\begin{figure}[H]
    \centering
    \includegraphics[width=0.32\linewidth]{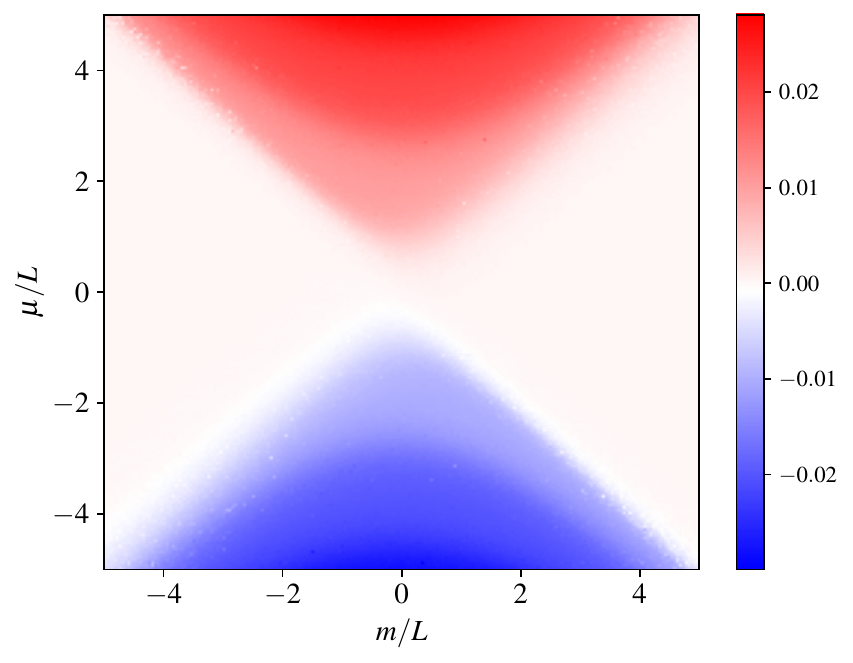}
    \includegraphics[width=0.32\linewidth]{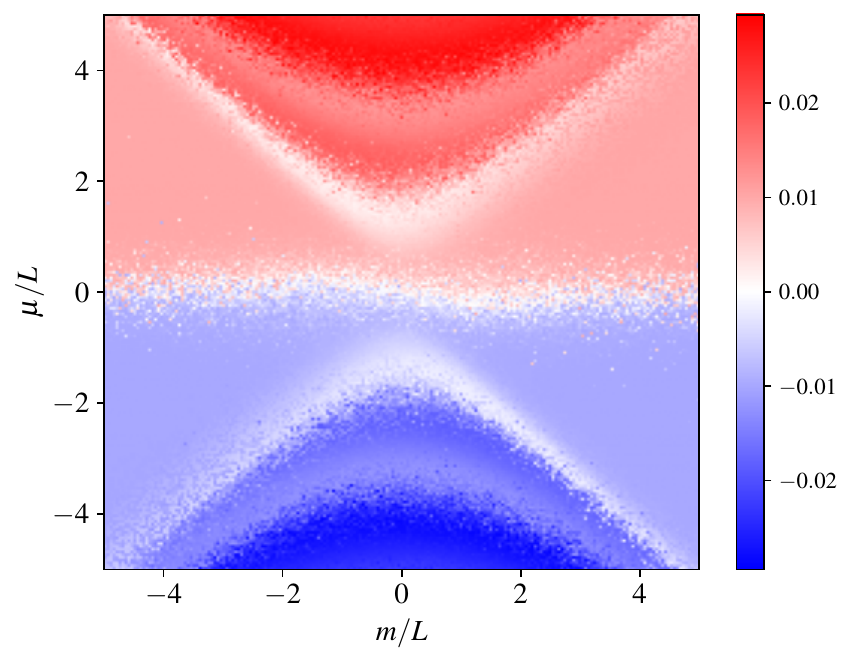}
    \includegraphics[width=0.32\linewidth]{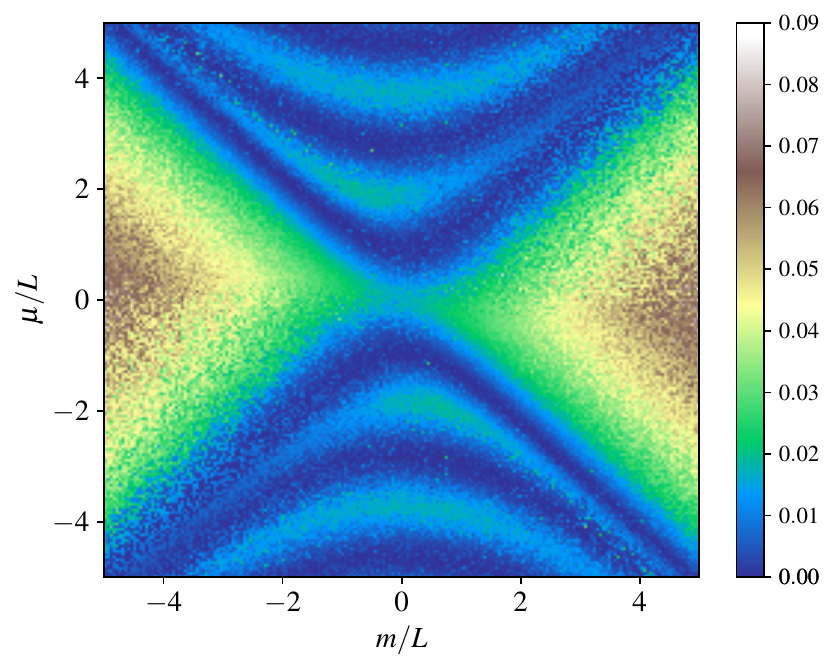}
    \caption{Ground state charge (left), first excited sate charge (middle) and the energy gap (right) with $N=100, r_h=50$.}
    \label{fig:enter-label}
\end{figure}
}\fi

\if{
\subsection{The Qubit Representation Hamiltonian}

The action for a Dirac fermion (two-component complex spinor) with mass $m$ coupled to a chemical potential $\mu$
in two-dimensional Minkowski space with local coordinates $(t,x)$ reads:
\begin{equation}
S = \int dt d x \, \bar{\psi} \left( i \gamma^\mu \partial_\mu - m + \mu \gamma^0 \right) \psi \ ,   \end{equation}
where $\bar{\psi} = \psi^\dagger \gamma^0$.
We will use the two-dimensional representation of the Clifford algebra:
\begin{equation}
\gamma^0 = \sigma_z, \gamma^1 = i \sigma_y, \gamma^5=\gamma^0\gamma^1 = \sigma_x \ ,
\label{gnotation}
\end{equation}
and $(\gamma^0)^{\dagger} = \gamma^0,  (\gamma^1)^{\dagger} = -\gamma^1, 
(\gamma^5)^{\dagger} = \gamma^5$.
The corresponding Hamiltonian is: 
\begin{equation}
 H = \int dx \, \psi^\dagger \left(-i\sigma_x \partial_x + m \sigma_z - \mu \right) \psi \ . 
 \label{HM}
 \end{equation}

To convert the Hamiltonian into the lattice Hamiltonian, we use the staggered fermion $\chi_n$, a single component Grassmann field, at each lattice site $n$~\cite{Kogut:1974ag,Susskind:1976jm}
where $a$ is the lattice size. They satisfy:
\begin{equation}
\{\chi_n,\chi_m\} = 0,~~~~\{\chi_n^{\dagger},\chi_m\} = \delta_{nm} \ .    
\end{equation}
The Hamiltonian (\ref{HM}) is mapped into: 
\begin{equation}
H=i\sum_{n=1}^{N-1}\frac{1}{2a}\left(\chi^\dagger_{n}\chi_{n+1} - \chi^\dagger_{n+1}\chi_n\right) + \sum_{n=1}^N\left(m(-1)^n-\mu\right)\chi_n^{\dagger}\chi_n
\ . 
\label{H2}
\end{equation}

The qubit-representation of the lattice Hamiltonian is obtained by Jordan-Wigner transformation~\cite{Jordan:1928wi}:
\begin{align}
\begin{split}
 \chi_n &= \frac{X_n-i Y_n}{2}\prod_{i=1}^{n-1}(-i Z_i) \ ,
 \\
 \chi^\dag_n &= \frac{X_n+i Y_n}{2}\prod_{i=1}^{n-1}(i Z_i) \ ,
\end{split}
\label{JW}
\end{align}
where $X_n,Y_n,Z_n$ are the Pauli matrices at the $n$-th site.
A dictionary to translate fields into Pauli operators is given in Table 1.
\begin{table}[H]
\begin{center}
\begin{tabular}{c|c|c}\toprule
Dirac Fermion Bilinears& Staggerd  & Pauli \\\hline
     $\overline{\psi}\psi$ & $\frac{(-1)^n}{a}\chi^\dagger_n\chi_n$ &  $\frac{(-1)^n}{2a}(Z_n+1)$ \\
     $\overline{\psi}\gamma_0\psi$ & $\frac{1}{a}\chi^\dagger_n\chi_n$ &  $\frac{1}{2a}(Z_n+(-1)^n)$ \\
     $\overline{\psi}\gamma_1\psi$ & $\frac{1}{2a}(\chi^\dagger_n\chi_{n+1}+\chi^\dagger_{n+1}\chi_{n})$ &  $\frac{1}{4a}(X_nY_{n+1}-Y_nX_{n+1})$ \\
    $\overline{\psi}\gamma_5\psi$ & $\frac{(-1)^n}{2a}(\chi^\dagger_n\chi_{n+1}-\chi^\dagger_{n+1}\chi_{n})$ &  $-\frac{i(-1)^n}{4a}(X_nX_{n+1}+Y_nY_{n+1})$ \\
    $\overline{\psi}\gamma_1\partial_1\psi$ & $-\frac{1}{2a^2}(\chi^\dagger_n\chi_{n+1}-\chi^\dagger_{n+1}\chi_{n})$ &  $-\frac{i}{4a^2}(X_nX_{n+1}+Y_nY_{n+1})$ \\
\end{tabular}
\end{center}
    \caption{The three different representations of the Dirac fermion field.}
    \label{tab:dic}
\end{table}

The Hamiltonian (\ref{H2}) in the Jordan-Wigner representation reads:
\begin{equation}
\label{eq:Ham_flat}
    H=\frac{1}{4a}\sum_{n=1}^{N-1}\left(X_nX_{n+1}+Y_nY_{n+1}\right)+\sum_{n=1}^N\frac{(m(-1)^n-\mu)Z_n}{2} \ ,
\end{equation}
where we take $N$ to be even, and we used Table~\ref{tab:dic} for the mass $M$ and the charge $Q$ operators:
\begin{equation}
    M = \frac{1}{2a}\sum_{n=1}^N (-1)^nZ_n,~~~Q=\frac{1}{2a}\sum_{n=1}^N Z_n \ .
    \label{charge}
\end{equation}
Note that the Hamiltonian (\ref{eq:Ham_flat}) has a symmetry:
$m\rightarrow -m$, since the first two terms in the Hamiltonian are independent of $m$ and $\mu$, while the last terms is invariant in the sense that changing the signs can be undone by a different enumeration of the qubits that does not affect the first two terms.
The set of eigenvalues of the Hamiltonian is also invariant
under the transformation: $\mu\rightarrow -\mu$, since it can be undone by
the mapping $Z_n\rightarrow -Z_n$, which has not affect on them.
The map of the eigenvectors of the Hamiltonian under this transformation is
simply $\ket{0} \leftrightarrow \ket{1}$.
As we will see, these are the underlying reasons for the symmetries of the energy, charge and entanglement properties of the system.

}\fi

\section{\label{sec:discussion}Discussion and Outlook}

In this work, we have established a minimal yet versatile lattice model of Dirac fermions on an $AdS$ black hole background, incorporating key gravitational ingredients—redshift, spin connection, and horizon structure—into qubit-ready Hamiltonians.  Our analysis has traversed spectral properties, entanglement measures, operator scrambling, spectral statistics, and disorder-driven localization, yielding several insights.
(i) Redshift and finite-size effects: The warp-factor weights imprint a spatially varying effective mass and hopping profile, leading to analytic corrections of order $O(\frac{1}{N^2})$  in the energy gap and affecting the transition in entanglement entropy relative to flat space.
(ii) Chiral gravitational effect on the lattice:  The spin connection in the JW-transformed Hamiltonian yields a unidirectional energy current at finite chemical potential — a boundary induced, curvature driven chiral gravitational effect analogue of the two-dimensional gravitational anomaly.
(iii) Operator scrambling without chaos: Horizons and spin-connection couplings enhance OTOC decay rates, yet the quadratic nature of our model precludes exponential Lyapunov growth.  This delineates clearly between kinematic scrambling and true quantum chaos.
(iv) Spectral crossover: Level-spacing ratios and Brody fits reveal a continuous drift from Poisson toward Wigner–Dyson statistics as the horizon enlarges, but saturation below the fully chaotic limit highlights integrability remnants in free theories.
(v) Interplay with disorder:  Intrinsic inhomogeneity from large  cooperates with external random fields to lower the threshold for many-body localization, suggesting gravity-inspired designs for tunable localization platforms.

Our lattice construction and findings open several avenues for further exploration:
(i) Interacting extensions:  Introducing quartic (Hubbard-like) interactions or coupling to a dynamical gauge field will break integrability and may generate genuine quantum chaos, enabling comparisons with SYK/JT predictions for scrambling and thermalization.
(ii) Quantum simulation:  The JW qubit mapping and explicit Hamiltonian terms lend themselves to digital or analog implementations on superconducting, trapped-ion, or cold-atom platforms, where one can directly probe redshift-induced transport and scrambling.
(iii)  Higher dimensions and spin:  Generalizing to higher dimensions, or including multiple spinor components may reveal richer anomaly structures (mixed gauge-gravitational anomalies) and edge-mode phenomena.
(iv) Entanglement dynamics:  Time-dependent studies of entanglement growth following quenches in mass, chemical potential, or horizon radius can shed light on post-quench thermalization and information spreading in curved-space settings.
(v)  Holographic benchmarks.  Comparing our free-fermion lattice results with continuum JT gravity and SYK-derived observables (e.g. spectral form factors, wormhole correlators) will help clarify the minimal ingredients necessary for emergent holographic behavior.

In summary, by bridging continuum gravitational physics and discrete quantum many-body models, our work provides a platform for systematic studies of how curvature, anomalies, and inhomogeneity sculpt quantum matter, paving the way toward engineered quantum simulations of holographic systems.

\section*{Acknowledgement}
This work is supported by the NSF under Grant No. OSI-2328774 (KI),
by the Israeli Science Foundation Excellence Center, the US-Israel Binational Science Foundation, the Israel Ministry of Science (YO).

\if{

\section{$S_N(\beta)$}

We analyze the asymptotic behavior of the sum:
\begin{equation}
S(N, \beta) := \sum_{n=1}^N \frac{1}{\sqrt{n(n + \beta)}} \ ,
\end{equation}
for large \(N\), where \(\beta > 0\) is a fixed parameter.
When $\beta=0$ we have:
\begin{equation}
\sum_{n=1}^N \frac{1}{n}
= \ln N + \gamma + \frac{1}{2N}
- \cdots  \ .
\end{equation}

To estimate the sum for fixed $\beta\neq 0$, we approximate it by an integral:
\begin{equation}
S(N, \beta) \approx \int_1^N \frac{1}{\sqrt{x(x + \beta)}} \, dx \ .
\end{equation}
Recall the identity:
\begin{equation}
\int \frac{1}{\sqrt{x(x + \beta)}} \, dx
= \ln\left( \sqrt{x} + \sqrt{x + \beta} \right) + C \ .
\end{equation}
Applying the bounds, we obtain:
\begin{equation}
\int_1^N \frac{1}{\sqrt{x(x + \beta)}} \, dx
= \ln\left( \sqrt{N} + \sqrt{N + \beta} \right)
- \ln\left( 1 + \sqrt{1 + \beta} \right) \ .
\end{equation}
Thus, the sum satisfies:
\begin{equation}
\sum_{n=1}^N \frac{1}{\sqrt{n(n + \beta)}}
\approx \ln\left( \sqrt{N} + \sqrt{N + \beta} \right)
- \ln\left( 1 + \sqrt{1 + \beta} \right) \ .
\end{equation}

We expand the leading term as \(N \to \infty\). First, write:
\begin{equation}
\sqrt{N + \beta} = \sqrt{N} \left( 1 + \frac{\beta}{2N} + \cdots \right) \ ,
\end{equation}
and 
\begin{equation}
\sqrt{N} + \sqrt{N + \beta} = 2\sqrt{N} + \frac{\beta}{2\sqrt{N}} + \cdots \ .
\end{equation}
Then,
\begin{equation}
\ln\left( \sqrt{N} + \sqrt{N + \beta} \right)
= \ln(2\sqrt{N}) + \frac{\beta}{4N} + O\left( \frac{1}{N^2} \right) \ .
\end{equation}
Therefore, the full asymptotic expansion is:
\begin{equation}
\sum_{n=1}^N \frac{1}{\sqrt{n(n + \beta)}}
= \frac{1}{2} \ln N + \ln 2
- \ln\left( 1 + \sqrt{1 + \beta} \right)
+ \frac{\beta}{4N} + O\left( \frac{1}{N^2} \right) \ .
\end{equation}

Consider next the case where $\beta = l N$, where $l$ is fixed:
\begin{equation}
S(N, \ell) := \sum_{n=1}^N \frac{1}{\sqrt{n(n + \ell N)}} \ .
\end{equation}

Define \(x = \frac{n}{N}\), so that \(n = Nx\) :
\begin{equation}
n(n + \ell N) = N^2 x(x + \ell),
\quad \Rightarrow \quad
\frac{1}{\sqrt{n(n + \ell N)}} = \frac{1}{N \sqrt{x(x + \ell)}} \ .
\end{equation}
Therefore, the sum becomes a Riemann sum:
\begin{equation}
S(N, \ell) = \sum_{n=1}^N \frac{1}{N} \cdot \frac{1}{\sqrt{\frac{n}{N} \left( \frac{n}{N} + \ell \right)}}
= \sum_{n=1}^N f\left( \frac{n}{N} \right) \cdot \frac{1}{N} \ ,
\end{equation}
with
\begin{equation}
f(x) := \frac{1}{\sqrt{x(x+\ell)}} \ .
\end{equation}
In the large \(N\) limit, the sum approaches an integral:
\begin{equation}
\sum_{n=1}^N \frac{1}{\sqrt{n(n + \ell N)}}
\approx \int_0^1 \frac{1}{\sqrt{x(x+\ell)}} \, dx \ .
\end{equation}
This integral reads:
\begin{equation}
\int_0^1 \frac{1}{\sqrt{x(x+\ell)}} \, dx
= \frac{2}{\ell} \sinh^{-1}\left( \frac{1}{\sqrt{\ell}} \right)
= \frac{2}{\ell} \ln\left( \frac{1 + \sqrt{1 + \ell}}{\sqrt{\ell}} \right) \ .
\end{equation}

Because \(f(x)\) has a square-root singularity at \(x = 0\), the Euler--Maclaurin expansion includes half-integer powers of \(1/N\). The asymptotic expansion form reads (Wong, Bleistein–Handelsman, or Bender–Orszag):
\begin{equation}
S(N, \ell) = \int_0^1 \frac{1}{\sqrt{x(x + \ell)}}\,dx + \sum_{k=1}^\infty \frac{c_k(\ell)}{N^{k - 1/2}} \  .
\end{equation}
Using:
\begin{equation}
c_1(\ell) = \frac{\zeta\left(\frac{1}{2}\right)}{\sqrt{\ell}} \approx -\frac{1.46035}{\sqrt{\ell}},
\end{equation}
\begin{equation}
c_2(\ell) = -\frac{\zeta\left(\frac{3}{2}\right)}{8 \ell^{3/2}} \approx -\frac{0.16587}{\ell^{3/2}} \ ,
\end{equation}
we get
the large-\(N\) asymptotic expansion:
\begin{equation}
S(N, \ell)
= \frac{2}{\ell} \sinh^{-1}\left( \frac{1}{\sqrt{\ell}} \right)
- \frac{1.46035}{\sqrt{\ell N}}
- \frac{0.16587}{\ell^{3/2} N^{3/2}}
+ O\left( \frac{1}{N^{5/2}} \right) \ .
\end{equation}

A second method.

Let \(\ell > 0\) be fixed. We consider the sum:
\begin{equation}
S(N, \ell) := \sum_{n=1}^N \frac{1}{n(n + \ell N)} \ .
\end{equation}
We write:
\begin{equation}
\frac{1}{n(n + \ell N)} = \frac{1}{\ell N} \left( \frac{1}{n} - \frac{1}{n + \ell N} \right) \ .
\end{equation}
Therefore,
\begin{equation}
S(N, \ell) = \frac{1}{\ell N} \sum_{n=1}^N \left( \frac{1}{n} - \frac{1}{n + \ell N} \right)
= \frac{1}{\ell N} \left( H_N - H_{N + \ell N} + H_{\ell N} \right) \ ,
\end{equation}
where \( H_k = \sum_{n=1}^k \frac{1}{n} \) is the \(k\)-th harmonic number.
The harmonic number has the asymptotic expansion:
\begin{equation}
H_k = \ln k + \gamma + \frac{1}{2k} - \frac{1}{12k^2} + \cdots \ ,
\end{equation}
where \(\gamma\) is the Euler–Mascheroni constant.
Apply this to:
\[
H_N,\quad H_{\ell N}, \quad H_{(\ell + 1)N},
\]
and compute:
\begin{align}
H_N - H_{(\ell + 1)N} + H_{\ell N}
&= \ln\left( \frac{\ell}{\ell + 1} \right)
+ \frac{1}{2N} \left( 1 + \frac{1}{\ell} - \frac{1}{\ell + 1} \right)
+ O\left( \frac{1}{N^2} \right) \ .
\end{align}
Substitute back into \(S(N, \ell)\), we obtain:
\begin{equation}
\sum_{n=1}^N \frac{1}{n(n + \ell N)}
= \frac{1}{\ell N} \ln\left( \frac{\ell}{\ell + 1} \right)
+ \frac{1}{2 \ell N^2} \left( 1 + \frac{1}{\ell} - \frac{1}{\ell + 1} \right) 
+ O\left( \frac{1}{N^3} \right) \ .
\end{equation}
Note that this formula diverges as $\ell \to 0^+$, and we need to take care of $\ell=0$ separately.

When only the mass term is considered, the ground state obeys this formula. Fig. \ref{fig:mass-term_r_is-0} shows the ground state energy per mass $E_0/m$ as a function of $N$.  
\begin{figure}[H]
    \centering
    \includegraphics[width=0.5\linewidth]{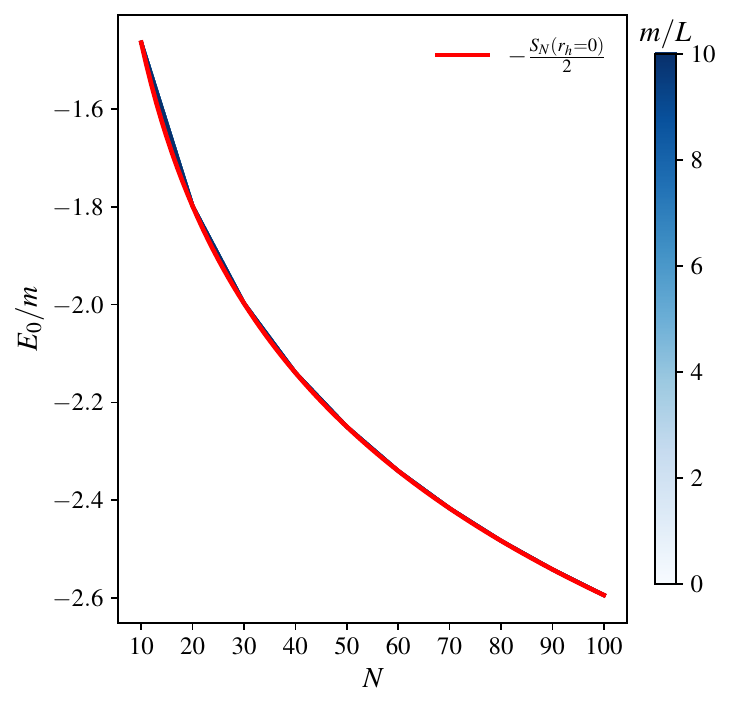}
    \caption{Mass term v.s. $N$-th harmonic series}
    \label{fig:mass-term_r_is-0}
\end{figure}

\begin{figure}[H]
    \centering
    \includegraphics[width=0.5\linewidth]{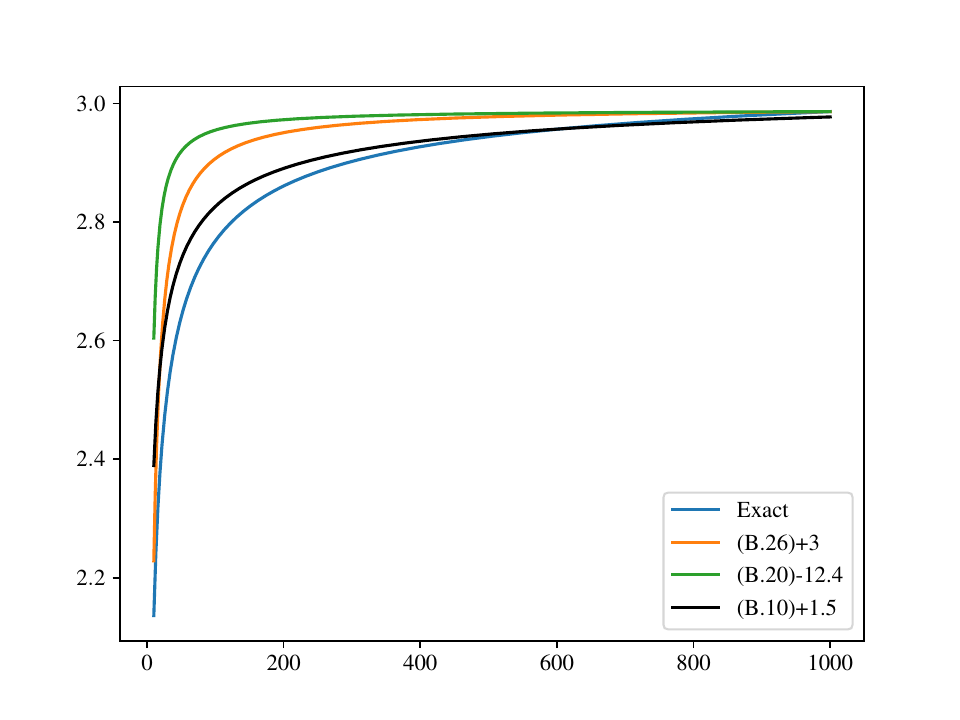}
    \caption{Comparison of asymptotic forms with some constants. \ki{is there any missing constants?}}
    \label{fig:enter-label}
\end{figure}

}\fi
\bibliographystyle{JHEP}
\bibliography{ref}

\end{document}